\documentclass{aastex62}

\usepackage{subfigure}
\usepackage{gensymb}
\usepackage{amssymb}
\usepackage{amsmath}
\usepackage{multirow}
\usepackage{bigdelim}

\received{September 10, 2018}
\revised{November 11, 2018}
\accepted{\today}
\submitjournal{ApJ}

\shorttitle{Limits of the Primitive Equations}
\shortauthors{Mayne et al.}

\begin{document}

\title{The Limits of the Primitive Equations of Dynamics for Warm,
  Slowly Rotating Small Neptunes and Super Earths.}

\correspondingauthor{Nathan Mayne}
\email{n.j.mayne@exeter.ac.uk}

\author[0000-0001-6707-4563]{N. J. Mayne}
\author{B. Drummond}
\affil{Astrophysics Group, \\
University of Exeter, \\
Exeter EX4 2QL, UK}

\author{F. Debras}

\author{E. Jaupart}
\affil{Ecole normale sup\'erieure de Lyon, \\
CRAL, UMR CNRS 5574, \\
69364 Lyon Cedex 07,  France}

\author{J. Manners}
\author{I. A. Boutle}
\affil{Met Office, \\
Exeter EX1 3PB, UK}

\author{I. Baraffe}
\affil{Ecole normale sup\'erieure de Lyon, \\
CRAL, UMR CNRS 5574, \\
69364 Lyon Cedex 07,  France}
\affil{Astrophysics Group, \\
University of Exeter, \\
Exeter EX4 2QL, UK}

\author{K. Kohary}
\affil{Astrophysics Group, \\
University of Exeter, \\
Exeter EX4 2QL, UK}



\begin{abstract}
  We present significant differences in the simulated atmospheric flow
  for warm, tidally--locked small Neptunes and super Earths (based on a nominal GJ~1214b) when solving the simplified, and commonly used,
  primitive dynamical equations or the full Navier--Stokes
  equations. The dominant prograde, superrotating zonal jet is
  markedly different between the simulations which are performed using
  practically identical numerical setups, within the same model. The differences arise due to the breakdown of the so--called `shallow--fluid' and traditional approximations, which worsens when rotation rates are slowed, and day--night temperature contrasts are increased. The changes in the zonal
  advection between simulations solving the full and simplified equations, give rise to significant differences in the atmospheric redistribution
  of heat, altering the position of the hottest part of the atmosphere
  and temperature contrast between the day and night sides. The
  implications for the atmospheric chemistry and, therefore,
  observations need to be studied with a model including a more
  detailed treatment of the radiative transfer and chemistry. 
  Small Neptunes and super Earths are extremely abundant and important, potentially
  bridging the structural properties (mass, radius, composition) of terrestrial and gas giant planets. Our
  results indicate care is required when interpreting the output of
  models solving the primitive equations of motion for such planets.
  \end{abstract}



\section{Introduction} \label{sec:intro}


The detection of exoplanets has revealed several classes of object
without direct analogues in our Solar system. One important example,
termed hot Jupiters, Jovian type planets in short period orbits
provide the strongest atmospheric signatures via observations such as
emission spectra \citep[e.g.,][]{todorov_2014}, transmission spectra
\citep[e.g.,][]{sing_2011} and emission as a function of orbital
phase, or phase curves \citep[e.g.,][]{knutson_2012}. However, these
same techniques are beginning to be applied to a second class of
planet, those between the size of Neptune and Earth, termed Super
Earths and small Neptunes. Importantly, this class of planet has been
shown to be the most abundant, when adopting the distinctions of
\citet{fressin_2013}, namely 1.25$R_\earth<R_{\rm p}<$4$R_{\earth}$,
where $R_{\earth}$ \& $R_{\rm p}$ are the Earth and planetary radii,
respectively. In addition, the Next Generation Transit Survey
\citep[NGTS,][]{wheatley_2013}, CHaracterising ExoPlanet Satellite
\citep[CHEOPS,][]{broeg_2013} and Transiting Exoplanet Survey
Satellite \citep[TESS,][]{rickler_2014} will significantly increase
the number of detections, and availability of characterisation
observations, in this size range in the near future. For example, the
recent detection of $\pi$ Mensae c (HD~39091c) using data from TESS
\citep{huang_2018}.


Aside from the abundance of potential targets, Super Earths and small
Neptunes also inhabit a potentially critical region of planetary
parameter space. These planets likely bracket the point at which
runaway accretion of a primary gaseous atmosphere occurs in the core
accretion model \citep{pollack_1996}. Therefore, they bridge the
structures of giant planets with thick hydrogen/helium dominated
atmospheres, to terrestrial planets with much thinner `secondary'
atmospheres \citep{lopez_2014}, as well as being in the size range
where irradiative evaporation becomes significant
\citep{owen_2012}. Finally, as these planets are generally cooler than
their hot Jupiter counterparts, the characteristic timescale to reach
chemical equilibrium for many species increases, leading to stronger
potential effects from chemical kinetics driven by transport and
photochemistry
\citep[see][for review]{madhusudhan_2016}.


Observations of several targets in this regime have demonstrated the
complex nature of their atmospheres. The Super Earth GJ~1214b
(equilibrium temperature $\sim$500\,K), has been intensively observed
returning a flat, featureless spectrum via both ground-based
\citep{bean_2010,bean_2011,crossfield_2011,demooij_2012,wilson_2014,caceres_2014}
and space-based instruments
\citep{desert_2011,berta_2012,fraine_2013,kreidberg_2014}. \citet{kreidberg_2014}
effectively ruled out a cloud--free atmosphere, but did so without
constraint on the bulk composition (i.e. metallicity) of the
gas--phase. Additionally, for the `warm Neptune' GJ~436b
\citep[slightly larger than the small Neptune limit at
$\sim$4.3$R_{\earth}$,][]{deming_2007} observations indicate a
condensate--rich atmosphere \citep{knutson_2014}.

The main focus of the theoretical modeling of exoplanets has been
either the well observed hot Jupiter objects, or potentially habitable
terrestrial planets. For terrestrial planets 3D General Circulation
Models (GCMs) adapted from those used to study Earth, have been
applied to explore the potential climates of, for example, Proxima
Centauri b \citep{turbet_2016, boutle_2017}, the response of an
Earth--like climate to weakened and intensified stellar irradiation
\citep[][respectively]{leconte_2013,charnay_2013}, alongside studies
including surface effects \citep{lewis_2018} and a dynamical ocean
\citep[e.g.,][]{wolf_2017,delgenio_2017}. For hot Jupiters, gas--phase
chemical equilibrium simulations have been performed, in 3D using
GCMs, across a range of targets
\citep[e.g.,][]{kataria_2016}. Treatments of the chemical kinetics
have also been included using simplified ``relaxation methods'' in 3D
\citep{cooper_2006,drummond_2018b,drummond_2018c} or more complete chemical networks
in pseudo-2D \citep{agundez_2012,agundez_2014}. Finally, the treatment
of clouds has been added both `diagnostically' \citep[where feedbacks
of the cloud presence on the atmosphere are
neglected,][]{parmentier_2016,helling_2016}, and `prognostically'
\citep[where the clouds evolve and interact with the
atmosphere,][]{lee_2016,lines_2018,lines_2018b}.

The atmosphere of the super Earth GJ 1214b is one of the most
extensively simulated atmospheres, using a range of different 3D
models \citep{menou_2012,kataria_2014,charnay_2015,drummond_2018} often
combining radiative transfer calculations with equilibrium
chemistry. \citet{zhang_2017} took a more simplified modeling approach
but explored the effect of a wide--range of bulk compositions. \citet{charnay_2015b} included a simplified treatment of condensation
of KCl and ZnS clouds.

Considering only the solution to the dynamical equations, much
progress has been made by GCMs solving the simplified or ``primitive''
equations of motion \citep[invoking hydrostatic balance, the
assumption of a thin or shallow atmosphere and a gravity constant with
height, see discussion in][]{mayne_2014}, such as derivatives of the
MIT--GCM \citep[with the main hot Jupiter adaptations presented
in][]{showman_2009} and the LMD$z$ \citep[used, for example,
in][]{charnay_2015b}. Concerns over the validity of the ``primitive''
equations for ``thick'' atmospheres, where the atmospheric scale
height becomes significant compared to the assumed total planetary
radius have been raised for Titan and Venus
\citep[see][]{tokano_2013}. Additionally, work has been done assessing
the limits of the primitive equations for Earth \citep[see][and
references therein]{tort_2015}, and the traditional approximation
specifically \citep{gerkema_2008}. Several 3D models have been applied
to exoplanets which solve the full Navier--Stokes equations, such as
the 3D radiation--hydrodynamics model of
\citet{dobbs_dixon_2013}\footnote{This code explicitly solves for the
  Centrifugal component, which is usually included in the
  gravitational term for GCMs \citep{showman_2008}}, the recently
developed THOR dynamical core \citep{mendonca_2016}, and of course our
own work \citep[e.g.,][]{mayne_2014b}. However, significant
differences between simulations of exoplanets using ``primitive'' and
more complete dynamical cores has yet to be found.

In our own work we have applied the Met Office Unified Model (or {\sc
  UM}) to exoplanets \citep[first in][]{mayne_2014}, which is able to
solve the dynamical equations adopting various levels of
simplification within the same numerical framework. In
\citet{mayne_2014b} we applied this capability to hot Jupiters,
finding differences in the evolution of the deep atmosphere between
the primitive and more complete equation set, but not in the
qualitative dynamics of the upper, observable atmosphere.

In this paper we continue our work using the {\sc UM} and investigate
the effect of the simplifications made to the dynamical equations in
the Super Earth or Small Neptune regime, for example, GJ~1214b, or
$\pi$ Mensae c (HD~39091c) the latter being recently detected using
the Transiting Exoplanet Survey Satellite (TESS)
\citep{huang_2018}. Focusing our work on GJ~1214b we find significant
differences in the resulting dynamical structure of the simulated
atmosphere between the ``primitive'' and more complete equations. This
is caused by the breakdown of the `shallow--fluid' and traditional
approximations, similar to that found for a terrestrial planet by
\citet{tort_2015}. The rest of this paper is structured as follows: in
Section \ref{sec:model} we introduce the model we are using, summarise
its evolution and describe the setups and parameters adopted. In
Section \ref{sec:results} we present results from the simulations,
starting with a ``standard'' setup for GJ~1214b (Section
\ref{subsec:standard_case}), followed by an exploration of the limits
of the primitive equations (Section \ref{subsec:prim_limits}, where
Section \ref{subsubsec:trad_approx} details the derivation of order of
magnitude estimates for the wind speeds, which are verified for our
simulations in Appendix \ref{app:equations}). We then discuss the
limitations of our work in Section \ref{sec:assumptions}, and conclude
in Section \ref{sec:conclusions}.

\section{Model Description} \label{sec:model}

For this study we use the Met Office {\sc Unified
  Model} (UM) which we have adapted to enable modelling of a variety
of planets and exoplanets. Initial adaptations and the implementation
of a Newtonian relaxation, or temperature forcing scheme where the
temperature is relaxed to a prescribed radiative equilibrium
temperature over a specified timescale, are detailed in
\citet{mayne_2014}, followed by adaptations for gas giants or planets
with extended atmospheres in \citet{mayne_2014b}. The radiative
transfer component was adapted in \citet{amundsen_2014}, and further
in \citet{amundsen_2017}, with simulations including full radiative
transfer used in \citet{helling_2016} and \citet{amundsen_2016}. The
model chemistry options include an analytical chemical equilibrium
scheme \citep{burrows_1999,amundsen_2016} as well as both a Gibbs
energy minimisation scheme \citep{drummond_2018} and chemical
relaxation scheme \citep{drummond_2018b,drummond_2018c}. We have also recently
coupled a chemical kinetics scheme \citep[originally developed in
1D][]{tremblin_2015,drummond_2016} to the UM which will be the focus
of a future work. The same model has been coupled to a `prognostic'
cloud scheme \citep{lines_2018,lines_2018b}, and used
to explore the evolution of the deep atmosphere of hot Jupiters,
combined with a 2D code \citep{tremblin_2017}, alongside exploration
of the dynamical acceleration mechanisms \citep{mayne_2017}. Finally, we have also adapted the surface schemes to model terrestrial exoplanets \citep{boutle_2017,lewis_2018}.

In this study we adopt a temperature forcing scheme
\citep[implementation described in][]{mayne_2014} where the
temperature is forced to a prescribed equilibrium temperature,
$T_{\rm eq}$, over a parameterised timescale, $\tau_{\rm rad}$. We
adopt the $\tau_{\rm rad}$ and $T_{\rm eq}$ used by \citet[their
Eqn. 8 and 9, respectively]{zhang_2017}, where $T_{\rm eq}$ is a
function of longitude, latitude, and pressure, and $\tau_{\rm rad}$ a
function of pressure only. The dayside--nightside temperature contrast
is controlled by the parameter $\Delta T_{\rm eq}$, also a function of
pressure ($p$), which \citet{zhang_2017} set at 600\,K at the top of
the atmosphere, decreasing linearly with $ln\,p$ toward zero at the
bottom of the atmosphere. In our height--based model we set
$\Delta T_{\rm eq}=600$\,K where $p\leq 10$\,Pa and decrease linearly
with $ln\,p$ to $\Delta T_{\rm eq}=0$\,K where
$p\geq 200\times 10^5$\,Pa. In Appendix \ref{app_sec:radiative} we
show that our conclusions are independent of the choice of this
temperature forcing scheme and hold when using a more complete radiative transfer
scheme. The planet is effectively modeled as a gas giant, as the inner
boundary does not include a surface treatment of the land or ocean and
radiative absorption/emission. However, the inclusion of a terrestrial
surface, in the case of vertically extended atmospheres, will not
likely change our main results. As in the study of \citet{mayne_2014b}
no `drag' is applied near the bottom boundary, as it is not required
for stability and if present (due to, for example, magnetic drag in the case of a gas giant or surface friction for a terrestrial planet) the
form and magnitude are poorly constrained.

The main equations solved by the {\sc UM} include the conservation of
momentum in each of the three directions, longitude, latitude and
height, coordinates $\lambda$, $\phi$ and $r$, respectively, with the
corresponding wind coordinates $u$ (zonal), $v$ (meridional) and $w$
(vertical), respectively. Additionally, the thermodynamic equation is
solved adopting potential temperature, $\theta$ and the Exner
Function, $\Pi$, where
$\Pi=\left(\frac{p}{p_0}\right)^{R/c_{p}} = \frac{T}{\theta}$ and
$c_p$ is the specific heat capacity, $R$ is the specific gas constant,
$p_0$ is a chosen reference pressure and $T$ the normal
temperature in Kelvin. This equation set is then closed via the ideal gas
equation, and heating injected into the system using the Newtonian
relaxation or temperature forcing scheme via the energy equation
\citep[see, for example][]{mayne_2014,mayne_2014b}. The equations are
detailed in full in several works, in particular
\citet{mayne_2014b}. As described in \citet{mayne_2014b} and
\citet{mayne_2017} a vertical `sponge' layer is used to damp vertical
velocities near the upper, low pressure boundary, and a diffusion
scheme is applied to remove grid--scale noise and aid numerical
stability. SI units are used throughout the code, and adopted
throughout this manuscript aside from when explicitly stated.

The {\sc UM}, as applied to exoplanets in \citet{mayne_2014b}, is
capable of solving varying levels of the dynamical equations within
the same model framework, from the most simplified `primitive'
equations which assume the atmosphere is in vertical hydrostatic
equilibrium, is a `shallow--fluid' and gravity is constant with
height, to the `full' equations, which do not make these same
assumptions. For this work, we adopt the nomenclature of
\citet{mayne_2014b}, i.e. ``primitive'' (most simplified) and ``full''
(least simplified), and refer the reader to \citet{mayne_2014b} (and
Section \ref{subsec:prim_limits}) for the detail of the exact terms
within the equations used. However, as discussed in
\citet{mayne_2014b}, the ``full'' equations still include
approximations. Notably the assumption of a spherical geoid, which is
valid for small values of $R_{\rm p}\Omega^2/g$, where $R_{\rm p}$ and
$\Omega$ are the planetary radius and rotation rate, respectively
\citep{benard_2014}. For our simulations of large radii,
slow--rotating planets, this parameter is small (see Section
\ref{subsec:prim_limits}). However, on Earth it can be larger and
errors associated with this approximation can be comparable to those
associated with the `shallow--fluid' and traditional approximations
\citep{benard_2014}.

Model parameters for our standard simulation are shown in Table \ref{basic_par} where the planet specific parameters correspond to those for GJ~1214b \citet{carter_2011}. The parameters relating to numerical settings are the same as those used in previous studies using the same model \citep[see][]{mayne_2014b,amundsen_2016,mayne_2017}. We note that while our model setup specifically corresponds to GJ~1214b we expect our results to be representative of other highly--irradiated, Super Earth/Warm Neptune atmospheres with similar properties.

In addition to the standard simulation of GJ~1214b, we design eight additional setups that investigate the effect of varying parameters such as the planet radius ($R_{\rm p}$) and rotation rate ($\Omega$), amongst others; the motivation for each of these setups will be described in a later section.  Each setup is described in  Table \ref{model_names}, where we show the parameters that are adjusted from the standard ones, shown in Table \ref{basic_par}. We also give each simulation setup a short name, which we use to refer to the setups in the text. For each setup we present simulations using both the full and primitive equation sets, while for one setup (dT$+$) we also present a simulation using the deep equation set, giving a total of 15 simulations in this work.

Each simulation is initialised at rest and in hydrostatic balance, with a zonally and meridionally homogeneous temperature--pressure profile. The substellar point is set at 180$^{\circ}$ longitude. For the initial temperature profile we use the mean temperature ($T_0$) profile from \citet[see their Fig. 2]{zhang_2017}. Each simulation was run for 1000\,days (throughout this work days refers to Earth days). For all simulations, the maximum zonal wind velocity and the mean zonal wind structure has ceased to evolve after a few hundred days, reaching a pseudo--steady state. The deep, high pressure atmosphere continues to slowly evolve after 1000\,days, however as shown for hot Jupiters by \citet{mayne_2017}, this does not appear to have a significant effect on the flow in the lower pressure regions. 

The pressure--altitude structure will vary between the full and primitive (or deep) equation sets due to the assumption of a gravity that is constant with height in the latter. In addition, the pressure--altitude structure also depends on various model parameters (e.g. $R$, $\Delta T_{\rm eq}$). Since our model is height--based, we adjust the top--of--atmosphere height ($z_{\rm top}$) to achieve a similar pressure range between each simulation. Each simulation encompasses the approximate pressure range $\sim$200$\times10^5$\,Pa to $\sim10$\,Pa. This means that the vertical resolution in {\it height} will be slightly different between different simulations though the vertical resolution in {\it pressure} will be approximately the same. However, these differences are typically small
($\sim$20\%), and have no effect on our conclusions (see test and discussion in Section \ref{subsubsec:trad_approx}). The value of $z_{\rm top}$ used for each simulation is shown in Table \ref{model_names}.

\begin{table*}
  \caption{Values of the standard parameters for simulations of GJ~1214b.}
\label{basic_par}
\centering
\begin{tabular}{lc}
  \tableline
  \tableline
  Quantity&Value\\
  \tableline
  Horizontal resolution, $N_{\lambda}$, $N_{\phi}$&$144_\lambda$, $90_\phi$\\
  Vertical resolution, $N_z$&66\\
  Dynamical Timestep, $\Delta t$ (s)&120\\
  Initial inner boundary pressure, $p_{\rm max}$ (Pa)&$200\times 10^5$\\
  Rotation rate, $\Omega$ (s$^{-1}$)&$4.60\times 10^{-5}$\\
  Radius, $R_{\rm p}$ (m)&$1.45\times10^7$\\
  Atmospheric height $z_{\rm top}$ (m)&$3.7\times 10^6$\\
  Surface gravity, $g_{\rm p}$ (ms$^{-2}$)&12.20\\
  Specific heat capacity (constant pressure), $c_{\rm p}$ (Jkg$^{-1}$K$^{-1}$)&12\,300\\
  Ideal gas constant, $R$ (Jkg$^{-1}$K$^{-1}$)&3573.5\\
  Temperature contrast, $\Delta T_{\rm eq}$ (K)&600\\
  Diffusion setting, $K_\lambda$ \citep[see][for details]{mayne_2017}&0.158\\
  Vertical, `sponge', damping coefficient $R_{w}$&0.15\\
  \tableline
\end{tabular}
\end{table*}

\begin{table*}
  \caption{Short names of the simulations presented in this work of
    GJ~1214b, with variations in parameters from those shown in Table
    \ref{basic_par}, atmospheric height and the dynamical equations set used
    \citep[see][for explanation and detail]{mayne_2014b}.}
\label{model_names}
\centering
\begin{tabular}{lllll}
  \tableline\tableline
  Short Name&\multicolumn{2}{l}{Adjusted parameters}&$z_{\rm top}$ (m)&Equation set\\
  \tableline
  Std Full&\multicolumn{2}{l}{-}&$3.70\times 10^6$&Full\\ 
  Std Prim&\multicolumn{2}{l}{-}&$3.00\times 10^6$&Primitive\\ 
  Std Hires Full&\multicolumn{2}{l}{$N_{\lambda}=288$, $N_{\phi}=180$, $N_{\rm z}=132$, $\Delta t=60$\,s}&$3.70\times 10^6$&Full\\ 
  Std Hires Prim&\multicolumn{2}{l}{$N_{\lambda}=288$, $N_{\phi}=180$, $N_{\rm z}=132$, $\Delta t=60$\,s}&$3.00\times 10^6$&Primitive\\ 
  $R_{\rm p}-$ Full&\multicolumn{2}{l}{R$_{\rm p}=6.0\times 10^6$\,m, $\Delta t=80$\,s}&$5.50\times 10^6$&Full\\ 
  $R_{\rm p}-$ Prim&\multicolumn{2}{l}{R$_{\rm p}=6.0\times 10^6$\,m, $\Delta t=80$\,s}&$3.00\times 10^6$&Primitive\\ 
  $R_{\rm p}+$ Full&\multicolumn{2}{l}{R$_{\rm p}=1.0\times 10^8$\,m}&$3.00\times 10^6$&Full\\ 
  $R_{\rm p}+$ Prim&\multicolumn{2}{l}{R$_{\rm p}=1.0\times 10^8$\,m}&$2.80\times 10^6$&Primitive\\ 
  CO$_2$ Full&\multicolumn{2}{l}{$c_{\rm p}=10\,123$\,Jkg$^{-1}$K$^{-1}$, $R=188.9$,Jkg$^{-1}$K$^{-1}$, $\Delta t=60$\,s}&$1.50\times 10^5$&Full\\ 
  CO$_2$ Prim&\multicolumn{2}{l}{$c_{\rm p}=10\,123$\,Jkg$^{-1}$K$^{-1}$, $R=188.9$\,Jkg$^{-1}$K$^{-1}$, $\Delta t=60$\,s}&$1.50\times 10^5$&Primitive\\ 
  $\Omega+$ Full&\multicolumn{2}{l}{$\Omega=9.2\times 10^{-5}$\,s$^{-1}$}&$3.70\times 10^6$&Full\\ 
  $\Omega+$ Prim&\multicolumn{2}{l}{$\Omega=9.2\times 10^{-5}$\,s$^{-1}$}&$3.00\times 10^6$&Primitive\\ 
  dT$+$ Full&\multicolumn{2}{l}{$\Delta T_{\rm eq}$ 800\,K}&$3.30\times 10^6$&Full\\ 
  dT$+$ Deep&\multicolumn{2}{l}{$\Delta T_{\rm eq}$ 800\,K, $g(r)=g_{\rm p}$}&$2.65\times 10^6$&Deep\\ 
  dT$+$ Prim&\multicolumn{2}{l}{$\Delta T_{\rm eq}$ 800\,K}&$2.65\times 10^6$&Primitive\\ 
  \tableline
\end{tabular}
\end{table*}

\section{Results} \label{sec:results}

In this section, we first present results from our baseline or standard  simulations (Std, see Table \ref{model_names}), with parameters matching those derived from observations of GJ~1214b (Section \ref{subsec:standard_case}). These simulations demonstrate a clear difference in the resolved atmospheric structure between the more simplified primitive equations and their more complete counterparts. Therefore, we follow this with an exploration of the fundamental limits of the primitive equations, and a demonstration of the effects of exceeding these limits on the simulated flows using our remaining simulation set (Section \ref{subsec:prim_limits}).

The majority of simulations of tidally--locked atmospheres return a
prograde, superrotating equatorial jet i.e. a coherent zonal flow in
the direction of the planetary rotation peaking in speed towards low latitudes
\citep[see discussion in][]{showman_2011,tsai_2014,mayne_2017}. Therefore, our analysis of the bulk dynamical structure of the atmosphere is performed via comparison of the zonal--mean, temporal--mean, zonal
wind as a function of latitude and pressure \citep[using
linear interpolation to convert quantities from a height to pressure
surfaces, as in][]{mayne_2014b, mayne_2017}. For tidally-locked planets in close orbits, with extended atmospheres,
the thermal structure in the upper, low pressure region of the
atmosphere is dominated by the radiative forcing. In the temperature
forcing setup, this is the regime of a very short radiative timescale,
where the temperature is rapidly relaxed to the equilibrium
temperature \citep{iro_2005}. As the radiative timescale increases
with pressure the advection can begin to alter the temperature
structure, and drive it from radiative equilibrium \citep[see for
example,][]{showman_2009,mayne_2014b,zhang_2017}. Therefore, changes
in the advection between simulations will lead
to effects on the thermal structure effectively weighted by the depth
at which the flow is occurring. Basically, faster flows from the day
to night side will lead to stronger homogenisation of the zonal
temperature structure which, in turn, will have an increasing effect
with increasing pressure.

To explore differences in the temperature structure between simulations we follow the approach of \citet{zhang_2017} and present simple, normalised thermal phase curves. This is simply the blackbody thermal emission from an isobaric surface integrated over the observable hemisphere as a function of phase \citep[see Eq. (18) of][]{zhang_2017}. The phase curve for a given pressure level is then normalised by \citep[][their Eq. (18)]{zhang_2017},
\begin{equation}
  \label{equation:norm_phase_curve}
    \bar{F}\left(\delta\right) = \frac{F\left(\delta\right)-\langle F\left(\delta\right)\rangle}{\langle F\left(\delta\right)\rangle},
\end{equation}
where $\bar{F}\left(\delta\right)$ is the normalised phase curve, $F\left(\delta\right)$ is the emission flux as a function of phase ($\delta$) and $\langle F\left(\delta\right)\rangle$ is the mean of the flux over the entire phase. We note that the normalised phase curve is negative where $F\left(\delta\right)<\langle F\left(\delta\right)\rangle$.

We use these simple phase curves as a tool to explore trends in the dayside--nightside temperature contrast as well as the longitudinal offset of the hot spot.  We stress that these phase curves are not intended to represent the real emission from the planet, which would require a full radiative transfer approach to capture the pressure and wavelength dependence of the emission flux.

\subsection{Standard Case} \label{subsec:standard_case}

\begin{figure*}
\begin{center}
  \subfigure[Std Prim: 800-1\,000\,days]{\includegraphics[width=8.5cm,angle=0.0,origin=c]{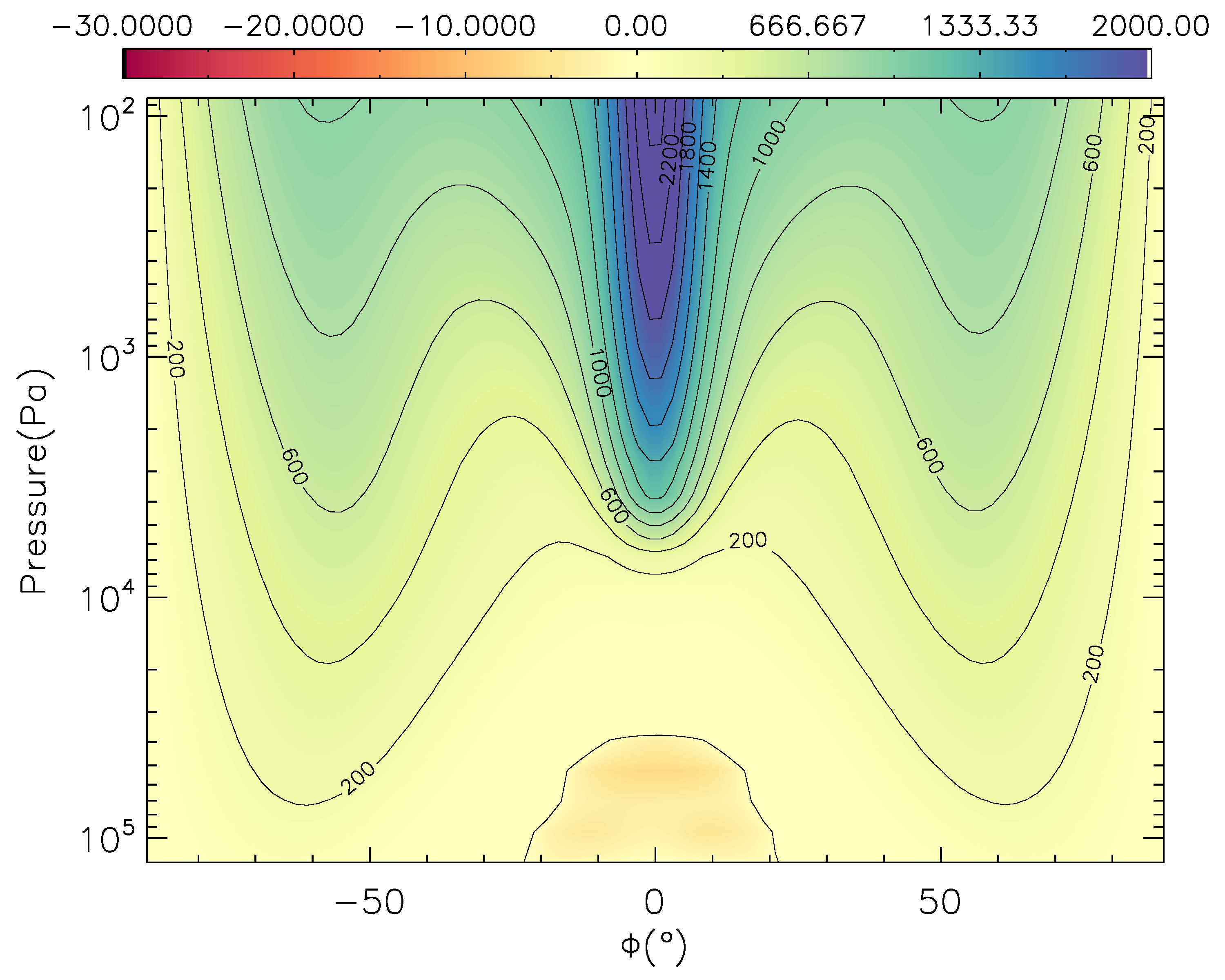}\label{std_prim_800_1000_uvel_bar}}
  \subfigure[Std Full: 800-1\,000\,days]{\includegraphics[width=8.5cm,angle=0.0,origin=c]{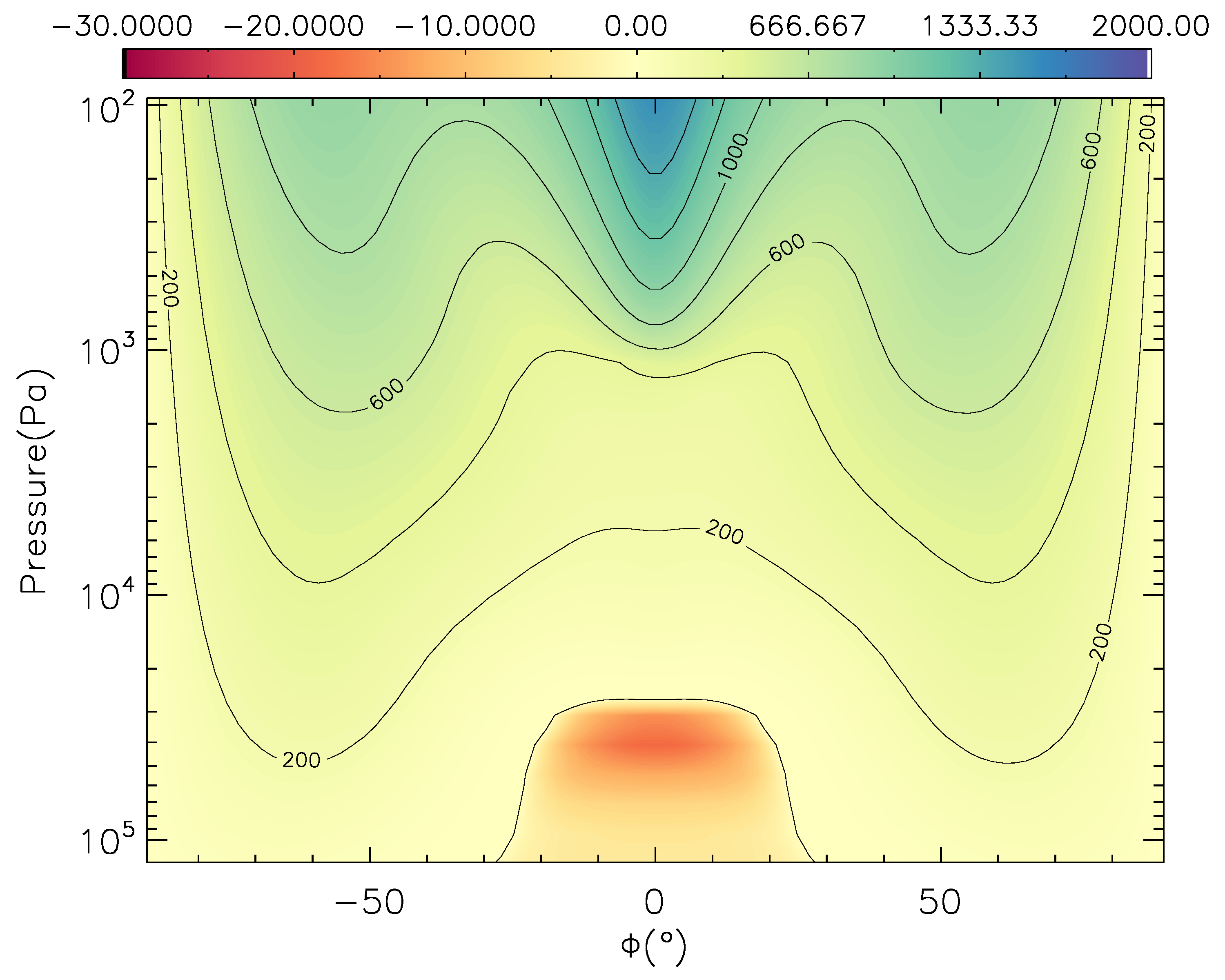}\label{std_full_800_1000_uvel_bar}}
  \subfigure[Std Hires Prim: 800-1\,000\,days]{\includegraphics[width=8.5cm,angle=0.0,origin=c]{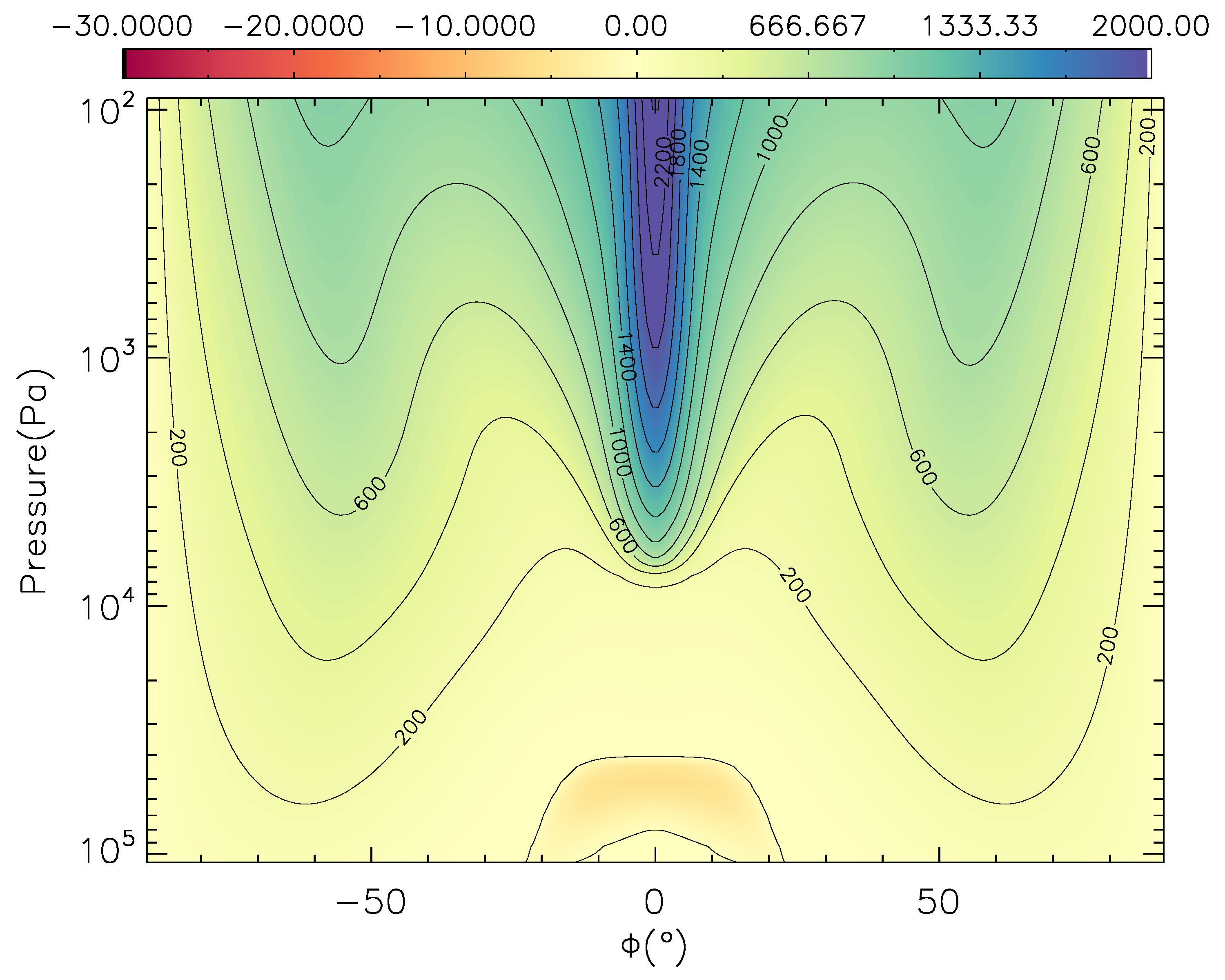}\label{hires_prim_800_1000_uvel_bar}}
  \subfigure[Std Hires Full: 800-1\,000\,days]{\includegraphics[width=8.5cm,angle=0.0,origin=c]{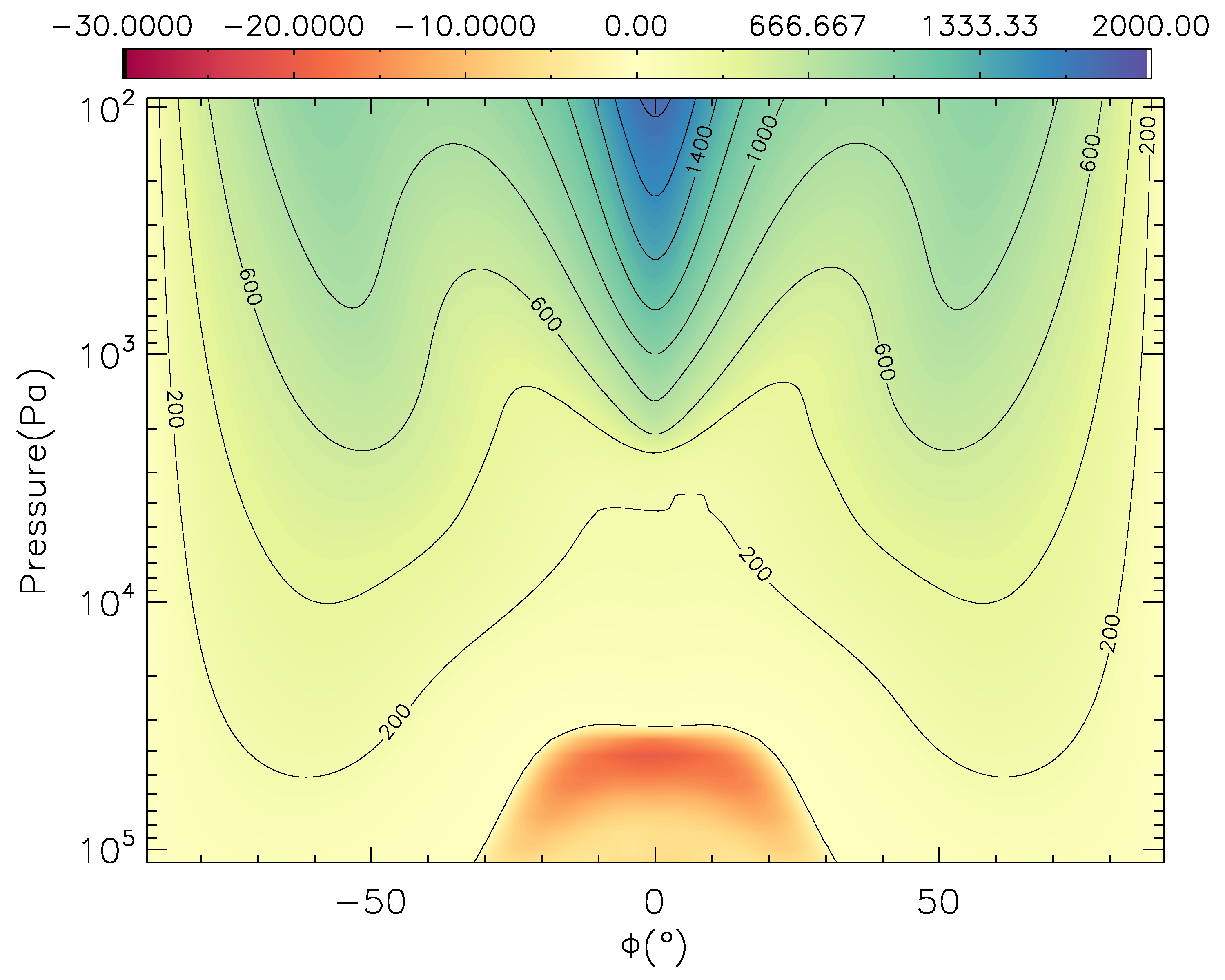}\label{hires_full_800_1000_uvel_bar}}
\end{center}
\caption{Figure showing the zonal and temporal mean of the zonal wind
  (ms$^{-1}$) as a function of latitude ($\phi$) and
  pressure ($\log_{10}(p\,[{\rm Pa}])$), for the primitive and full versions of the Std and Std Hires simulations. Note the modeled pressure
  domain extends down to $200\times 10^5$\,Pa but only the,
  relatively, dynamically active region of the atmosphere is shown
  here. The simulation short name (see Table \ref{model_names} for
  explanation of simulation names) and temporal averaging period are
  given below each subfigure. \label{std_uvel_bar}} 
  \end{figure*}

Figures \ref{std_uvel_bar}, shows the zonal--mean, temporal--mean, zonal
wind as a function of latitude and pressure, for the Std Prim, Std Full, and their high spatial
resolution counterparts (Std Hires Prim and Std Hires Full, respectively, where the horizontal and vertical resolution
have been doubled). Note, for these and subsequent similar figures the contour lines for all figures are the same, but the colour scales are varied between
simulation `pairs' i.e. each matching pair of primitive and full
simulations have the same colour scales. Figure \ref{std_uvel_bar} shows a clear difference between the simulation using the primitive equations and the more complete full versions (\textit{top panels}). The zonal prograde flow
is decelerated, and spread across a larger latitude range and
shallower pressure/height range for the simulation using the full
equations. The difference is marked with a $\sim 1.5\times$ increase
in the maximum zonal wind speed, and significant prograde zonal
velocities penetrating an order of magnitude deeper in pressure for
the primitive case, but `sharpened' to a significantly more peaked
latitudinal profile. This difference is also recovered in the higher
spatial resolution simulations (\textit{middle panels}), demonstrating
that this is not an issue of poor resolution or resolution differences
between simulation pairs (to the author's knowledge these are the highest resolution GCM simulations published for an exoplanet atmosphere to date). For this work, all of our simulations have been run using a simplified temperature forcing scheme to model the atmospheric heating. However, as shown in Appendix \ref{app_sec:radiative} our conclusions are unchanged when moving to the use of a more sophisticated radiative transfer scheme.

\begin{figure*}
\begin{center}
  \subfigure[Std Full: 100\,Pa]{\includegraphics[width=8.5cm,angle=0.0,origin=c]{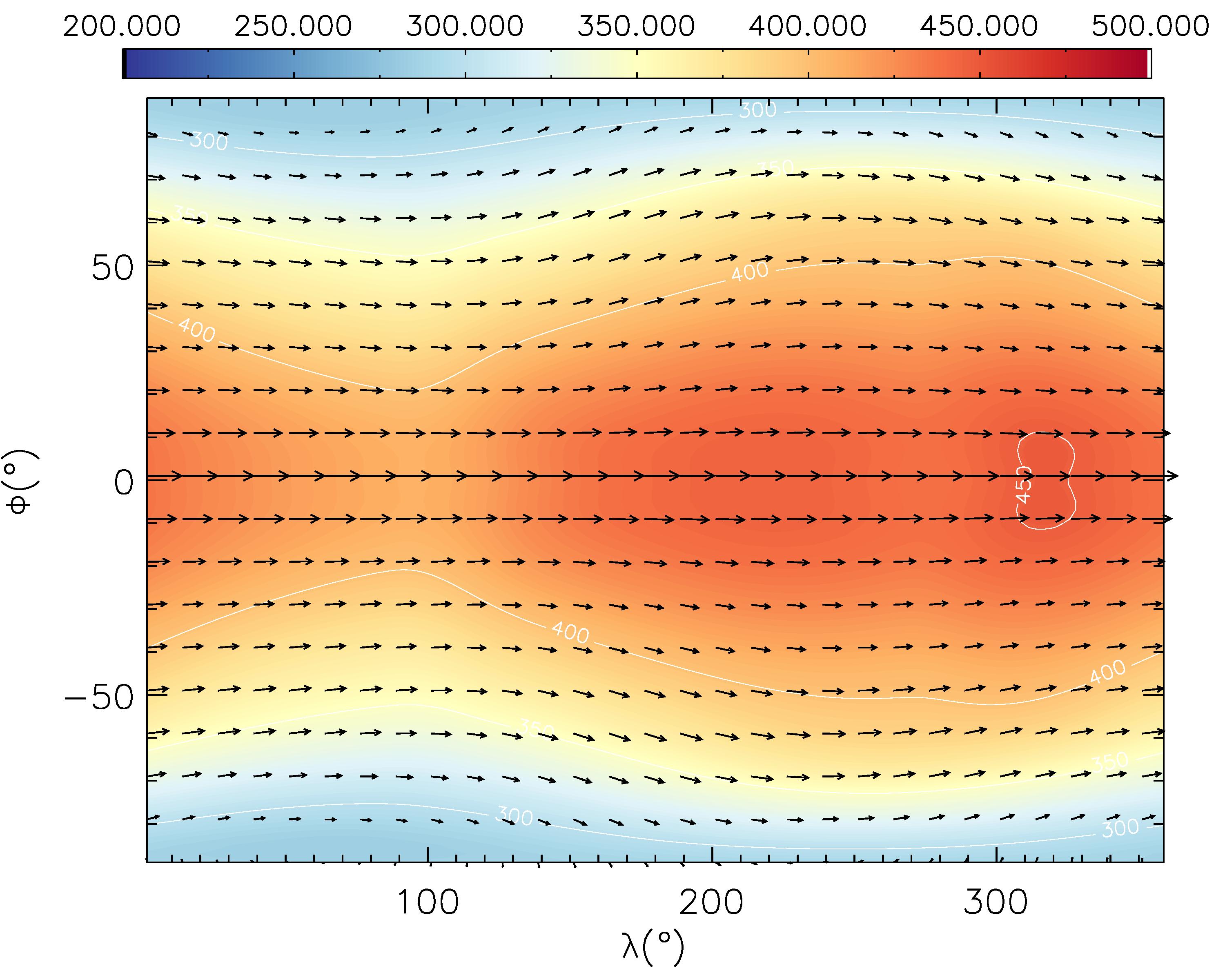}\label{std_full_1000_100_slice}}
  \subfigure[Std Full: 3\,000\,Pa]{\includegraphics[width=8.5cm,angle=0.0,origin=c]{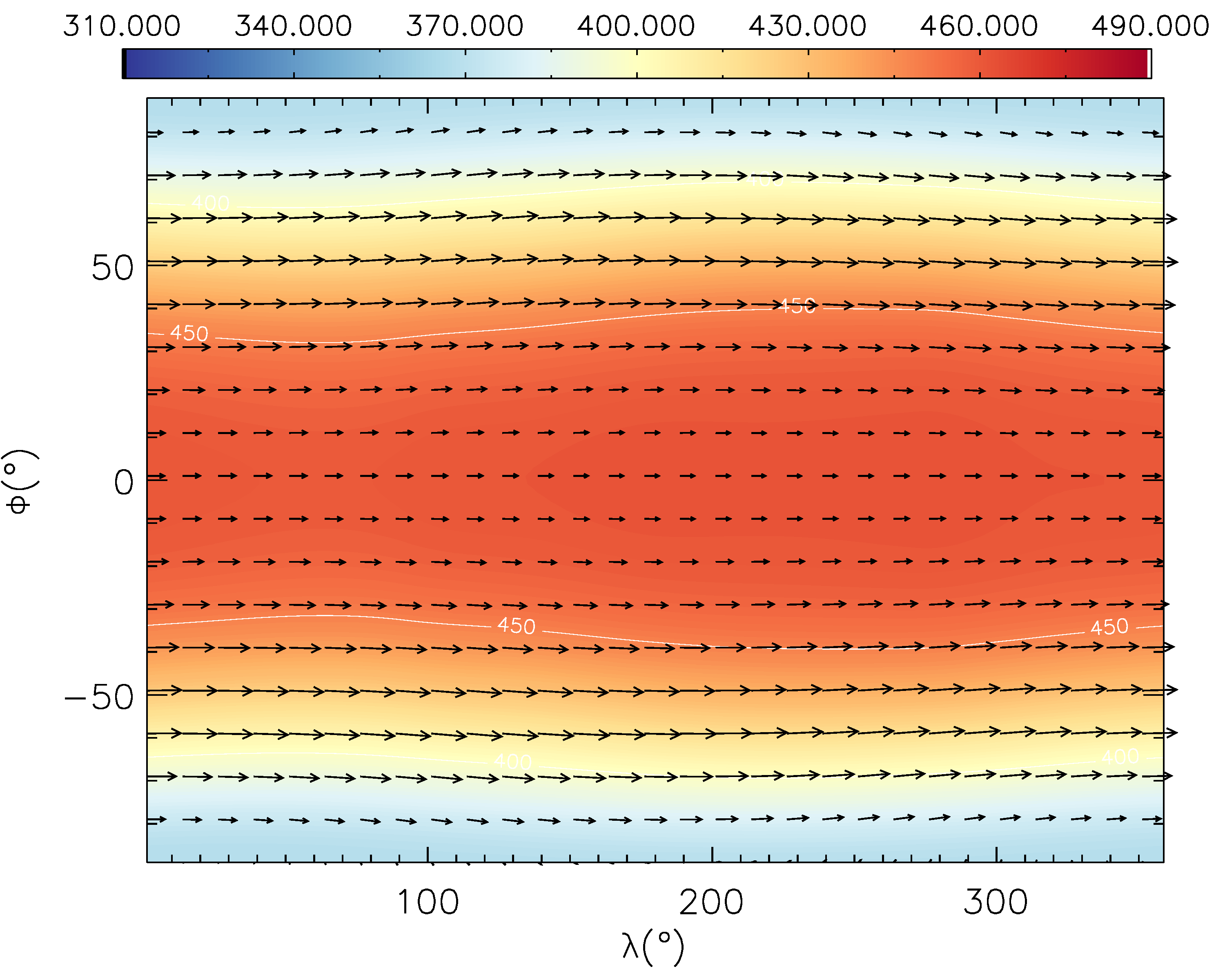}\label{std_full_1000_3000_slice}}
  \subfigure[Std Full-Prim: 100\,Pa]{\includegraphics[width=8.5cm,angle=0.0,origin=c]{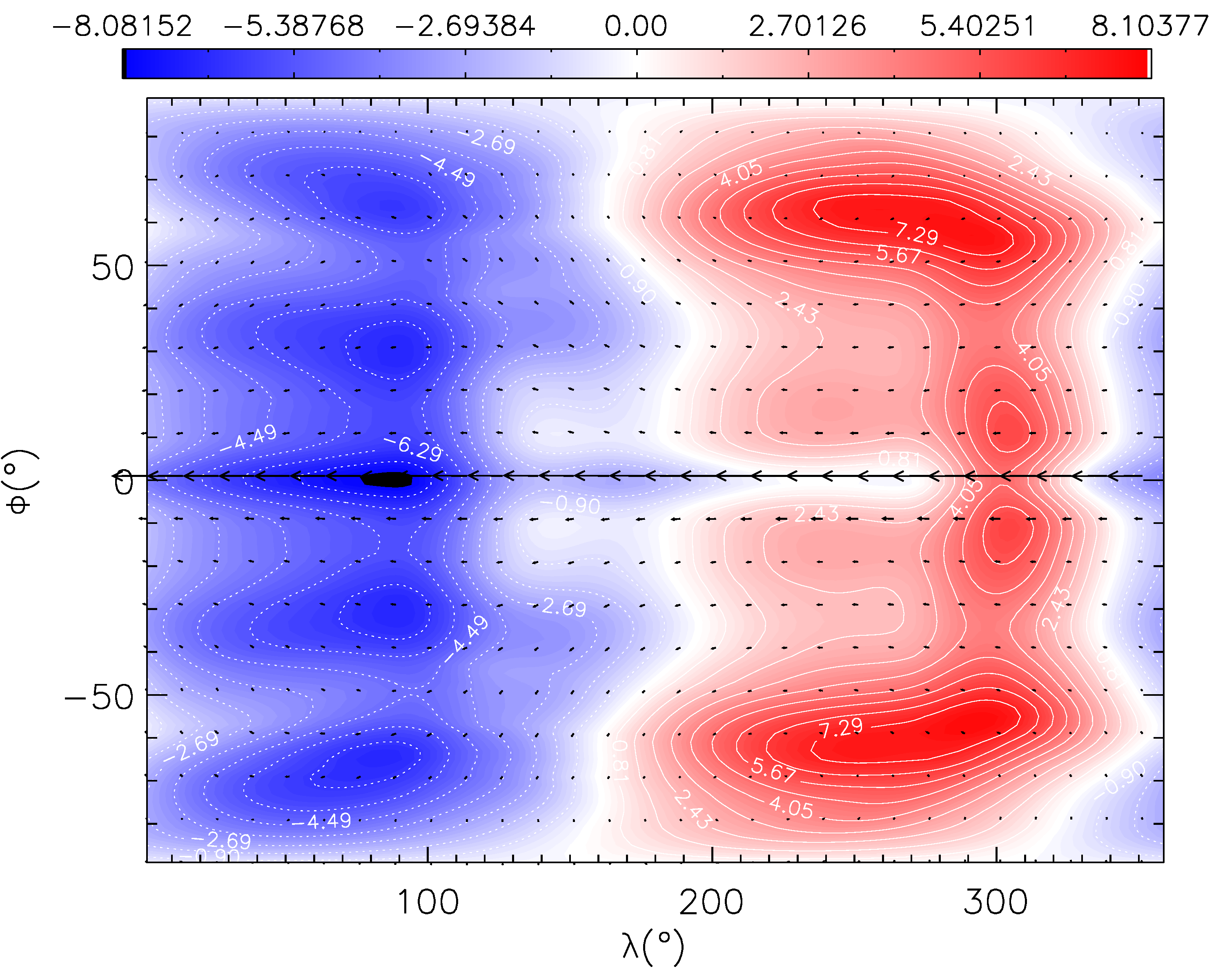}\label{std_diff_1000_100_slice}}
  \subfigure[Std Full-Prim: 3\,000\,Pa]{\includegraphics[width=8.5cm,angle=0.0,origin=c]{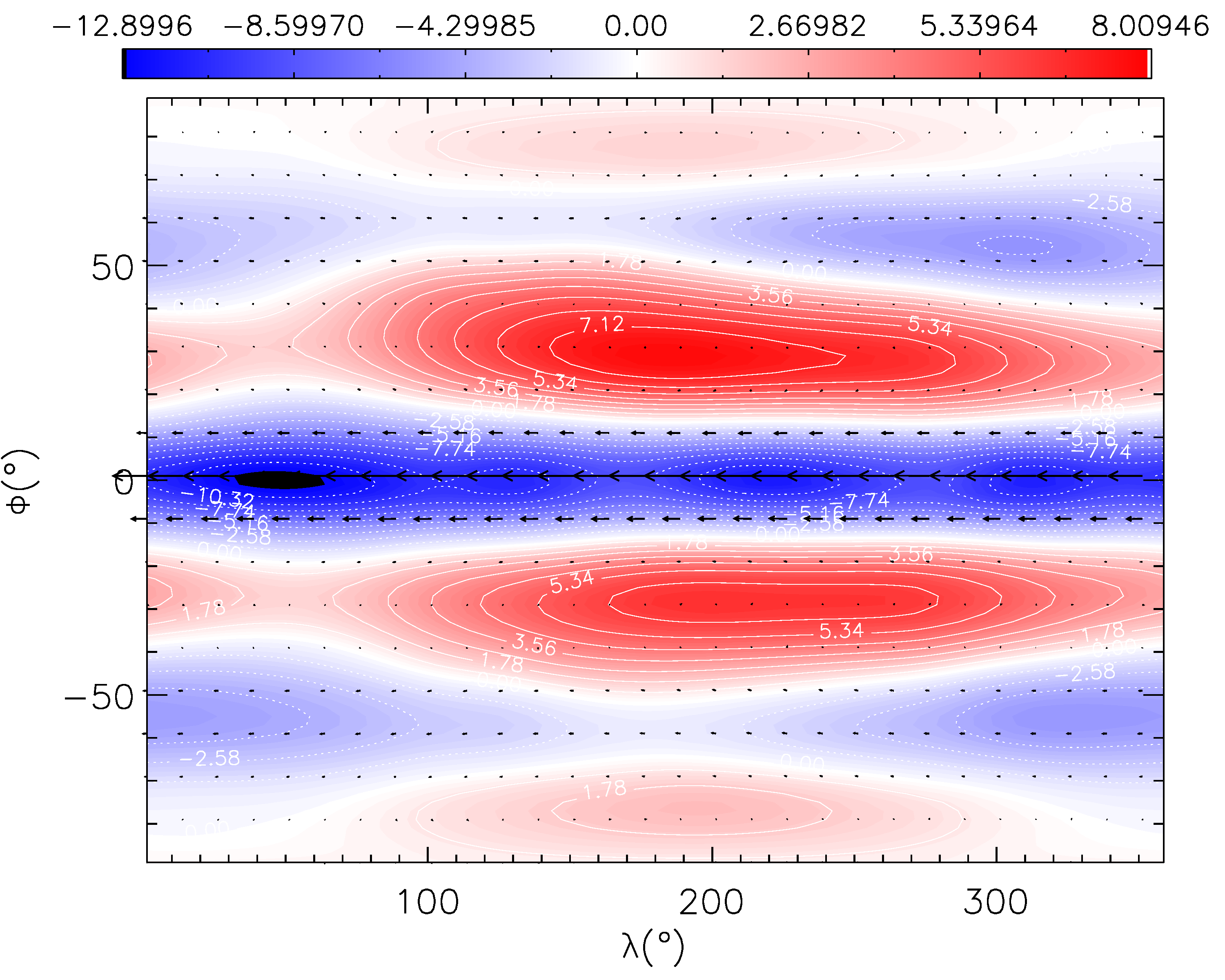}\label{std_diff_1000_3000_slice}}
  \subfigure[Std Prim/Full: 100\,Pa]{\includegraphics[width=8.5cm,angle=0.0,origin=c]{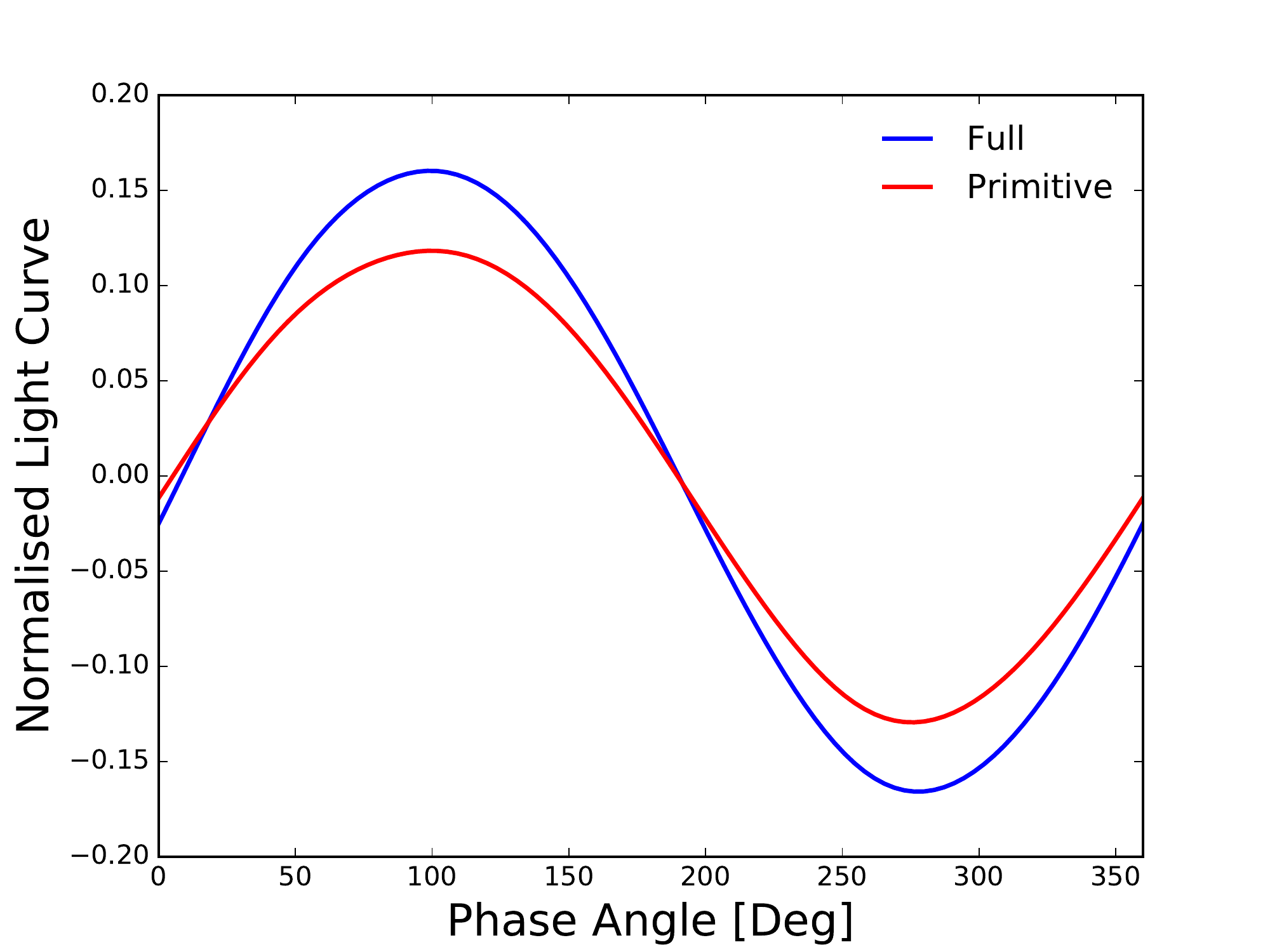}\label{phase_std_100}}
  \subfigure[Std Prim/Full: 3\,000\,Pa]{\includegraphics[width=8.5cm,angle=0.0,origin=c]{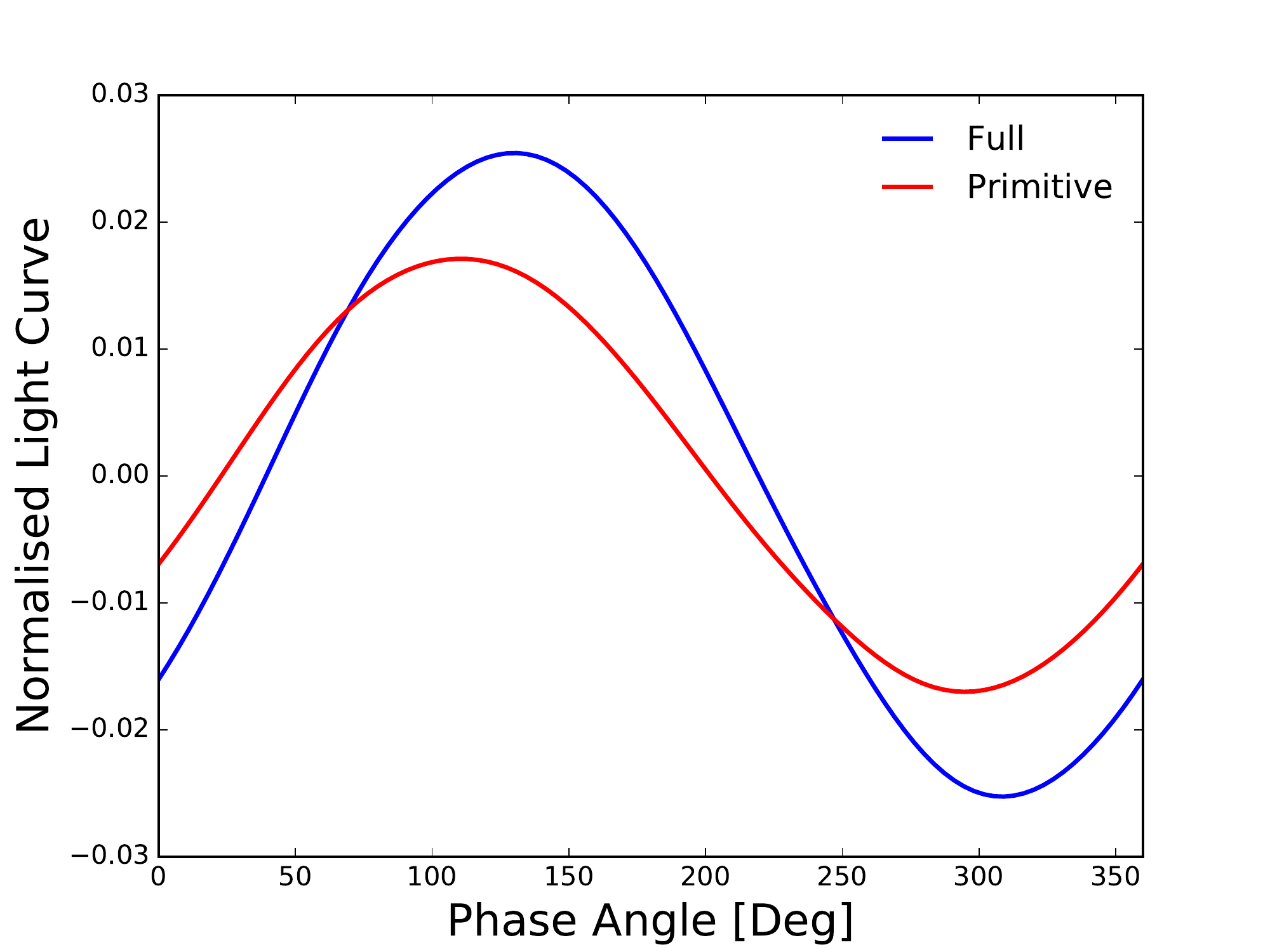}\label{phase_std_3000}}
\end{center}
\caption{Figure showing the temperature (colour scale, K) and
  horizontal wind (vector arrows) as a function of $\lambda$ and
  $\phi$, for isobaric surfaces at 100 and 3\,000\,Pa at 1\,000\,days for the Std Full
  simulation and the difference with the Std Prim (in the sense Std
  Full$-$Std Prim), as well as the simplified thermal phase curves at
  both pressures for both simulations (see Table \ref{model_names} for
  explanation of simulation names). Note the change in the vertical
  axes for the \textit{bottom panels}. \label{thermal_std}}
\end{figure*}

Figure \ref{thermal_std} shows the temperature (colour scale) and
horizontal wind (vector arrows) at isobaric surfaces (in $\lambda$ and
$\phi$) of 100 and 3\,000\,Pa at 1\,000 days for the Std Full
simulation (\textit{top panels}), and the differences in these fields
for the Std Prim simulation (\textit{middle panels}, where all
differences are in the sense full simulation minus primitive
simulation). Additionally, the simplified thermal phase curves for
both simulations at the two depths are shown (\textit{bottom panels}). 
  
Figure \ref{thermal_std} shows significant
changes in the thermal structure, and subsequent phase curve between
the primitive and full equations for our standard setup. 
The changes in the zonal flow (see Figure \ref{std_uvel_bar}) clearly translate to alterations in the temperature structure. For the standard simulations at 100\,Pa (\textit{left column}, Figure \ref{thermal_std}), the full equations result in an increase in temperature near and to the east of the
sub--stellar point (180$^{\degree}$), with cooler temperatures to the west, effectively driven by a weaker superrotating jet, as can be seen from the difference in the horizontal velocity (vector arrows, and see Figure \ref{std_uvel_bar}). This results, overall, in a warmer dayside and cooler nightside and, therefore, an increased amplitude in the simplified phase curve. For 3\,000\,Pa (\textit{right column}), the regions flanking the equatorial jet, in latitude, and at the poles are warmed, with the most significant warming on the dayside. The jet, and mid-latitude regions are cooled, with the strongest cooling on the night side. The amplitude of the simplified phase curve is proportional to $T^4\cos\phi$, heavily weighted to the equatorial
regions\footnote{We note that we show normalised phase curves (see Eqn. \eqref{equation:norm_phase_curve}) which means that relative differences in the amplitude of the normalised phase curve between simulations can be larger than might be expected based on the relative temperature differences.}. Therefore, the overall amplitude is again increased, however here the peak of the curve is also shifted. As discussed the radiative timescale is longer
at deeper pressures so changes in the flow can more easily affect the
longitudinal temperature structure. The peak of the warming is close
to the sub--stellar point, shifting the peak amplitude of the
simplified phase curve back toward this point.

Clearly, for the case of a slowly rotating warm Small Neptune (or potentially Super Earth) such as GJ~1214b, care must be taken when interpreting results derived from simulations solving the primitive equations. As discussed there are several additional assumptions made when simplifying from the full to the primitive equations so it is important to isolate how each of them contributes to the differences in the resolved flow.

\subsection{Limits of the Primitive Equations}
\label{subsec:prim_limits}

In \citet{mayne_2014b} we detail the terms included for each equation set, but here we restate only the dominant zonal momentum equation to explore the impacts of the approximations made in the primitive equations. 

The equation of zonal momentum conservation in the full equations is:
\begin{equation}
\frac{Du}{Dt}=\frac{uv\tan\phi}{r}-\frac{uw}{r}+fv-f^{'}w-\frac{c_p\theta}{r \cos\phi}\frac{\partial \Pi}{\partial \lambda}+\textbf{D}(u),\\
\label{zon_full}
\end{equation}
where $\frac{D}{Dt}$ is the material derivative and $\textbf{D}$ the
diffusion operator \citep[see][]{mayne_2017}. $f$ and $f^{'}$ are
the Coriolis parameters defined as,
\begin{equation}
f=2\Omega\sin\phi,
\end{equation}
and
\begin{equation}
f^{'}=2\Omega\cos\phi.
\end{equation}

The terms in Eqn. \eqref{zon_full} can be `grouped' into those
associated with rotation, i.e. $f$ and $f^{\prime}$, so--called
`metric' or advective terms associated with flow along a curved
surface, i.e. $\frac{uv\tan\phi}{r}$ and $\frac{uw}{r}$, and the
pressure gradient term, here
$\frac{c_p\theta}{r \cos\phi}\frac{\partial \Pi}{\partial
  \lambda}$. As discussed in Section \ref{sec:model} the full
equations include the spherical--geoid approximation which is only
justified when $R_{\rm p}\Omega^2/g$ is small \citep{benard_2014}. For
our simulations, Tables \ref{basic_par} and \ref{model_names} show
$R_{\rm p}\Omega^2/g$ is $\lesssim 0.01$ for all cases, and
$\sim 0.0025$ for the standard case, meaning this approximation does
not lead to significant errors. For faster rotating, and smaller radii
planets this may, however, begin to introduce more significant errors
\citep{benard_2014}.

The key underlying assumption made in constructing the primitive equations is that the aspect ratio of the motion is small i.e. the flow in the vertical direction is significantly smaller in scale than that in the horizontal direction. This validates the three major assumptions \citep[see, for example,][]{vallis_2006,mayne_2014b}. (1) \textit{The hydrostatic approximation:} the atmosphere is in vertical hydrostatic balance, regulated by fast acoustic waves, (2) \textit{The shallow--fluid approximation:} the atmosphere is thin in relation to the size of the planet meaning that $r$ can be replaced with $R_{\rm p}$ and $\partial / \partial r$ can be replaced by $\partial / \partial z$, and (3) \textit{The traditional approximation:} where the smaller metric and Coriolis terms are neglected. In the case of the zonal momentum equation, for example, this requires that $w \ll v \tan \phi$ such that the term $\frac{uw}{r}$ is ensured to be negligible when compared to the term $\frac{uv\tan\phi}{r}$ \citep[see Eqn. \eqref{zon_full} and][]{mayne_2014b}. This condition also ensures that the term $2\Omega w\cos\phi$ is small compared to $2\Omega v\sin \phi$ and can similarly be dropped. The assumption of a shallow--fluid and the traditional approximation must be taken together to ensure that the equation set conserves angular momentum 
\citep{white_1995}, and also allow the further assumption; (4) gravity is constant with height. The approximations of hydrostasy and a gravity constant with height directly impact the vertical momentum equation, and indirectly impact the zonal momentum \citep[see][for explicit equations]{mayne_2014b}. However, the shallow--fluid and traditional approximations directly impact the zonal (and meridional) momentum equation, evident when comparing the zonal momentum equation for the primitive case,
\begin{equation}
\frac{Du}{Dt}=\frac{uv\tan\phi}{R_{\rm p}}+fv-\frac{c_p\theta}{R_{\rm p} \cos\phi}\frac{\partial \Pi}{\partial \lambda}+\textbf{D}(u),\\
\label{zon_prim}
\end{equation}
with the full equation (Eq. \eqref{zon_full}). Specifically, for Eqns \eqref{zon_full} and \eqref{zon_prim}, the shallow--fluid approximation results in $r$ being replaced with $R_{\rm p}$ for the first (metric) and final (pressure gradient) terms. The traditional approximation then leads to the omission of the second (metric) $\frac{uw}{r}$ term and $f^{\prime}$ (or $2\Omega w\cos\phi$, a rotational or Coriolis term) as discussed. 

In this section we review each of the assumptions made in constructing the primitive equations, for the regime of warm small Neptunes (or Super Earths), and explore manifestations of their limitations using the simulation set presented in Table \ref{model_names}.

\subsubsection{The Hydrostatic Approximation}\label{subsubsec:hydro_balance}

Clearly, hydrostasy is enforced in the primitive simulations, however
for all our simulations solving the full equations vertical
hydrostatic equilibrium remains a good, global, approximation. Small
departures from hydrostatic balance are found in small regions of the
highest, low pressure, atmospheric layers, but do not significantly
alter the bulk flow. Typically, hydrostasy will hold for flows with a
horizontal scale larger than the vertical scaleheight,
$H=RT/g\sim 1.5\times 10^5$\,m, in our case, which is less than 0.2\%
of the planetary circumference.

\subsubsection{The Shallow--fluid Approximation}\label{subsubsec:shallow_fluid}

The shallow--fluid approximation is evidently contingent on the condition that 
the height above the inner boundary of the dominant zonal flow, or the vertical extent of the dynamically active atmosphere ($z_{U}$) is much smaller than the planetary radius i.e. in order to replace $r$ with $R_{\rm p}$, where $r$ can be expressed as $R_{\rm p}+z_{U}$ we require $z_{U} \ll R_{\rm p}$. For a Super Earth $R_{\rm p}$ is fixed as it has a solid surface, but for a small Neptune $R_{\rm p}$ could simply be shifted to lower pressures if regions of the deep atmosphere were quiescent and not affecting the overall dynamics. However, from our simulations we see that the flow extends throughout the majority of our modelled domain, and the inner boundary could not be raised significantly without affecting the flow. In the Std Full and Std Prim cases, shown in Figure \ref{std_uvel_bar}, the vertical extent of the dynamically active atmosphere is $\sim 20\%$ of the planetary radius. We have performed simulations matching the standard setup yet with an increased, or decreased planetary radius ($R_{\rm p}+$ and $R_{\rm p}-$, respectively, see Table {\ref{model_names}}), to explore this effect. Additionally, we perform a simulation adopting the standard planetary radius, but with a CO$_2$ dominated atmosphere, following \citet{zhang_2017}, leading to an increased mean molecular weight, and subsequent vertical compression of the atmosphere itself.

\begin{figure*}
\begin{center}
\subfigure[$R_{\rm p}-$ Prim: 800-1\,000\,days]{\includegraphics[width=8.5cm,angle=0.0,origin=c]{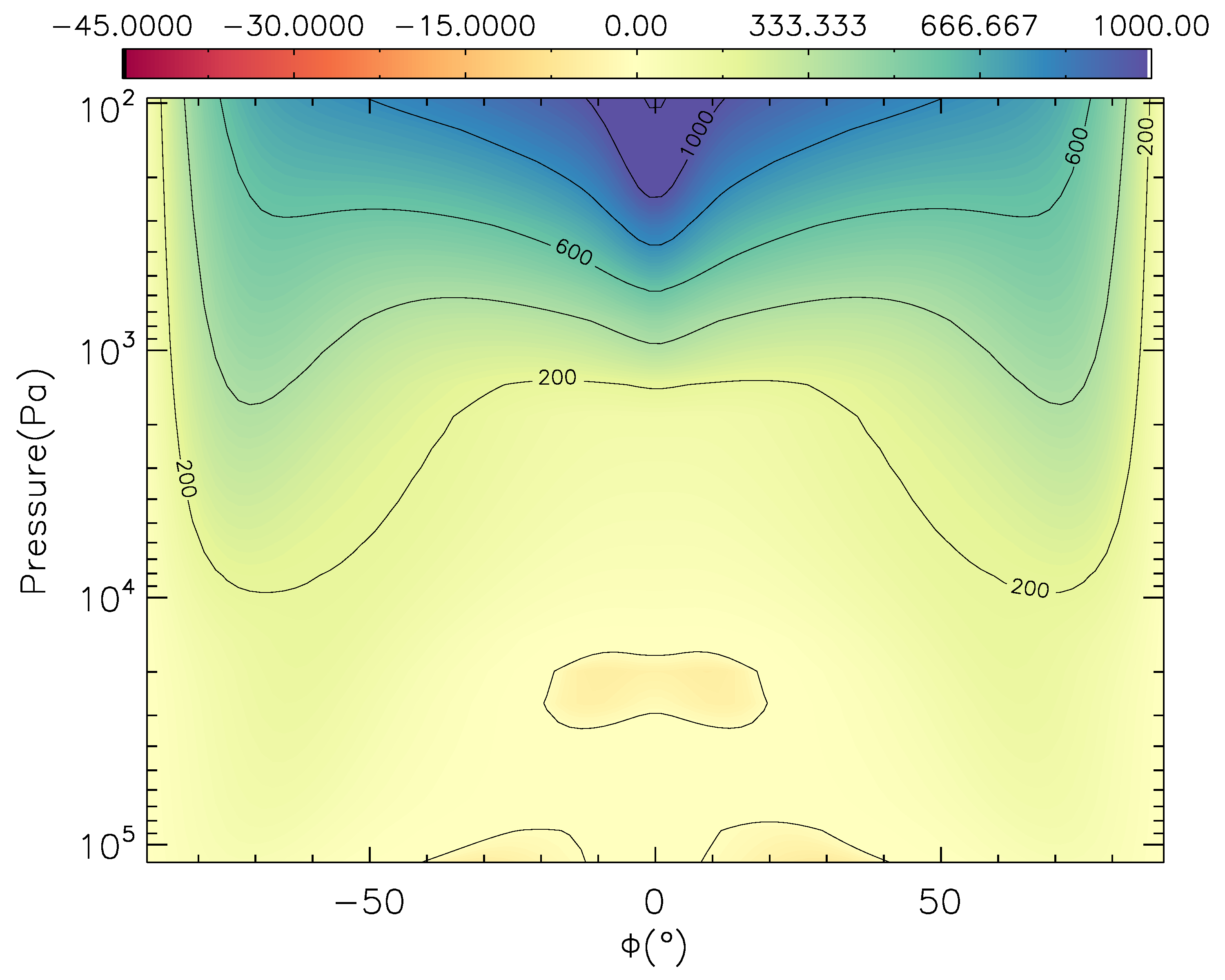}\label{redr_prim_800_1000_uvel_bar}}
\subfigure[$R_{\rm p}-$ Full: 800-1\,000\,days]{\includegraphics[width=8.5cm,angle=0.0,origin=c]{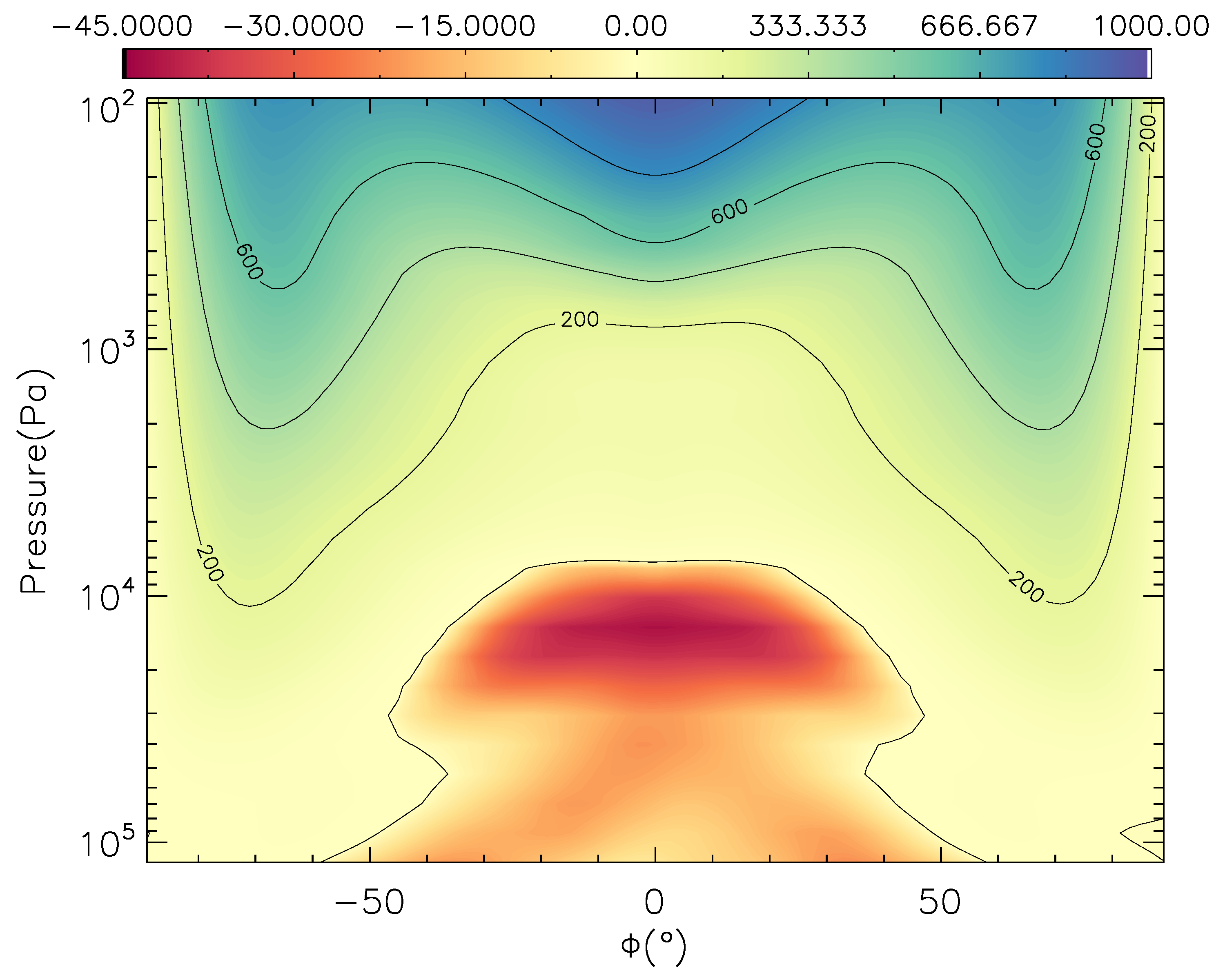}\label{redr_full_800_1000_uvel_bar}}
\subfigure[$R_{\rm p}+$ Prim: 800-1\,000\,days]{\includegraphics[width=8.5cm,angle=0.0,origin=c]{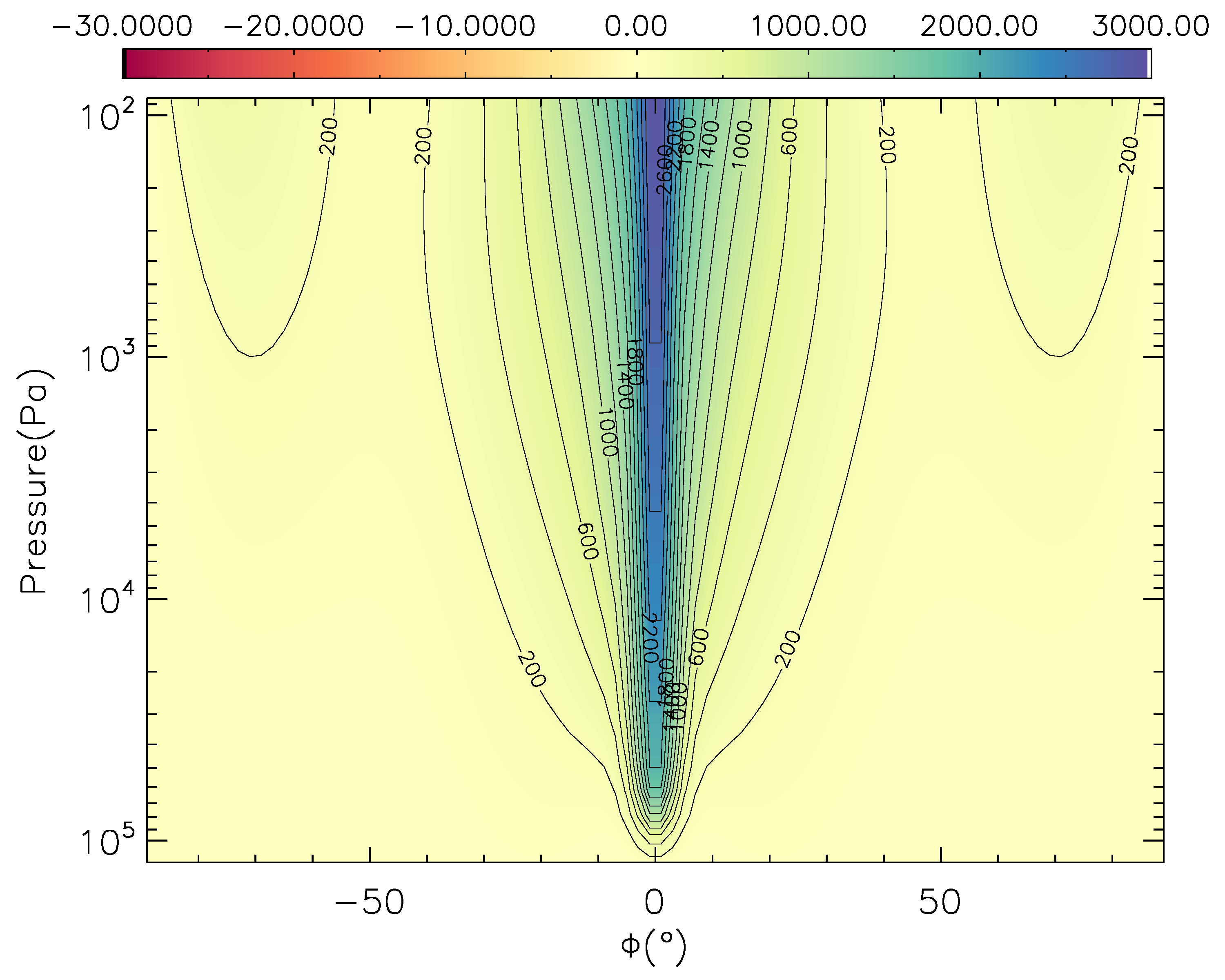}\label{incr_prim_800_1000_uvel_bar}}
\subfigure[$R_{\rm p}+$ Full: 800-1\,000\,days]{\includegraphics[width=8.5cm,angle=0.0,origin=c]{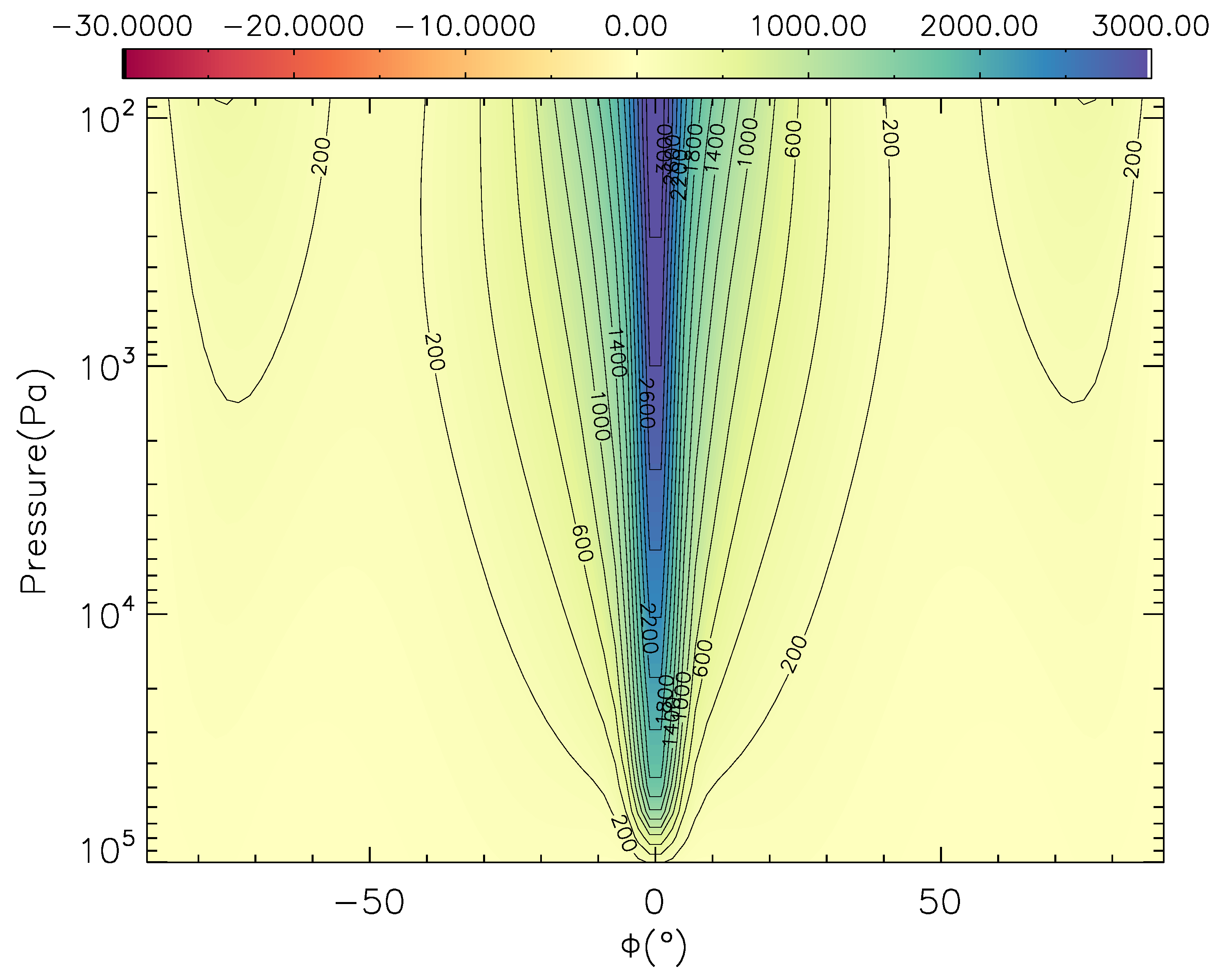}\label{incr_full_800_1000_uvel_bar}}
\subfigure[CO$_{\rm 2}$ Prim: 800-1\,000\,days]{\includegraphics[width=8.5cm,angle=0.0,origin=c]{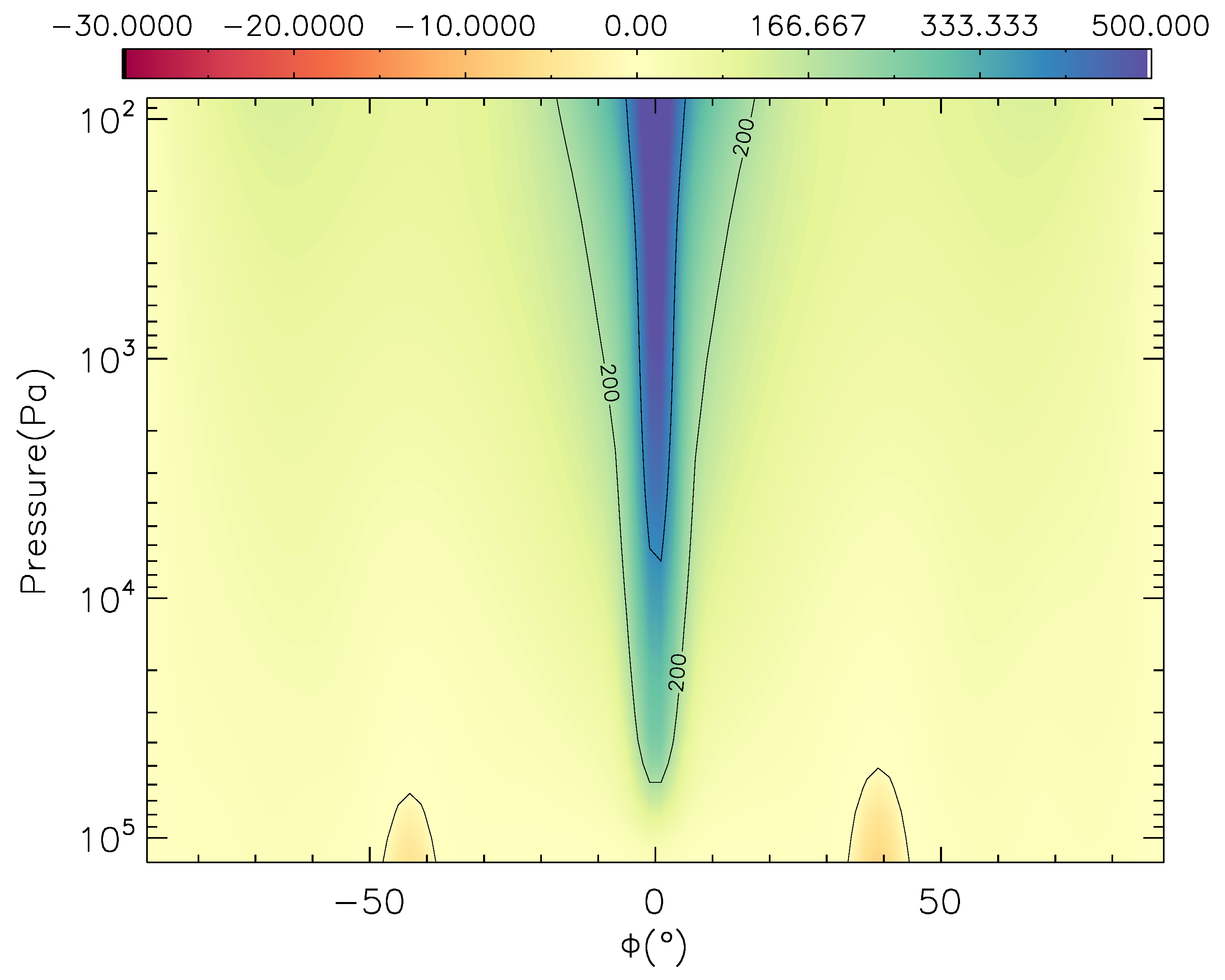}\label{co2_prim_800_1000_uvel_bar}}
  \subfigure[CO$_{\rm 2}$ Full: 800-1\,000\,days]{\includegraphics[width=8.5cm,angle=0.0,origin=c]{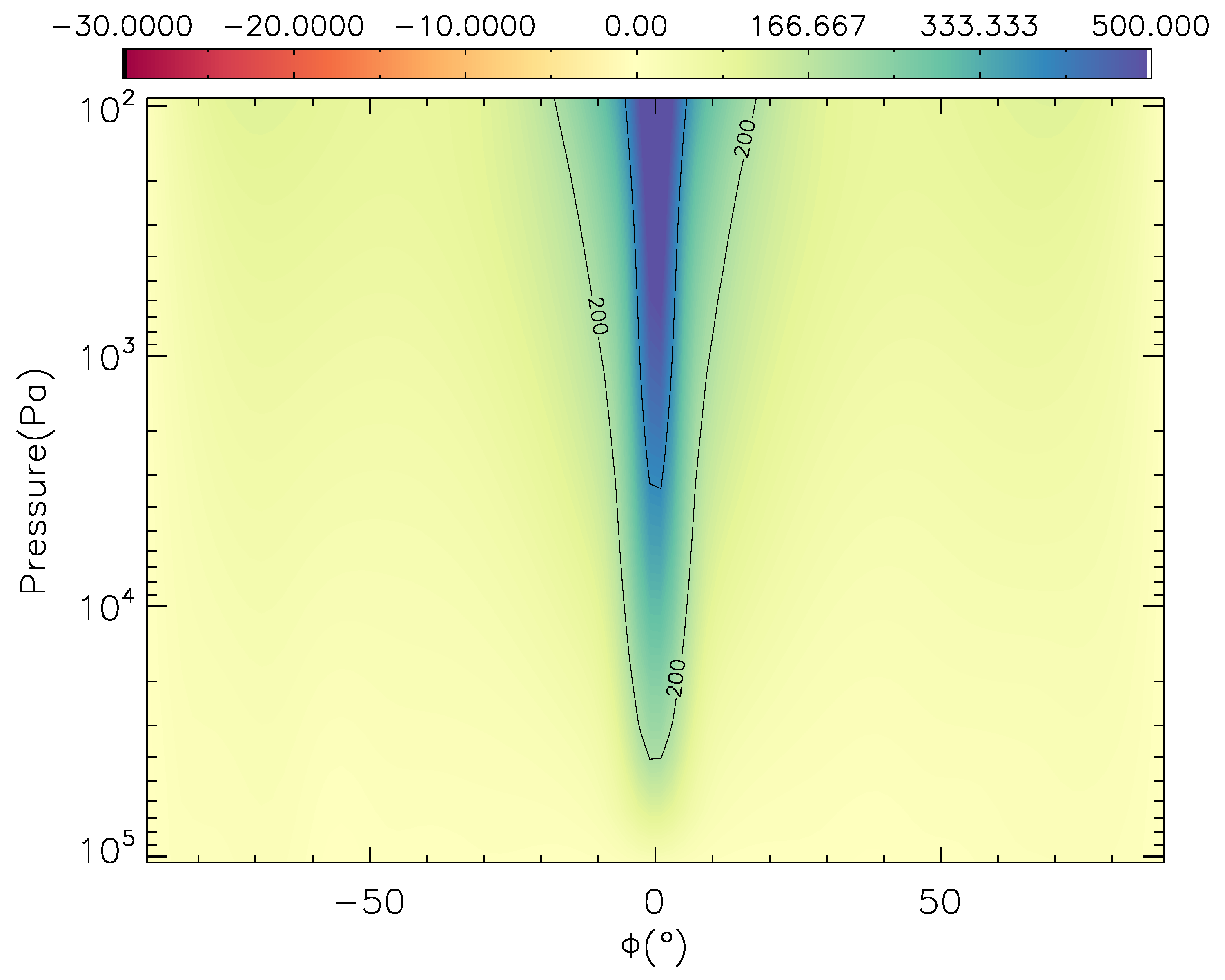}\label{co2_full_800_1000_uvel_bar}}
\end{center}
\caption{Figure similar to Figure \ref{std_uvel_bar}
  but for simulations adopting a decreased ($R_{\rm p}-$) and increased ($R_{\rm p}+$) planetary radius, and a CO$_2$ dominated atmosphere leading to a significantly reduced atmospheric scaleheight \label{shallow_uvel_bar}}
\end{figure*}

Figures \ref{shallow_uvel_bar} shows the same information, in the same
format as Figure \ref{std_uvel_bar} but for the $R_{\rm p}-$,
$R_{\rm p}+$ and CO$_{\rm 2}$ simulations. Clearly, for the
$R_{\rm p}-$ simulation differences remain between the primitive and
full simulations, whereas these differences are almost completely
removed when the planetary radius is increased ($R_{\rm p}+$). They
are also significantly reduced when the atmospheric scaleheight is
reduced (CO$_2$) compared with the standard setup. However, some
differences remain, which are discussed in Section
\ref{subsubsec:trad_approx}. It is important to note that due to the
assumption of a constant gravity in the primitive equations, as
opposed to reducing proportional to $(R_{\rm p}/r)^2$ in the full
case, differences in $g$ between the primitive and full cases will be
reduced as the planetary radius is increased, or atmospheric
scaleheight reduced. This could be affecting our results and
contributing to the increasing similarity in the resolved flow between
the full and primitive cases for the $R_{\rm p}+$ and CO$_2$
simulations. However, we show in Section \ref{subsubsec:trad_approx}
that for the standard case the difference in gravity between the
primitive and full case does not significantly affect the flow (see
Figure \ref{dt_uvel_bar}), where this differences is, at its maximum,
$\sim 20$\% (compared to a maximum of $10$\% in the $R_{\rm p}+$
simulation).

\begin{figure*}
\begin{center}
  \subfigure[$R_{\rm p}-$ Full: 100\,Pa]{\includegraphics[width=8.5cm,angle=0.0,origin=c]{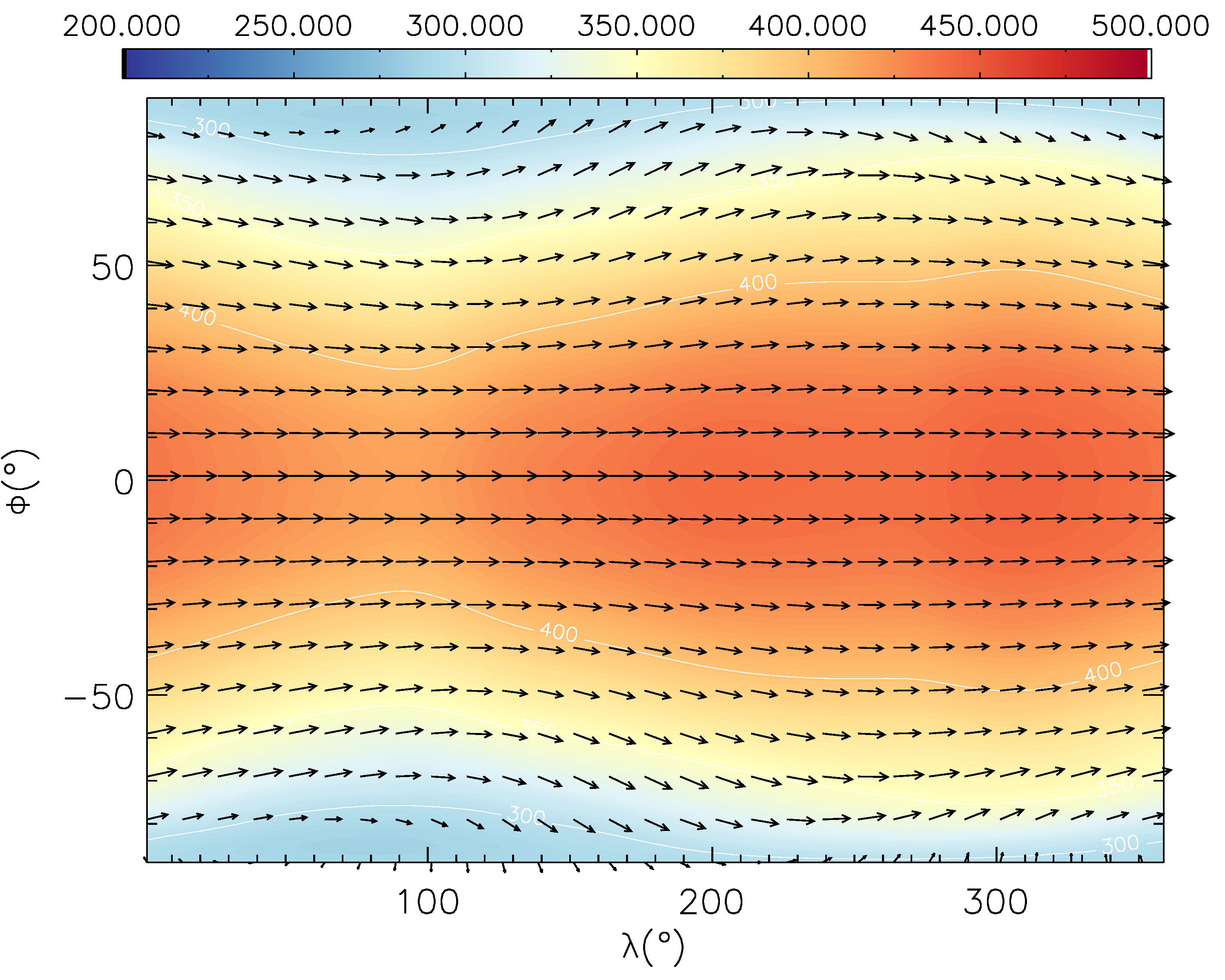}\label{redr_full_1000_100_slice}}
  \subfigure[$R_{\rm p}-$ Full: 3\,000\,Pa]{\includegraphics[width=8.5cm,angle=0.0,origin=c]{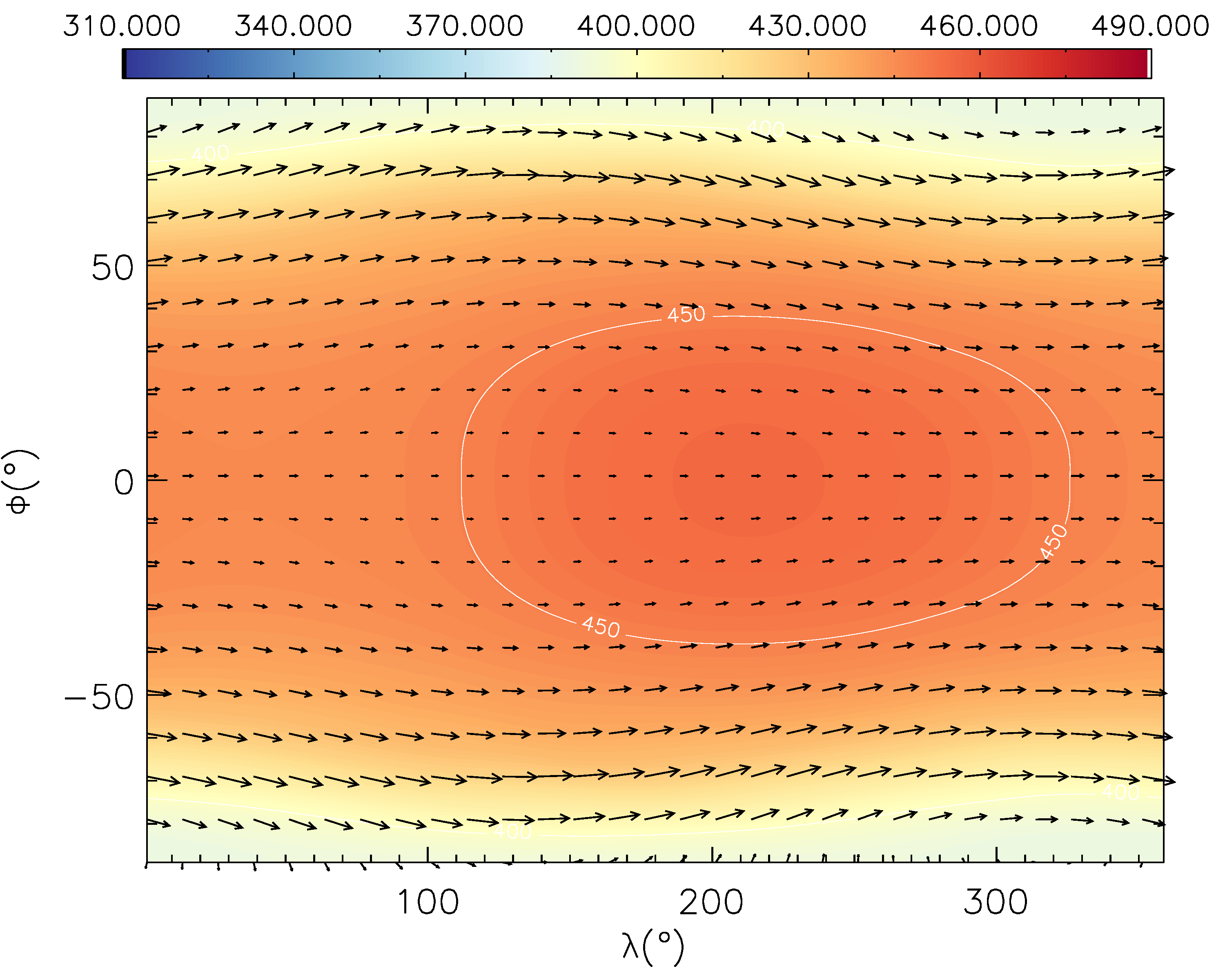}\label{redr_full_1000_3000_slice}}
  \subfigure[$R_{\rm p}-$ Full-Prim: 100\,Pa]{\includegraphics[width=8.5cm,angle=0.0,origin=c]{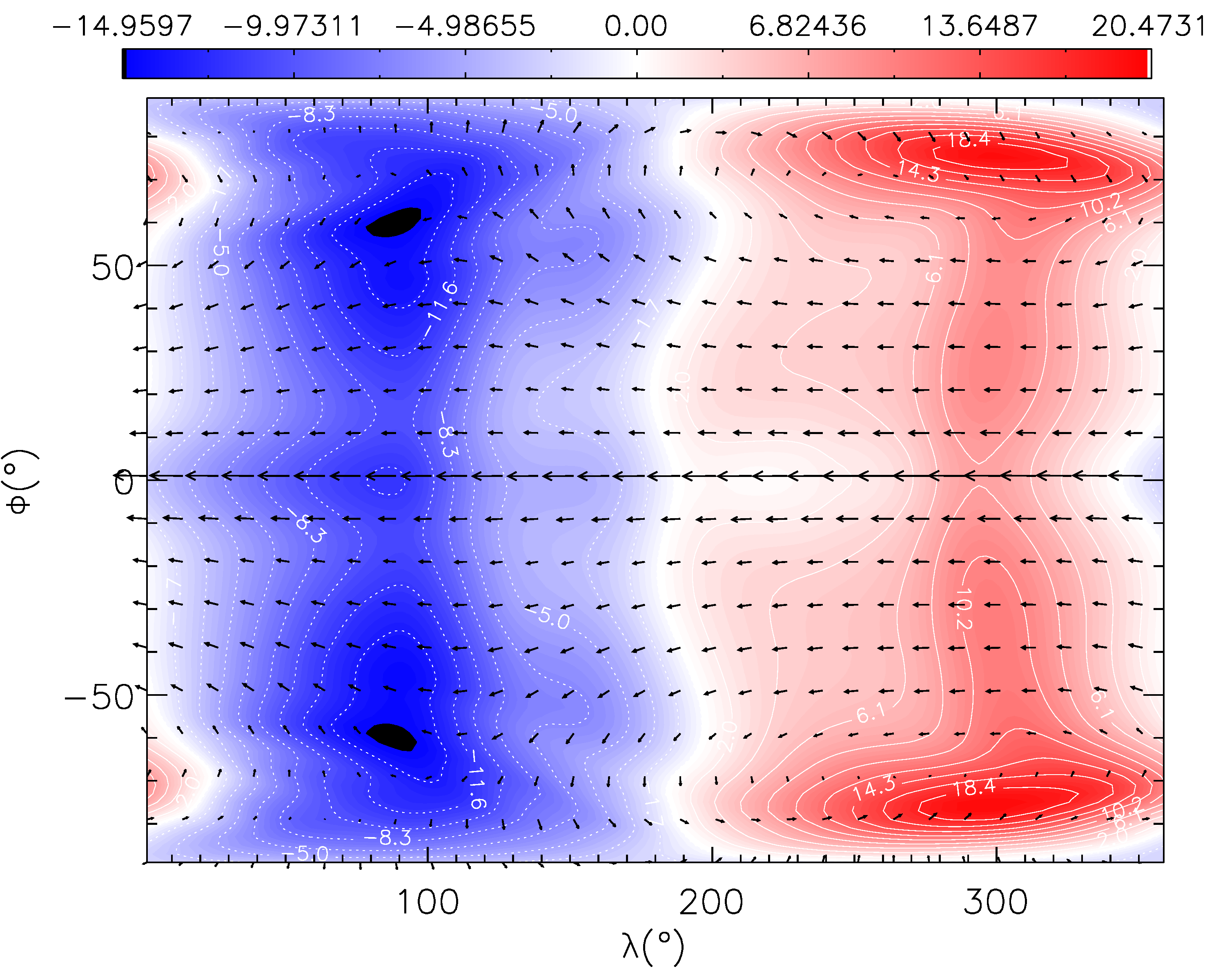}\label{redr_diff_1000_100_slice}}
  \subfigure[$R_{\rm p}-$ Full-Prim: 3\,000\,Pa]{\includegraphics[width=8.5cm,angle=0.0,origin=c]{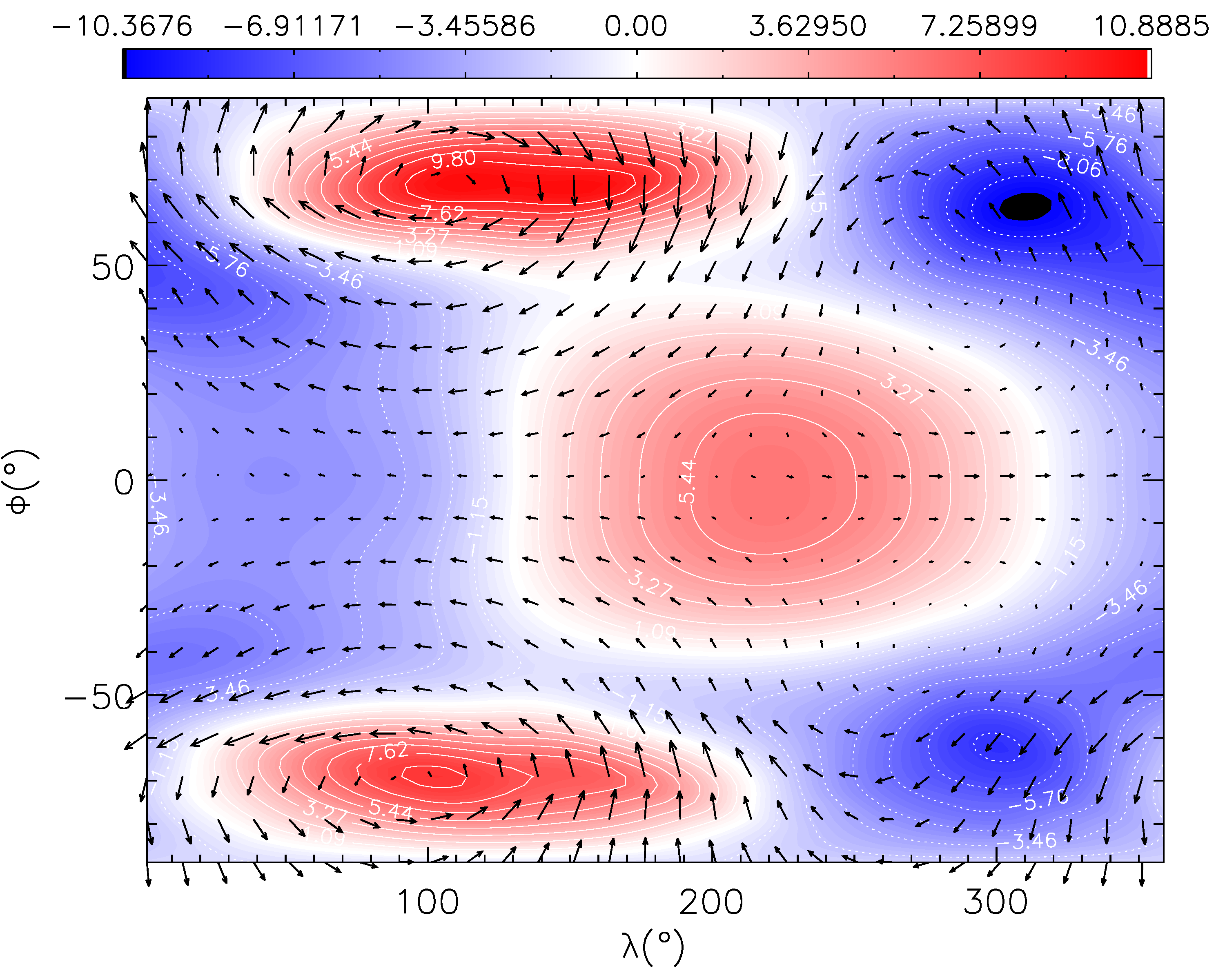}\label{redr_diff_1000_3000_slice}}
  \subfigure[$R_{\rm p}-$ Prim/Full: 100\,Pa]{\includegraphics[width=8.5cm,angle=0.0,origin=c]{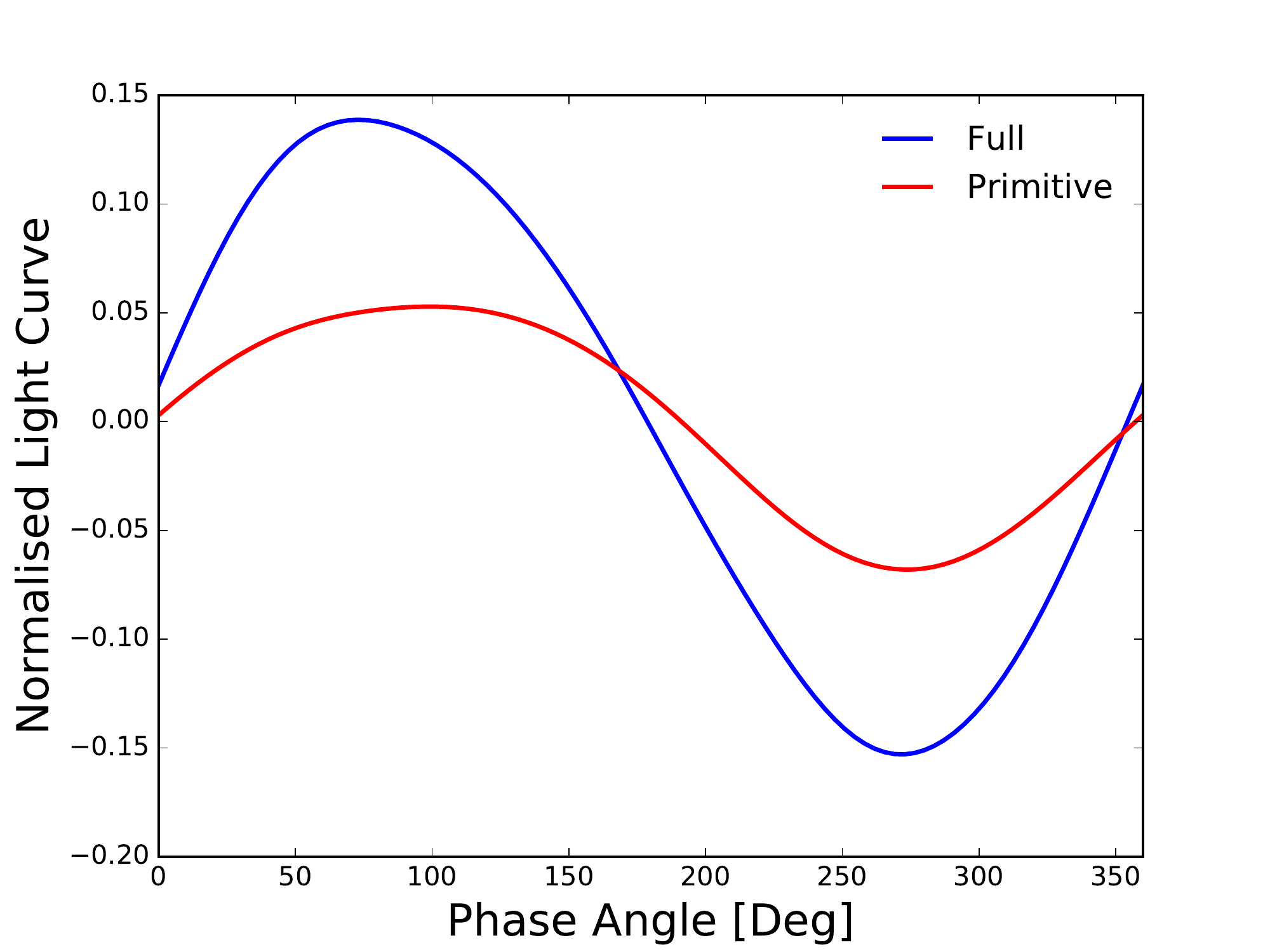}\label{phase_redr_100}}
  \subfigure[$R_{\rm p}-$ Prim/Full: 3\,000\,Pa]{\includegraphics[width=8.5cm,angle=0.0,origin=c]{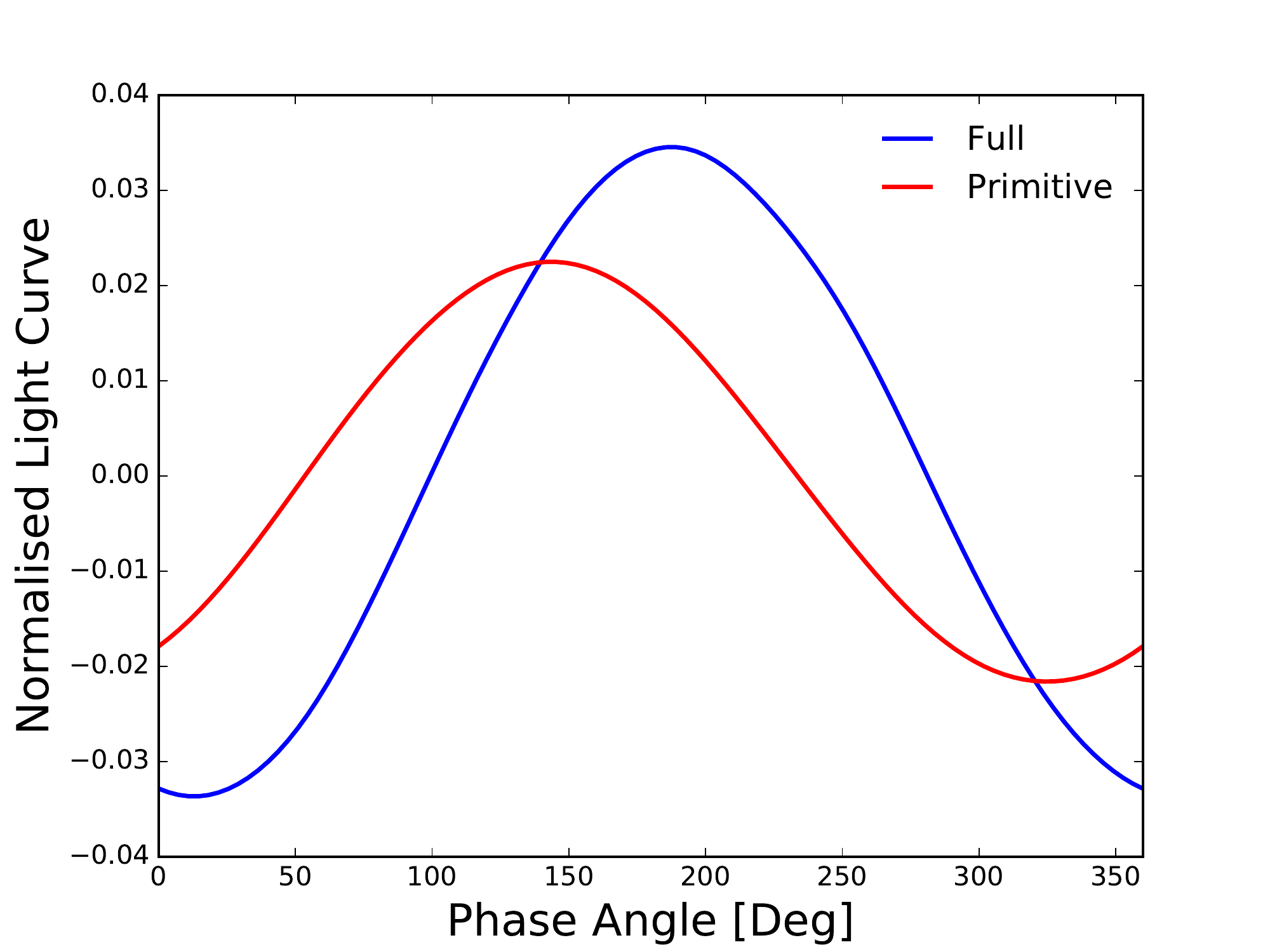}\label{phase_redr_3000}}
\end{center}
\caption{As Figure \ref{thermal_std} but for the reduced planetary radius simulations, $R_{\rm p}-$ (see Table \ref{model_names} for
  explanation of simulation names). Note the change in the vertical
  axes for the \textit{bottom panels}. \label{thermal_redr}}
\end{figure*}

\begin{figure*}
\begin{center}
  \subfigure[CO$_2$ Full: 100\,Pa]{\includegraphics[width=8.5cm,angle=0.0,origin=c]{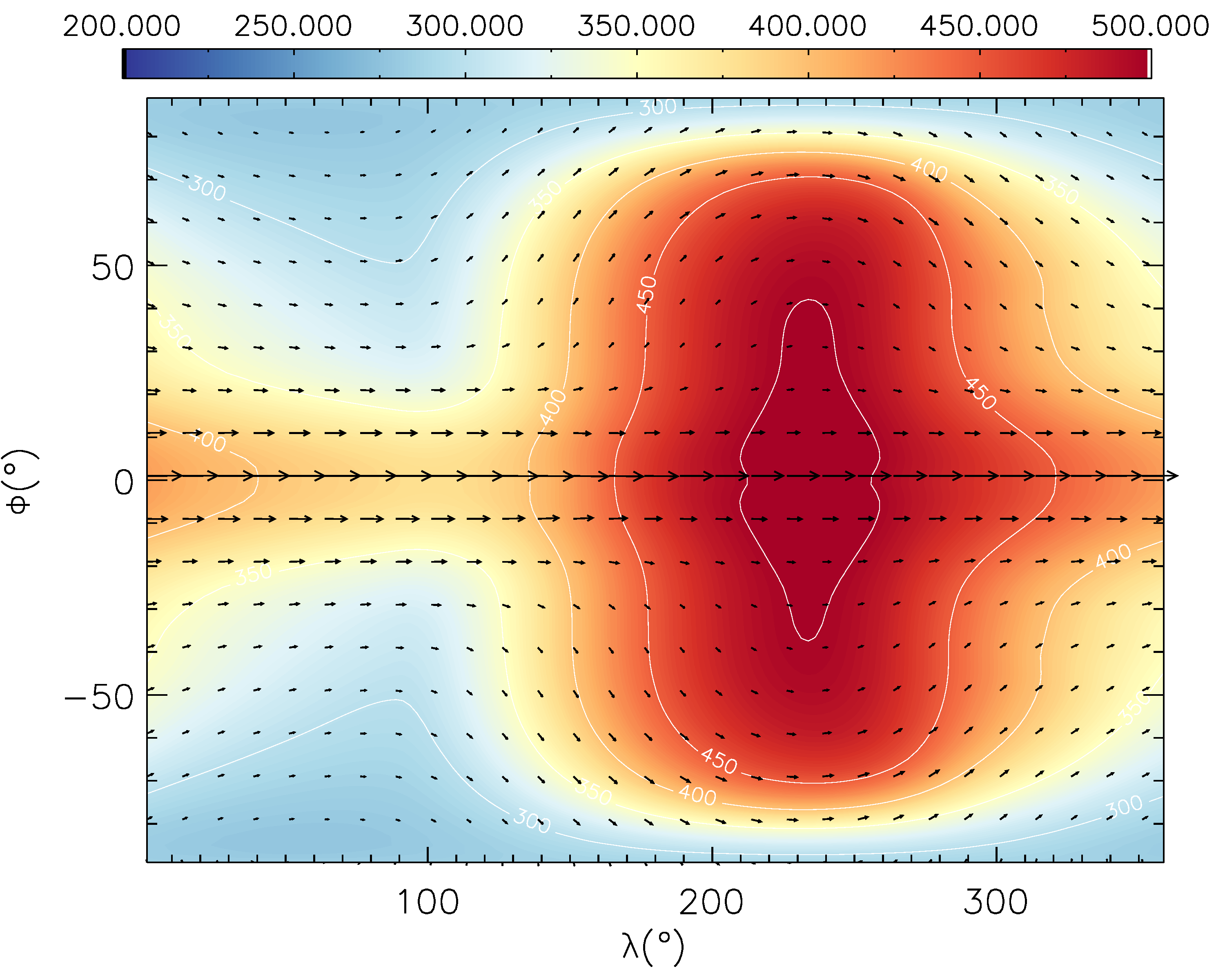}\label{co2_full_1000_100_slice}}
  \subfigure[CO$_2$ Full: 3\,000\,Pa]{\includegraphics[width=8.5cm,angle=0.0,origin=c]{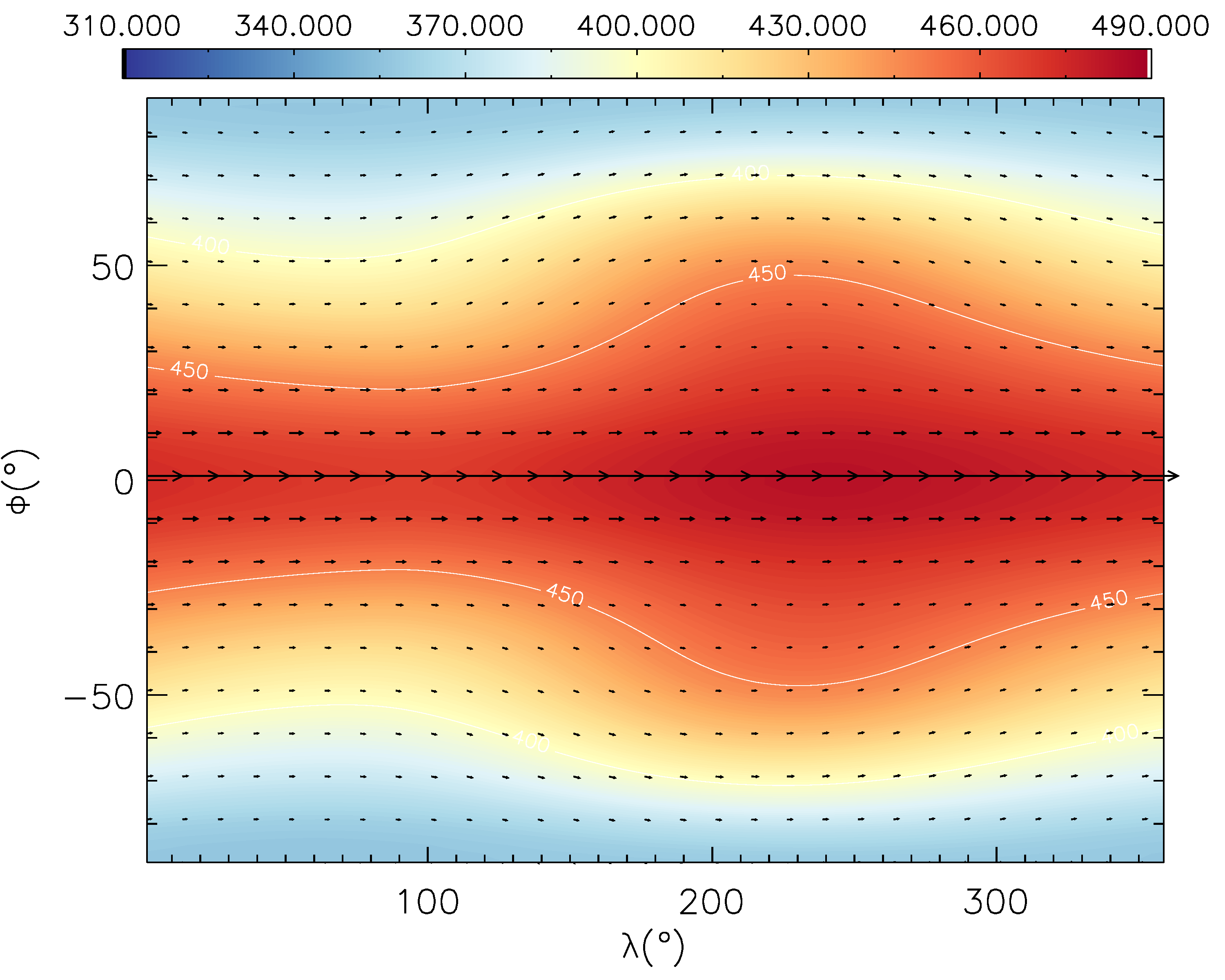}\label{co2_full_1000_3000_slice}}
  \subfigure[CO$_2$ Full-Prim: 100\,Pa]{\includegraphics[width=8.5cm,angle=0.0,origin=c]{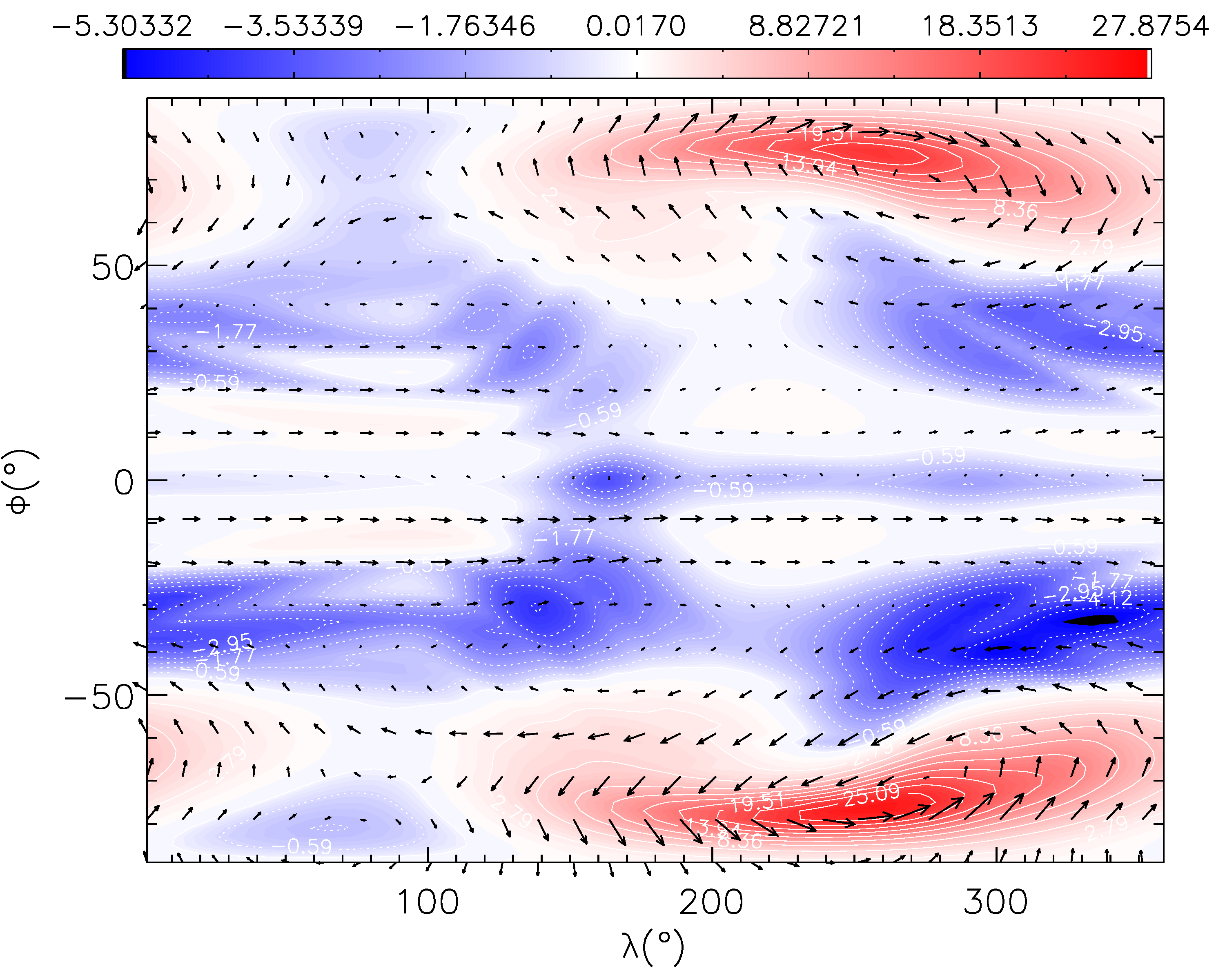}\label{co2_diff_1000_100_slice}}
  \subfigure[CO$_2$ Full-Prim: 3\,000\,Pa]{\includegraphics[width=8.5cm,angle=0.0,origin=c]{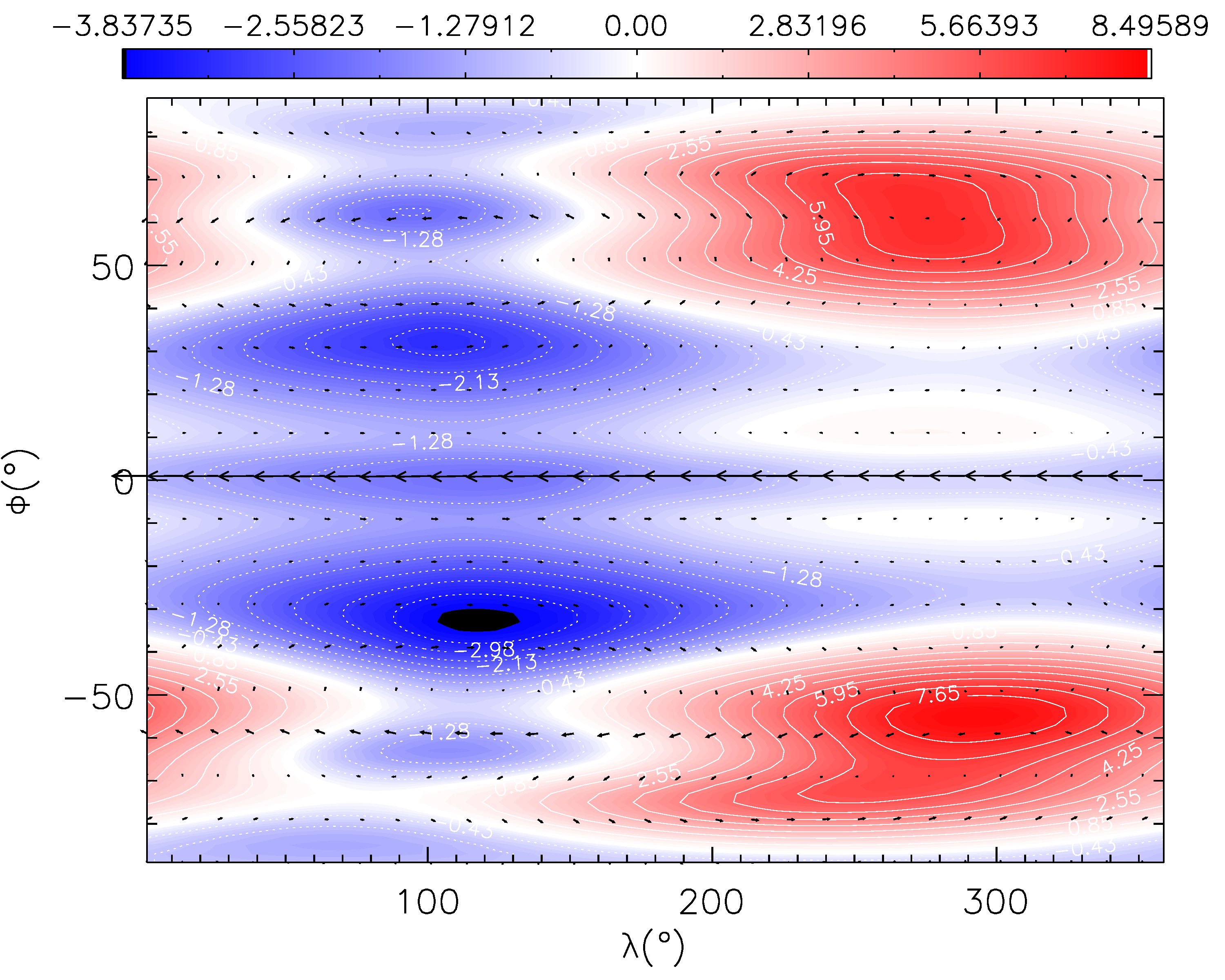}\label{co2_diff_1000_3000_slice}}
  \subfigure[CO$_2$ Prim/Full: 100\,Pa]{\includegraphics[width=8.5cm,angle=0.0,origin=c]{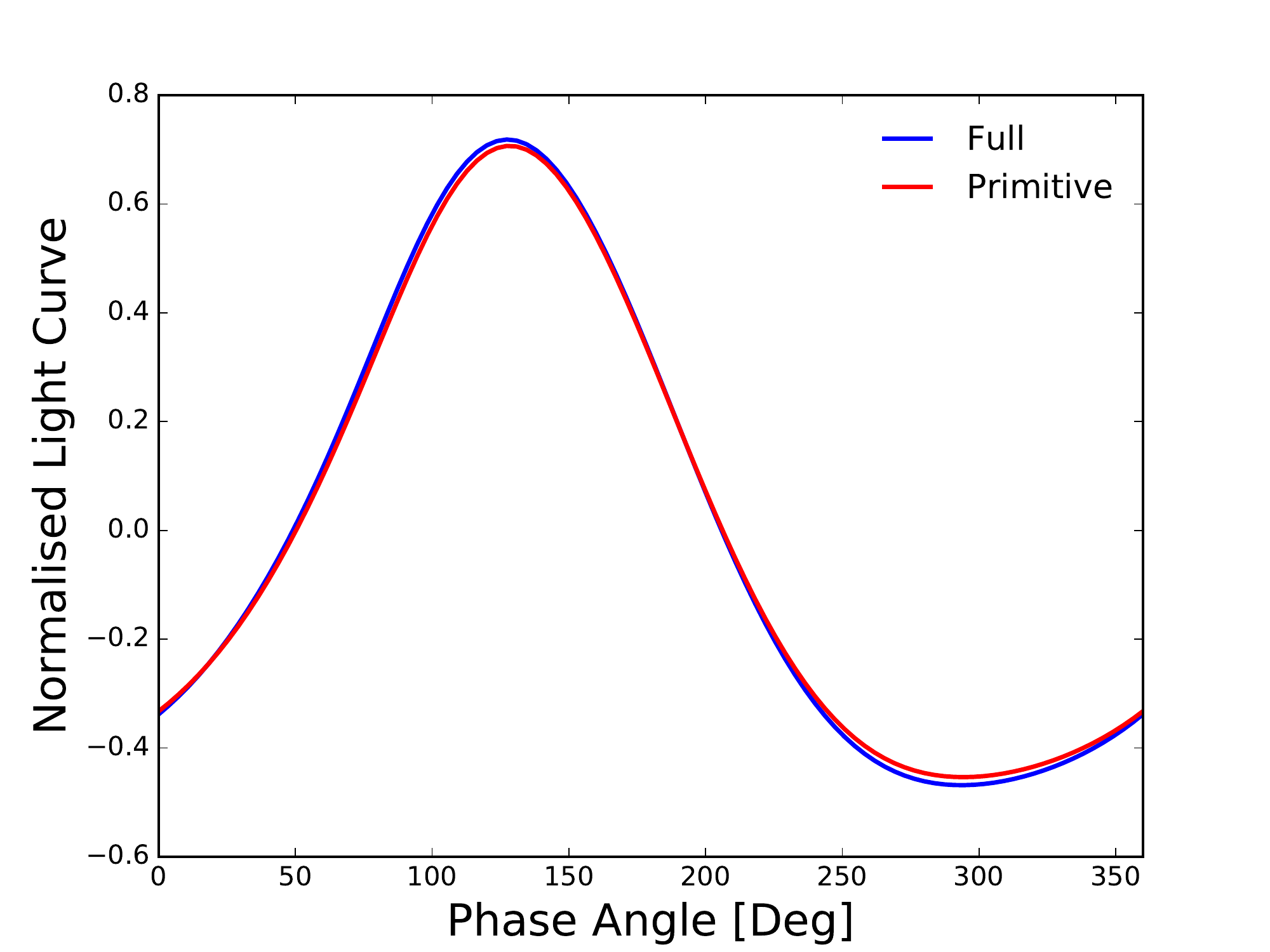}\label{phase_co2_100}}
  \subfigure[CO$_2$ Prim/Full: 3\,000\,Pa]{\includegraphics[width=8.5cm,angle=0.0,origin=c]{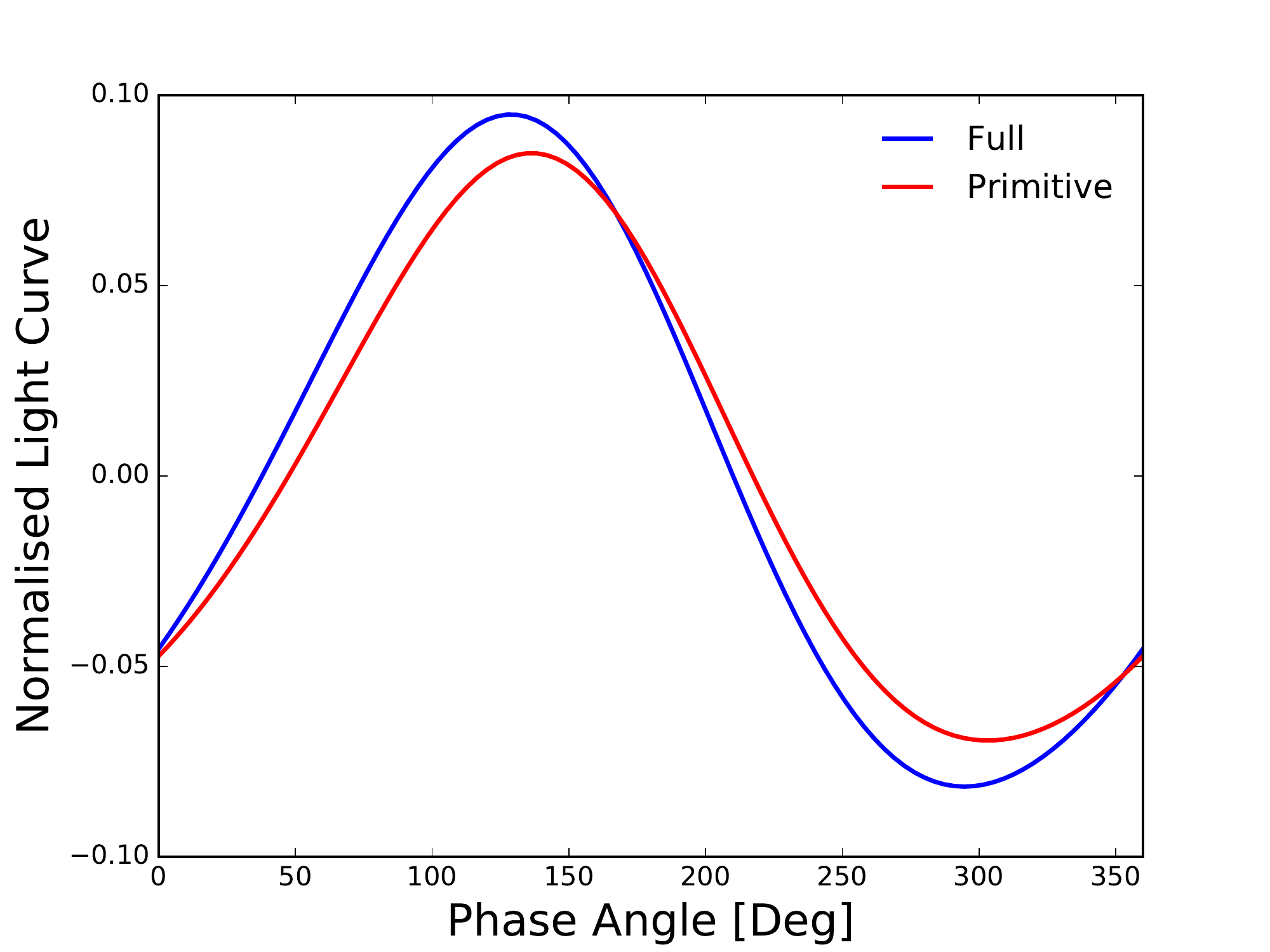}\label{phase_co2_3000}}
\end{center}
\caption{As Figure \ref{thermal_std} but for the CO$_2$ simulations (see Table \ref{model_names} for explanation
  of simulation names). Note the change in the vertical axes for the
  \textit{bottom panels}. \label{thermal_co2}}
\end{figure*}

For the $R_{\rm p}+$ the
temperature changes between the primitive and full simulations, at
both 100 and 3\,000\,Pa, are generally less than $\sim
2$\,K. Therefore, Figures of the temperature structure for these
simulations are omitted. In the case of the CO$_2$ simulation, the differences in the zonal wind structure resolved using the primitive or full equations are reduced, over the standard setup, but are not negligible. However, the differences between the two flows does reduce towards lower pressures. Figures \ref{thermal_redr} and \ref{thermal_co2} shows the same information as Figure \ref{thermal_std} but for the $R_{\rm p}-$ and CO$_2$ simulations.

The pattern of the change, at 100\,Pa, when moving to the more
complete equations of motion for the $R_{\rm p}-$ case is similar to
our standard GJ~1214b setups, but enhanced. The regions east of the
sub--stellar point are heated and those to the west are cooled, with a
net day side heating and night side cooling. For the higher pressure,
3\,000\,Pa, the pattern is slightly different, with heating close to
the poles to the west of the substellar point, and east of the
substellar point at the equator. Again, as the simplified phase curve
amplitude is weighted by $\cos\phi$ we have an increase in the peak
day side amplitude, and reduction in the night side minimum. The
offset between the peak at the substellar point also reduces, as for
the standard simulations, caused by overall less efficient
redistribution of heat in the more complete equation case. This is
shown, at 100\,Pa, by a net reduction in the prograde or superrotating
flow. For the CO$_2$ simulation, in Figure \ref{thermal_co2},
more significant temperature changes are found at 100\,Pa, but as
these occur very close to the poles they contribute negligibly to the
resulting simplified phase curve, and the full and primitive versions
closely match. For the deeper pressure level, the absolute changes are smaller, but distributed closer to the equator resulting in a slight shift in the simplified phase curve. The change in the horizontal wind, shown by the vector arrows, is again a reduction in the strength of the superrorating jet at the
equator, although some more complex changes occur at high latitudes
for the lower pressure surface of the CO$_2$ simulation.

The simulations in this section demonstrate that the flow recovered from simulations adopting either the primitive or full equations of motion differ significantly for `thick atmospheres', where $z_{U}/R_{\rm p}\sim20\%$ or more. The limitation of the shallow--fluid approximation, within the primitive equations, is well known, but here we demonstrate that an important class of exoplanet, warm small Neptunes or Super Earths, potentially inhabit this problematic regime. 

\subsubsection{Traditional
  Approximation} \label{subsubsec:trad_approx} As discussed in
\citet{mayne_2014b} the physical justification for the traditional
approximation is relatively weak, and concerns have been raised over
its validity for thick atmospheres \citep{tokano_2013,mayne_2014b},
and terrestrial planets \citep{gerkema_2008,tort_2015}. Also as
discussed the condition $w \ll v \tan \phi$ indicates whether the
terms omitted by this approximation are negligible or not. It is hard
to diagnose this condition \textit{a priori} as we don't know the
magnitude of the winds. However, this condition can clearly never be
met at the equator where $\tan\phi=0$. Therefore, for this
approximation to hold, and the primitive equations to correctly
capture the flow, the equatorial region where $w \gtrsim v \tan \phi$
must be sufficiently restricted, and connected smoothly to the flow at
mid--latitudes where the condition can more easily be met.

In order to develop an order of magnitude estimate for the typical
$v\tan\phi$ and $w$ values we invoke several assumptions, before
exploring the results from the simulations themselves. The resulting
equations are highly simplified, but allow us to explore the
approximate behaviour of the atmosphere and understand the simulation
results. Firstly, we assume that our standard case is a reasonable
approximation of the global dynamic and thermodynamic structure of
GJ~1214b and similar warm small Neptunes or Super Earths. Given this
assumption, within the dynamically `active' region from
$10^2-10^5$\,Pa (or 1\,mbar to 1\,bar), it can reasonably be assumed
that:
\begin{itemize}
    \item The atmosphere is globally superrotating: this always holds at the equator, and in most cases is also true elsewhere in the atmosphere.
    \item The longitudinal variation in temperature is smaller than the equilibrium day--night temperature contrast (i.e. the jet is acting to redistribute heat and homogenise the temperature).
    \item The meridional winds are predominantly driven by the Coriolis effect on the zonal wind, which follows from our previous two assumptions. 
    \item The gas is incompressible, and in vertical hydrostatic equilibrium where the former is reasonable as the flow speed is negligible compared to the sound speed for the majority of the atmosphere (see Section \ref{sec:assumptions}), and as we have discussed the former holds for all but the lowest pressure regions (Section \ref{subsubsec:hydro_balance}). 
\end{itemize}

For an inviscid incompressible atmospheric flow in a steady state the Euler, mass continuity, energy and ideal gas equations can be written as:
\begin{gather}
(\vec{v} \cdot  \vec{\nabla}) \vec{v} - 2 \vec{\Omega} \wedge \vec{v} = -\dfrac{1}{\rho} \vec{\nabla} p + \vec{g},\label{eq:NS} \\
\dfrac{\mathrm{D} \rho}{\mathrm{D} t} = 0, \label{eq:continuity}\\
\dfrac{\mathrm{D} T}{\mathrm{D} t} = \dfrac{T_{\mathrm{eq}}-T}{\tau_\mathrm{rad}}, \label{eq:energy}
\end{gather}
and
\begin{gather}
p = \rho R T, \label{eq:ideal}
\end{gather}
respectively. Where $\vec{v}$, $\vec{\nabla}$, $\vec{\Omega}$ and $\vec{g}$ are merely the vector forms of the variables and operator previously defined. Using the vector identity $\vec{\nabla} (v^2/2) = (\vec{v} \cdot  \vec{\nabla}) \vec{v} + \vec{v} \wedge (\vec{\nabla} \wedge \vec{v})$ and taking the scalar product of Eqn.\eqref{eq:NS} with $\vec{v}$ we obtain
\begin{equation}
\vec{v} \cdot \vec{\nabla}\bigg(\dfrac{v^2}{2}\bigg) = -\dfrac{1}{\rho} \dfrac{\mathrm{D} p}{\mathrm{D} t} - w g.
\label{eq:kinetik}
\end{equation}
This equation simply shows that the advection of kinetic energy is balanced by the advection of pressure and bouyancy in the atmosphere. 

For an incompressible gas $\vec{\nabla} \cdot \vec{v} = 0$ and therefore $\dfrac{1}{r \cos \phi} \dfrac{\partial u}{\partial \lambda} + \dfrac{1}{r} \dfrac{\partial v}{\partial \phi} +\dfrac{\partial w}{\partial r} = 0$ suggesting the standard order of magnitude estimate of
\begin{gather}
    \dfrac{U}{L_U} \sim \dfrac{W}{H_W},
    \label{eq:U-W}
\end{gather}
where $U$ is an order of magnitude for $u$, $W$ for $w$, $L_U$ is a characteristic horizontal length for $u$ and $H_W$ a characteristic scale height for $w$. Hydrostatic balance then implies that the vertical component of the advection of $p$ is balanced by $wg$,
\begin{equation}
    -\dfrac{1}{\rho}\dfrac{D p}{D t} - wg = -\dfrac{1}{\rho} (\vec{v} \cdot \vec{\nabla}_{\bot}) p \sim \dfrac{1}{\rho}\dfrac{D p}{D t},
    \label{eq:advection_P}
\end{equation}
where $\vec{\nabla_{\bot}}$ is the horizontal gradient. The second order of magnitude equality arises from the fact that $U$ is dominating the other winds, and therefore we do not expect the total advection to differ strongly from the horizontal advection. We have verified this point in our simulations. Further by substituting for $p$ using Eqn. \eqref{eq:ideal} using \eqref{eq:energy} we obtain
\begin{equation}
\dfrac{Dp}{Dt} =  \rho R \dfrac{DT}{Dt} = \rho R  \dfrac{(T_\mathrm{eq}-T)}{\tau_\mathrm{rad}}.
\label{eq:P-T}
\end{equation}
Combining Eqns.\eqref{eq:kinetik}, \eqref{eq:advection_P} and \eqref{eq:P-T} yields
\begin{equation}
\vec{v} \cdot \vec{\nabla}\bigg(\dfrac{v^2}{2}\bigg) \sim R\dfrac{(T_\mathrm{eq}-T)}{\tau_\mathrm{rad}} \\, 
\end{equation}
which can be simplified to
\begin{equation}
\dfrac{U^3}{2L_U} \sim R\dfrac{\Delta T}{\tau_\mathrm{rad}},
\label{eq:U_est}
\end{equation}
where $\Delta T$ is the order of magnitude of the difference between the actual temperature in the atmosphere and the corresponding temperature at radiative equilibrium.  As we assume that the temperature variations with longitude are negligible compared to the equilibrium day--night contrast, we can further state that the steady--state temperature around the planet is approximately constant at  $T\sim(T_\mathrm{eq, day}+T_\mathrm{eq, night})/2$, where $T_\mathrm{eq, day}$ and $T_\mathrm{eq, night}$ are the average day and night side equilibrium temperatures. Therefore, on average the difference between the steady state temperature and equilibrium temperature is  $\Delta T \sim (T_\mathrm{eq, day} - T_\mathrm{eq, night})/2$. 


For our simulations the characteristic length scale for the zonal jet is about half the planetary circumference or $\sim\pi R_{\rm p}$ and we expect $\Delta T\sim 0.5 \Delta T_{\rm eq}$ (or half the total day--night contrast, see Section \ref{sec:model}) which is $300$\,K and $400$\,K for the Std and dT$+$ simulations, respectively. At 100\,Pa, where the maximum zonal velocity is found the radiative timescale is $\sim 10^4s$. Therefore, using these values in Eqn. \eqref{eq:U_est} yields
\begin{equation}
U \sim \sqrt[3]{{2 \pi R_{\rm p}} R \dfrac{\Delta T}{\tau_\mathrm{rad}}} \sim 1.4 \times 10^3 {\rm ms}^{-1}.
\label{eq:U_estimate}
\end{equation}
Our maximum zonal velocities are close to, or within approximately a factor 2 of this value which is a remarkable level of agreement given the approximations we have invoked to derive the estimated wind speeds, giving us confidence in our expressions. Eqn. \eqref{eq:U_estimate} can then be substituted into Eqn. \eqref{eq:U-W} to provide an estimate for $W$, 
\begin{equation}
W \sim \dfrac{H}{L} \sqrt[3]{{2 \pi R_{\rm p}} R\dfrac{\Delta T}{\tau_\mathrm{rad}}},
\label{eq:estimate_w}
\end{equation}
where $H$ is now the typical height scale for $W$, $L$ the typical horizontal scale for the typical zonal velocity, $U$ (i.e. we have dropped the subscripts used in Eqn. \eqref{eq:U-W}).

To estimate $V$, as stated we assume that the meridional motions are primarily driven by the Coriolis force. The meridional component of the momentum equation, Eqn. \eqref{eq:NS}, assuming that the Coriolis term is dominant can be written as
\begin{equation}
    \dfrac{Dv}{Dt} \sim 2 \Omega U \sin{\phi}, \\
\end{equation}
which can be approximated by
\begin{equation}
    \dfrac{UV}{L_V} \sim 2 \Omega U \sin{\phi}, \\
\end{equation}
where $L_V$ is a characteristic scale for $V$, and this expression can further be simplified to
\begin{equation}
\dfrac{UV}{L_V} \sim 2 \Omega U \sin{\phi} \\
V \sim 2 L_V \Omega \sin \phi.
\end{equation}
In our simulations which $L_V \sim \pi/4 R_{\rm p}$,  yielding
\begin{equation}
V \sim \dfrac{\pi}{2} R_{\rm p} \Omega \sin \phi \sim 400 \sin{\phi} \ \  {\rm ms}^{-1}.
\label{eq:v_estimate}
\end{equation}
Again this estimate matches the simulation results remarkably well to within a factor $\sim$2. A final, simple manipulation provides an estimate for $V\tan\phi$,
\begin{equation}
V \tan \phi \sim \dfrac{\pi}{2} R_{\rm p} \Omega \dfrac{\sin \phi^2}{\cos \phi}, 
\label{eq:estimate_v}\\
\end{equation}
which can be combined with Eqn. \eqref{eq:estimate_w}, to explore the global behaviour of the $w \ll v \tan \phi$ condition.

The equations we have derived (Eqns. \eqref{eq:estimate_w} and \eqref{eq:estimate_v}) to estimate the magnitude of the flow are valid in the upper atmosphere where the radiative timescale is short and temperature gradient high (and friction is low). This is also the region, as we have shown, where the zonal flow is dominant and the largest differences between the primitive and full equations of motion are found. Eqns. \eqref{eq:estimate_v} and \eqref{eq:estimate_w} are linked by a common assumption namely that the flow is dominated by the $U$ component which is balancing the pressure (temperature) gradient. Eqn. \eqref{eq:estimate_v} then requires the further assumption that the $V$ component is driven, chiefly, by the Coriolis forces, where Eqn. \eqref{eq:estimate_w} requires the assumption that the flow is divergence free. The derivation of these equations, however, does not rely on adopting either the primitive or full equations explicitly. In Appendinx \ref{app:equations} we explicitly verify these equations using our simulation outputs. 

Although very simplified, without a proper prescription of $H$, $L$ or
$\Delta T$, these Eqns. \eqref{eq:estimate_w} and
\eqref{eq:estimate_v} allow us to draw insight into the nature of the
flow. Firstly, as the rotation rate ($\Omega$) increases $V\tan\phi$
increases, strengthening the validity of the traditional
approximation. This is the same effect observed in the simulations
presented in Section \ref{subsubsec:shallow_fluid}. Essentially, the
faster the rotation the more valid the primitive equations will
become, all else being equal. The second insight provided by the order
of magnitude estimates, Eqns \eqref{eq:estimate_v} and
\eqref{eq:estimate_w}, is that as the radiative forcing is increased
i.e. $\Delta T$, which acts to increase the horizontal advection, the
$W$ velocities will increase and the primitive equations become less
valid. Additionally, the presence of the ideal gas constant in the
estimate for $W$ shows that the composition of the atmosphere, and
mean molecular weight will also play a role in the validity of the
traditional approximation. Finally, the aspect ratio of the
atmosphere, $H/L$, also appears in Eqn. \eqref{eq:estimate_w},
confirming the well known limit of the primitive equations to flows
with small aspect ratios.

In addition \citet{tort_2015} have shown that the traditional
approximation becomes increasingly less valid as, in their case,
$\Omega$ is increased, yet $\Omega R_{\rm p}$ remains constant, for
the terrestrial regime. If $\Omega R_{\rm p}$ is constant, Eqn
\eqref{eq:estimate_v} is clearly constant for a given $\phi$. Eqn
\eqref{eq:estimate_w} can be expressed as
$W\propto \frac{R_{\rm p}^{1/3}}{L}$ which, using $L\sim R_{\rm p}$
becomes $W\propto \frac{1}{R_{\rm p}^{2/3}}$ which at constant
$\Omega R_{\rm p}$ can be expressed as $W\propto
\Omega^{2/3}$. Therefore, at constant $\Omega R_{\rm p}$ as $\Omega$
is increased $W$ increases, while $V$ is constant, meaning the
condition $w \ll v \tan \phi$ becomes increasingly less valid.

Eqns. \eqref{eq:estimate_v} and \eqref{eq:estimate_w} also make interesting predictions for the atmospheric interaction with condensates. Larger vertical velocities are predicted for increasing levels of irradiation which could potentially lead to larger particles sizes being `lofted' to high altitudes compared to less irradiated planets. Additionally, the meridional velocities increase with rotation, potentially  leading to more meridional mixing of cloud material as found by \citet{lines_2018}.

In the remainder of this section, we explore the results from the simulations where the rotation rate, forcing or temperature contrast, mean molecular weight and gravity are varied demonstrating the regimes where the primitive equations become inaccurate. 

\paragraph{Composition} results from the CO$_2$ simulation, where the mean molecular weight of the atmosphere has been increased, have already been introduced in Figures \ref{shallow_uvel_bar} and \ref{thermal_co2}, as part of the shallow--fluid discussion (Section \ref{subsubsec:shallow_fluid}). For the CO$_2$ simulation the ratio of scaleheight of the atmosphere, to the planetary radius is similar to that of the $R_{\rm p}+$ simulation, being around 1\% and 3\%, respectively (see Tables \ref{basic_par} and \ref{model_names}). Therefore, the residual differences between the resolved primitive and full flows for the CO$_2$ case, which are not observed to the same extent in the $R_{\rm p}+$ simulation, are likely caused by the change in the mean molecular weight itself, driving a change in the $W$ field, as show in Eqn \eqref{eq:estimate_w}. Figure \ref{shallow_uvel_bar} shows the peak velocities of prograde equatorial jet extending down to shallower depths, and therefore higher pressures in the full case, compared to that of the primitive case. Although the value of $R$ has changed by an order of magnitude moving from the standard simulation (compare values from Table \ref{basic_par} with those from Table \ref{model_names}), Eqn. \eqref{eq:estimate_w} shows that $W\propto \sqrt{R}$, suggesting a less significant change in the flow is expected. 

\begin{figure*}
\begin{center}
  \subfigure[$\Omega +$ Prim: 800-1\,000\,days]{\includegraphics[width=8.5cm,angle=0.0,origin=c]{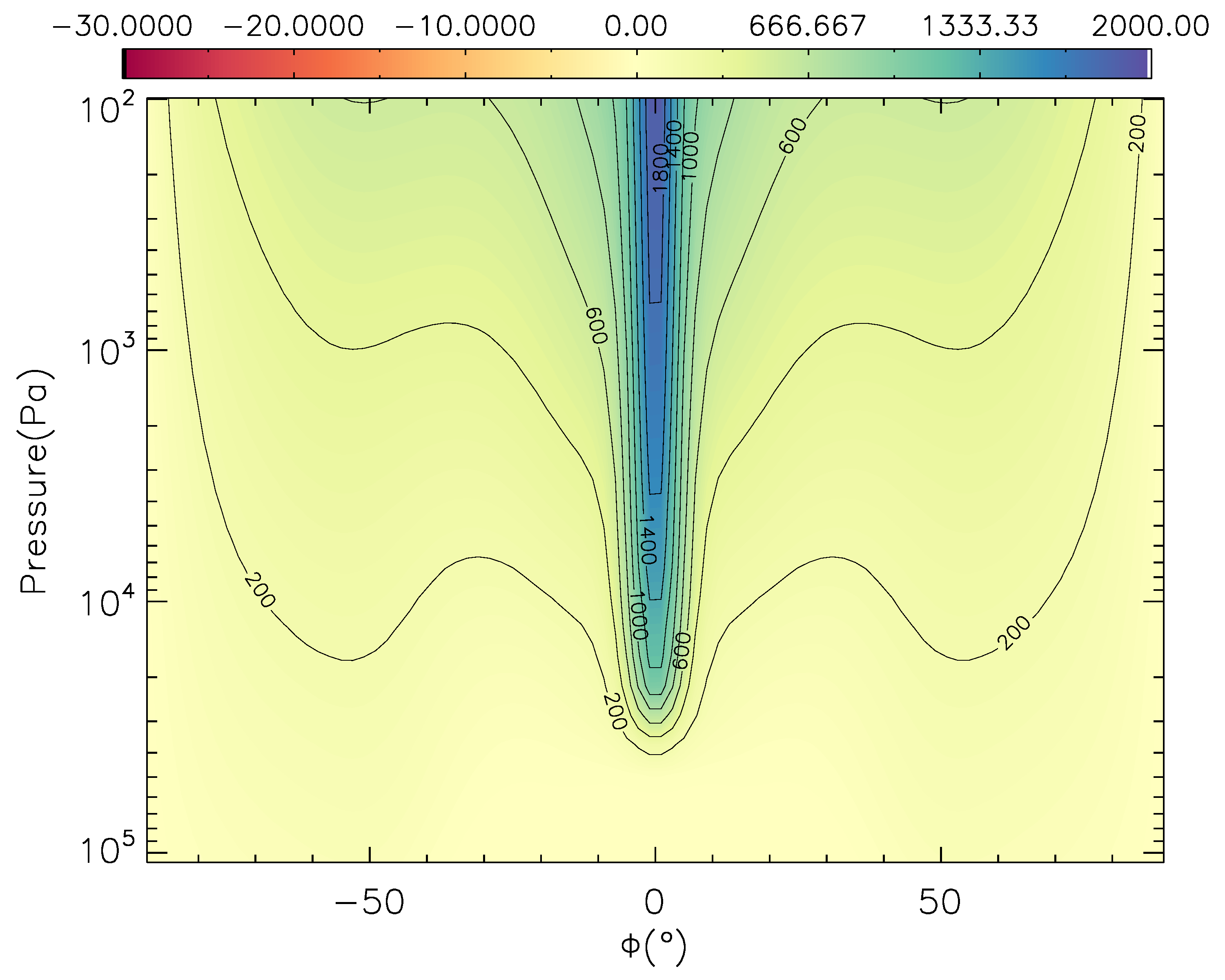}\label{omega_prim_800_1000_uvel_bar}}
  \subfigure[$\Omega +$ Full: 800-1\,000\,days]{\includegraphics[width=8.5cm,angle=0.0,origin=c]{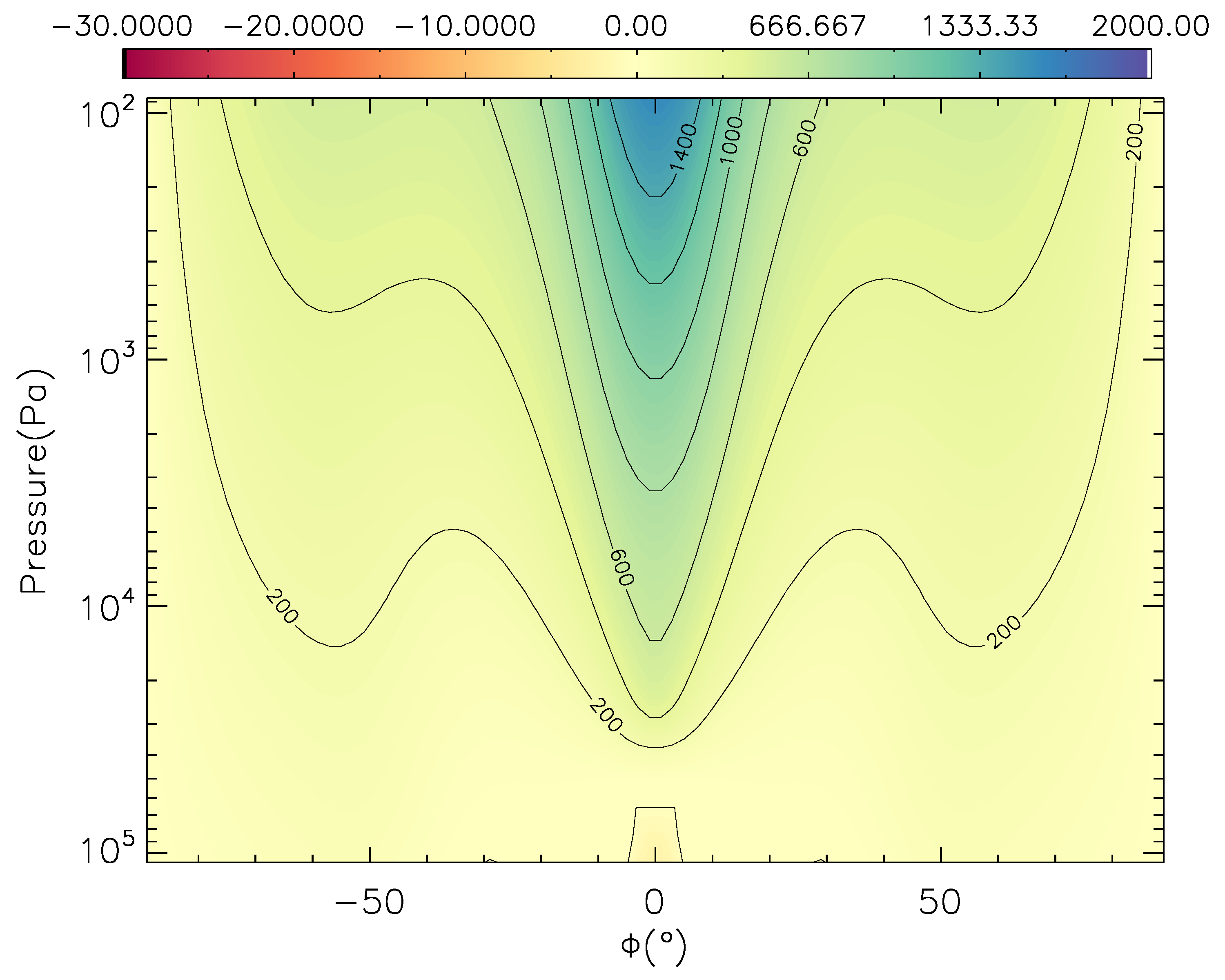}\label{omega_full_800_1000_uvel_bar}}
 \end{center}
\caption{Figure similar to Figure \ref{std_uvel_bar} but showing the
  relative lack of difference between the ``primitive'' and ``full''
  equations for a faster rotating planet (see
  Table \ref{model_names} for explanation of simulation
  names). \label{omega_uvel_bar}}
\end{figure*}

\paragraph{Rotation rate} Figure \ref{omega_uvel_bar} shows the zonal--mean, temporal--mean zonal wind for the $\Omega +$ Prim and $\Omega +$ Full simulations. For these simulations the setup is the same as the standard case but with an increased rotation rate (see Table \ref{model_names}). As predicted both by our order of magnitude analysis the difference in the flow, between the primitive and full equations, is reduced when increasing the rotation rate, as the flow becomes dominated by Coriolis forces.

\begin{figure*}
\begin{center}
  \subfigure[$\Omega+$ Full: 100\,Pa]{\includegraphics[width=8.5cm,angle=0.0,origin=c]{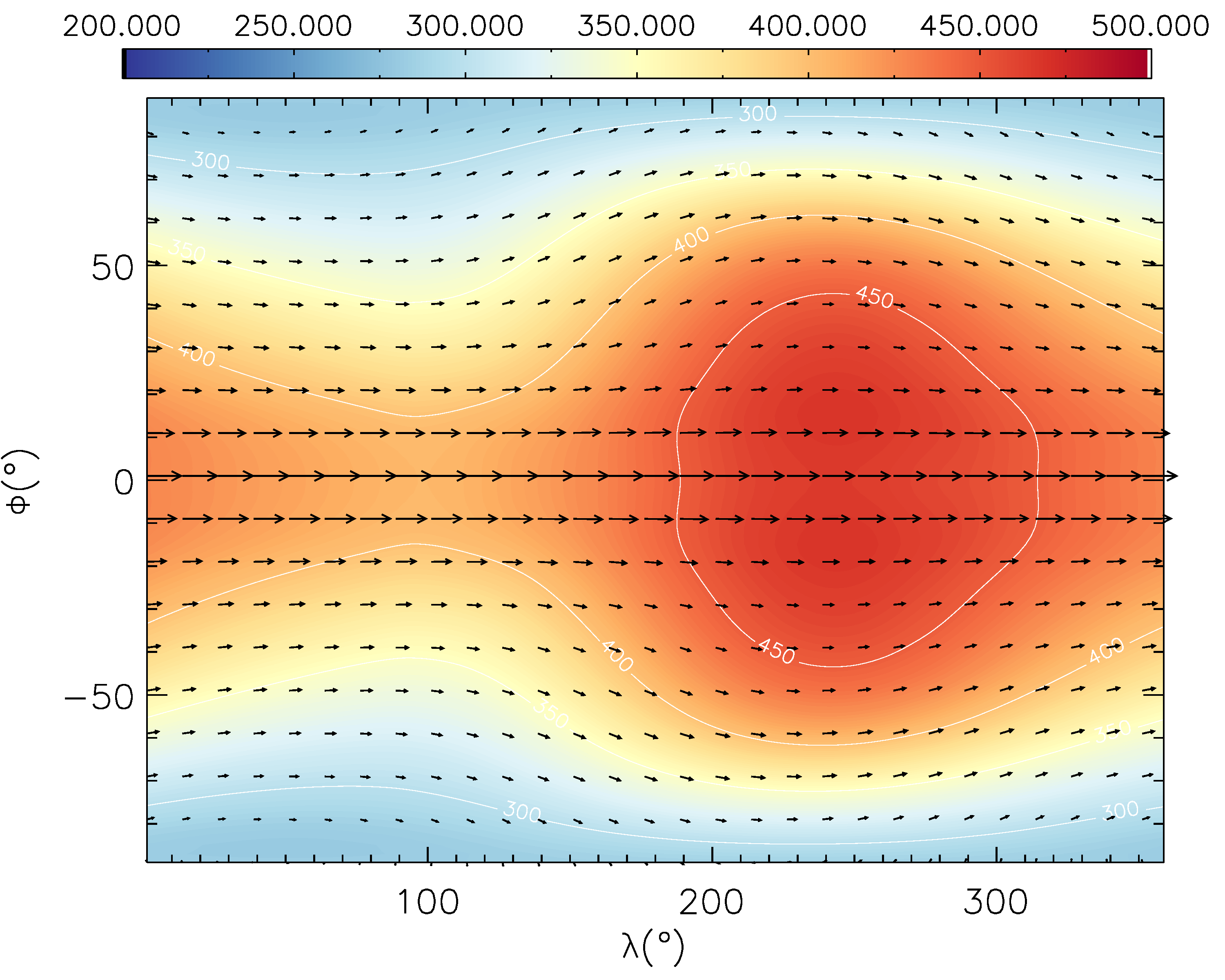}\label{omega_full_1000_100_slice}}
  \subfigure[$\Omega+$ Full: 3\,000\,Pa]{\includegraphics[width=8.5cm,angle=0.0,origin=c]{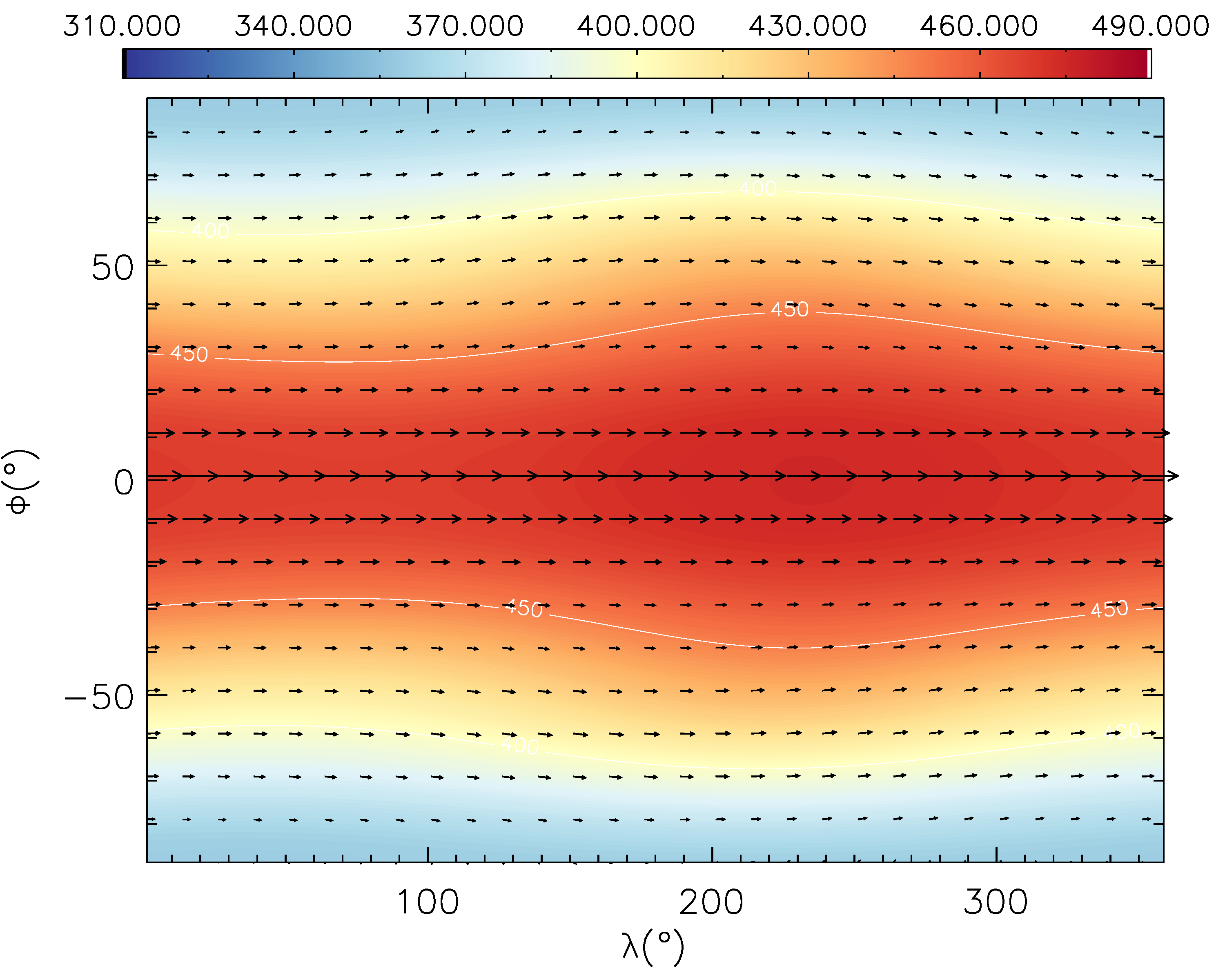}\label{omega_full_1000_3000_slice}}
  \subfigure[$\Omega+$ Full-Prim: 100\,Pa]{\includegraphics[width=8.5cm,angle=0.0,origin=c]{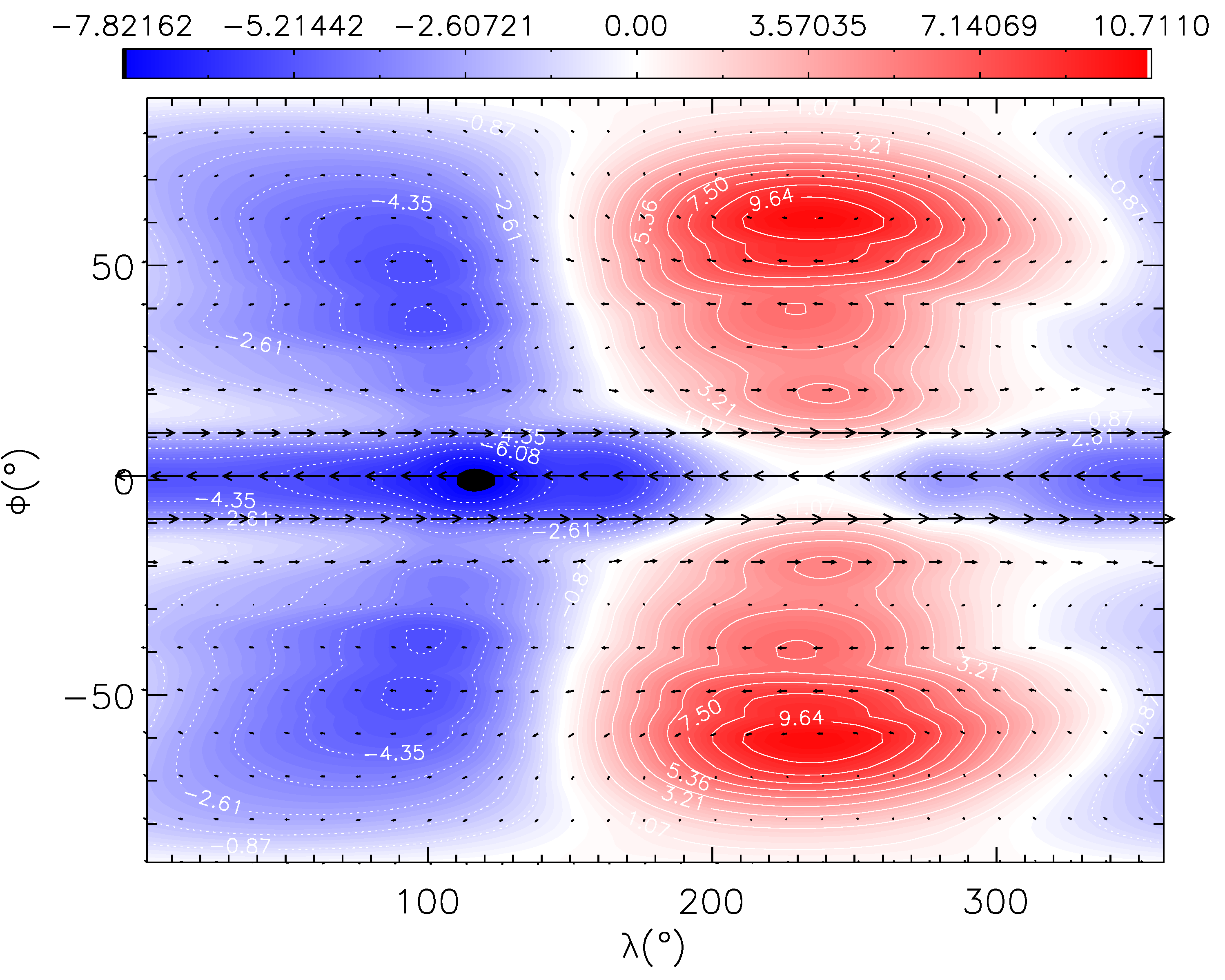}\label{omega_diff_1000_100_slice}}
  \subfigure[$\Omega+$ Full-Prim: 3\,000\,Pa]{\includegraphics[width=8.5cm,angle=0.0,origin=c]{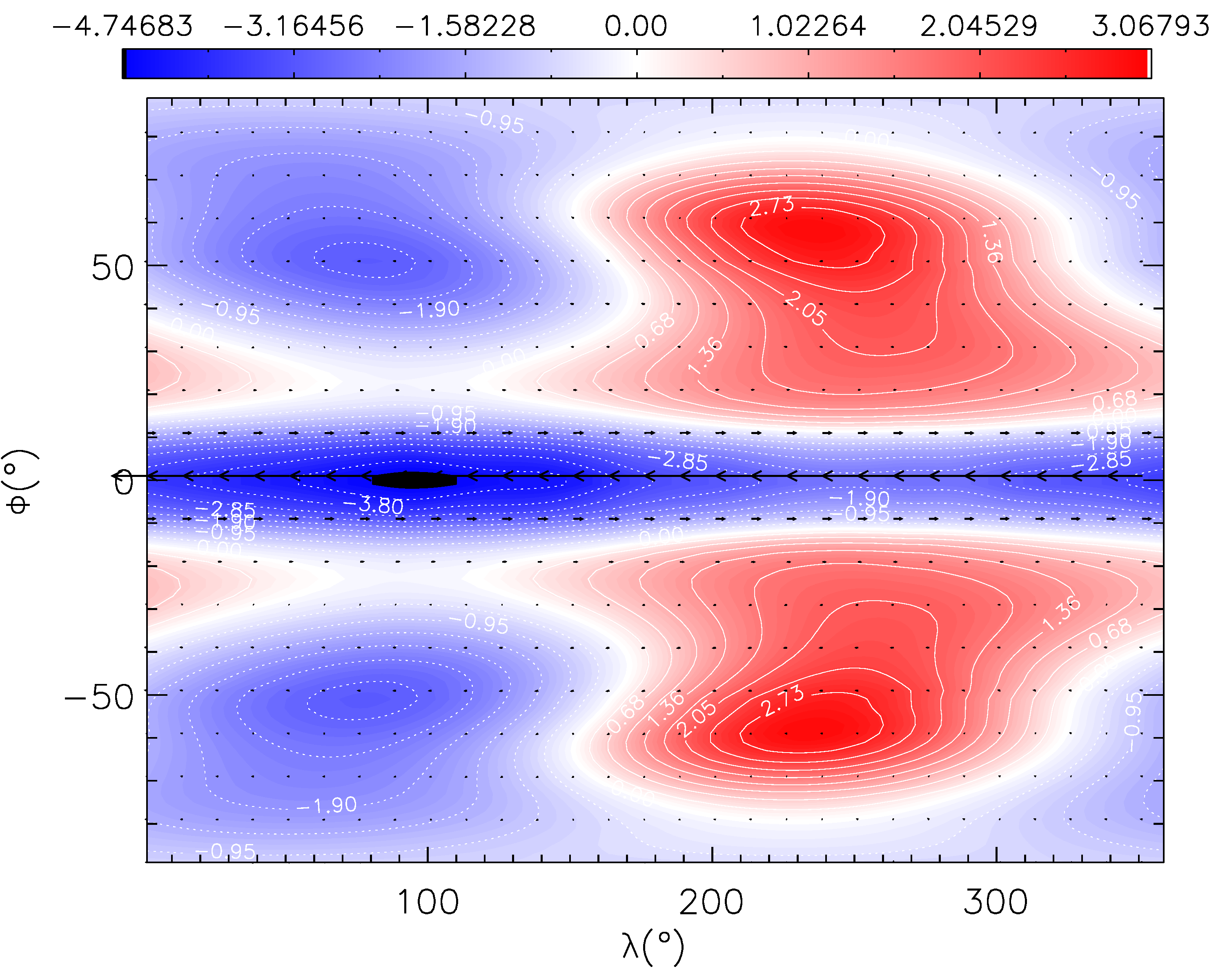}\label{omega_diff_1000_3000_slice}}
  \subfigure[$\Omega+$ Prim/Full: 100\,Pa]{\includegraphics[width=8.5cm,angle=0.0,origin=c]{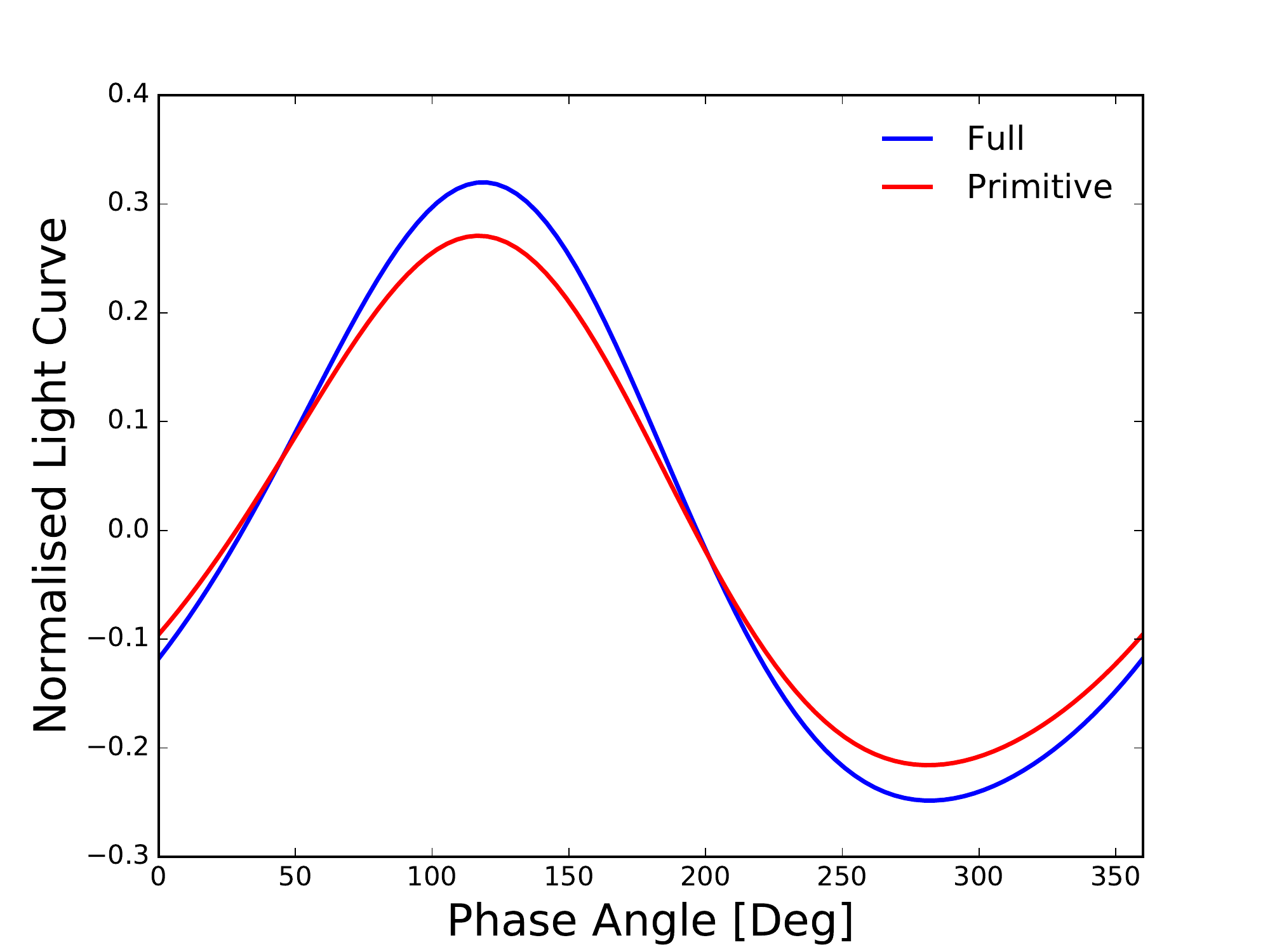}\label{phase_omega_100}}
  \subfigure[$\Omega+$ Prim/Full: 3\,000\,Pa]{\includegraphics[width=8.5cm,angle=0.0,origin=c]{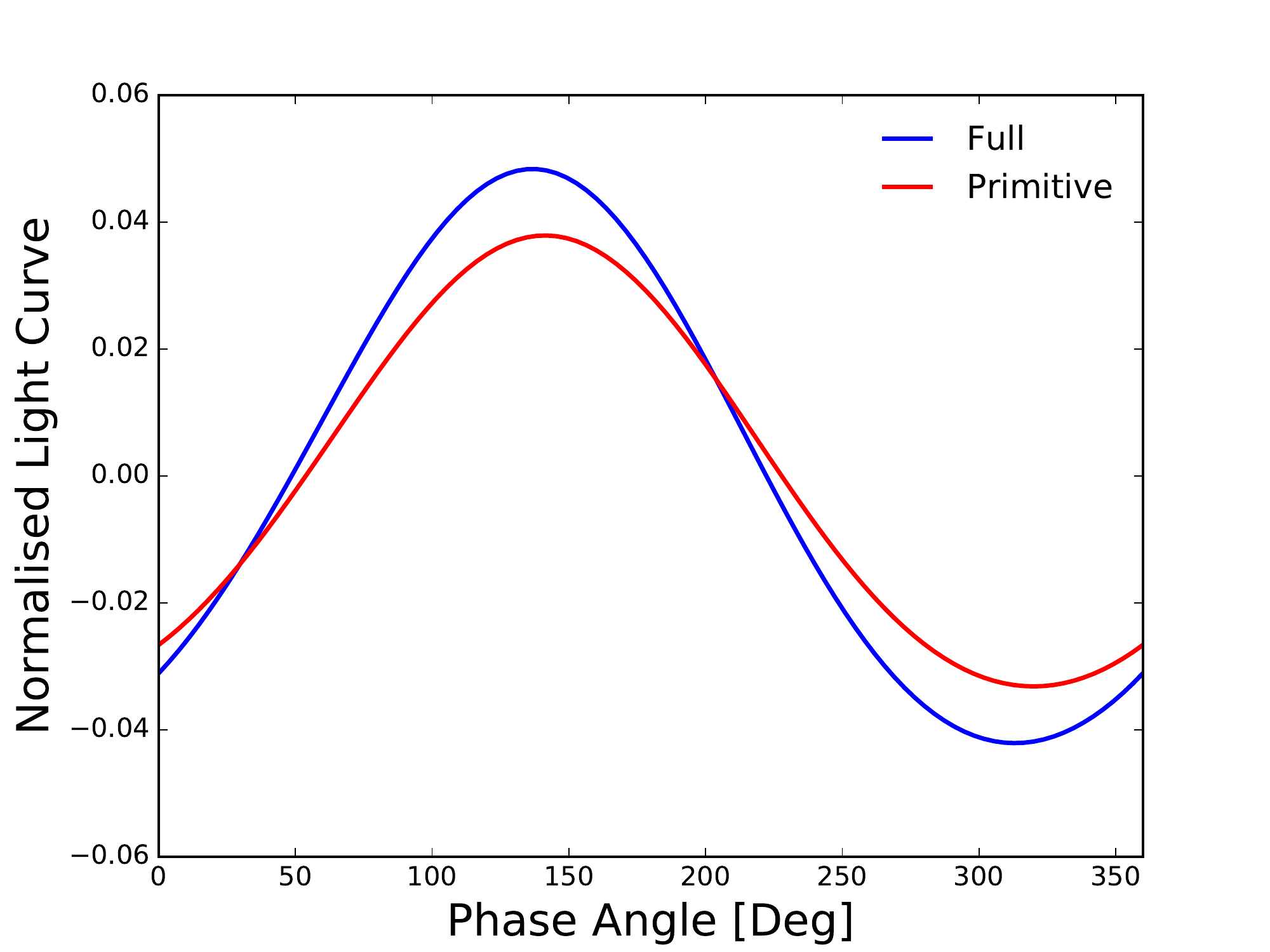}\label{phase_omega_3000}}
\end{center}
\caption{As Figure \ref{thermal_std} but for the $\Omega+$ simulations (see Table \ref{model_names} for
  explanation of simulation names). Note the change in the vertical
  axes for the \textit{bottom panels}. \label{thermal_omega}}
\end{figure*}

The differences in zonal flow between the primitive and full $\Omega +$ simulations (see Figure \ref{omega_uvel_bar}), result
in changes to the temperature structure and simplified thermal phase
curve which are shown in Figure \ref{thermal_omega}. As
for the standard case (Figure \ref{thermal_std}) a net warming is
shown on the dayside and a net cooling on the nightside for the
$\Omega+$ simulations, at both pressure levels, albeit at a reduced
level when moving to the more complete equations. For the higher
pressure, a very small shift in the peak amplitude of the simplified
phase curve is found, as temperature changes at this level are around
a few degrees. 

\paragraph{Forcing/advection and gravity} Eqn \eqref{eq:estimate_w}
shows that as the temperature contrast is increased $W$ increases,
leading to a weakening of the $w \ll v \tan \phi$
condition. Additionally, the acceleration due to gravity does not
appear in our estimate of $V\tan\phi$, but does enter the estimate for
$W$ via the scaleheight ($H\propto RT/g$) (Eqns \eqref{eq:estimate_v}
and \eqref{eq:estimate_w}, respectively). To demonstrate, and test the
impact of this, we have run a set of simulations (dT$+$, see Table
\ref{model_names}), matching the standard setup, but with an increased
temperature contrast solving the primitive, deep and full equations of
motion \citep[see][for explicit definition of equation
sets]{mayne_2014b}. The deep and full equations differ only by the
assumption of a gravity constant with height in the former.

\begin{figure}
\begin{center}
  \subfigure[dT$+$ Deep: 800-1\,000\,days]{\includegraphics[width=8.5cm,angle=0.0,origin=c]{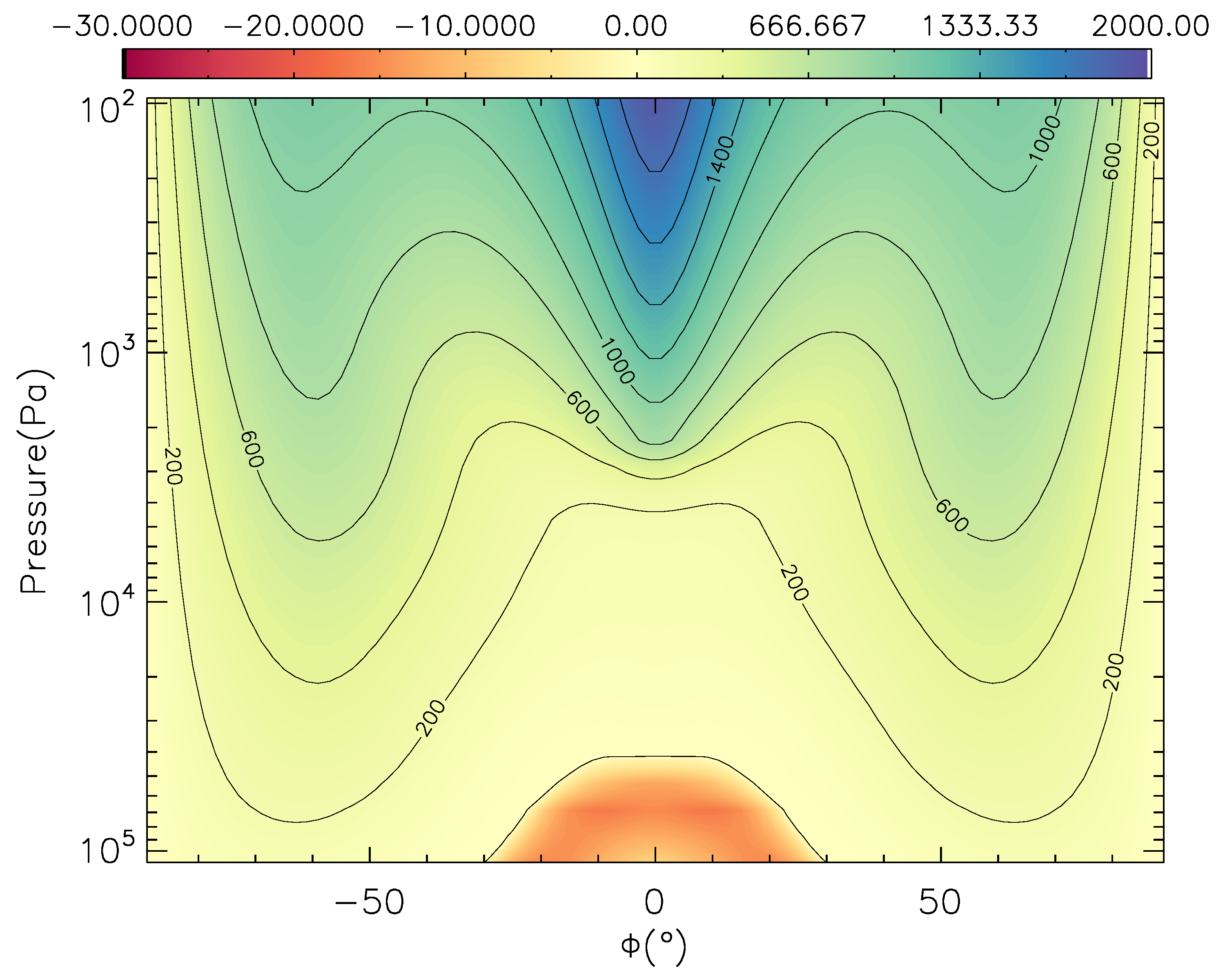}\label{dt_deep_800_1000_uvel_bar}}
  \subfigure[dT$+$ Prim: 800-1\,000\,days]{\includegraphics[width=8.5cm,angle=0.0,origin=c]{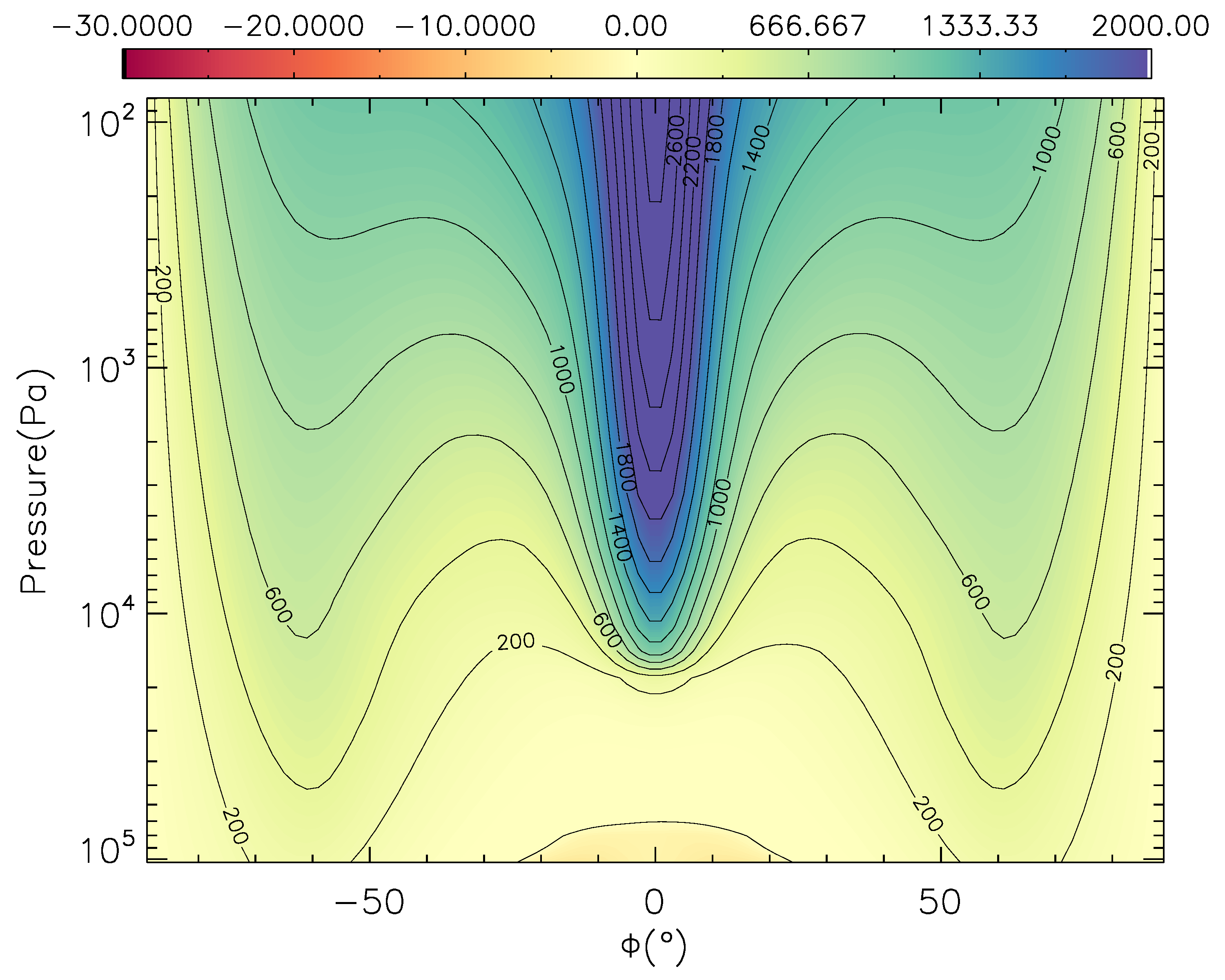}\label{dt_prim_800_1000_uvel_bar}}
  \subfigure[dT$+$ Full: 800-1\,000\,days]{\includegraphics[width=8.5cm,angle=0.0,origin=c]{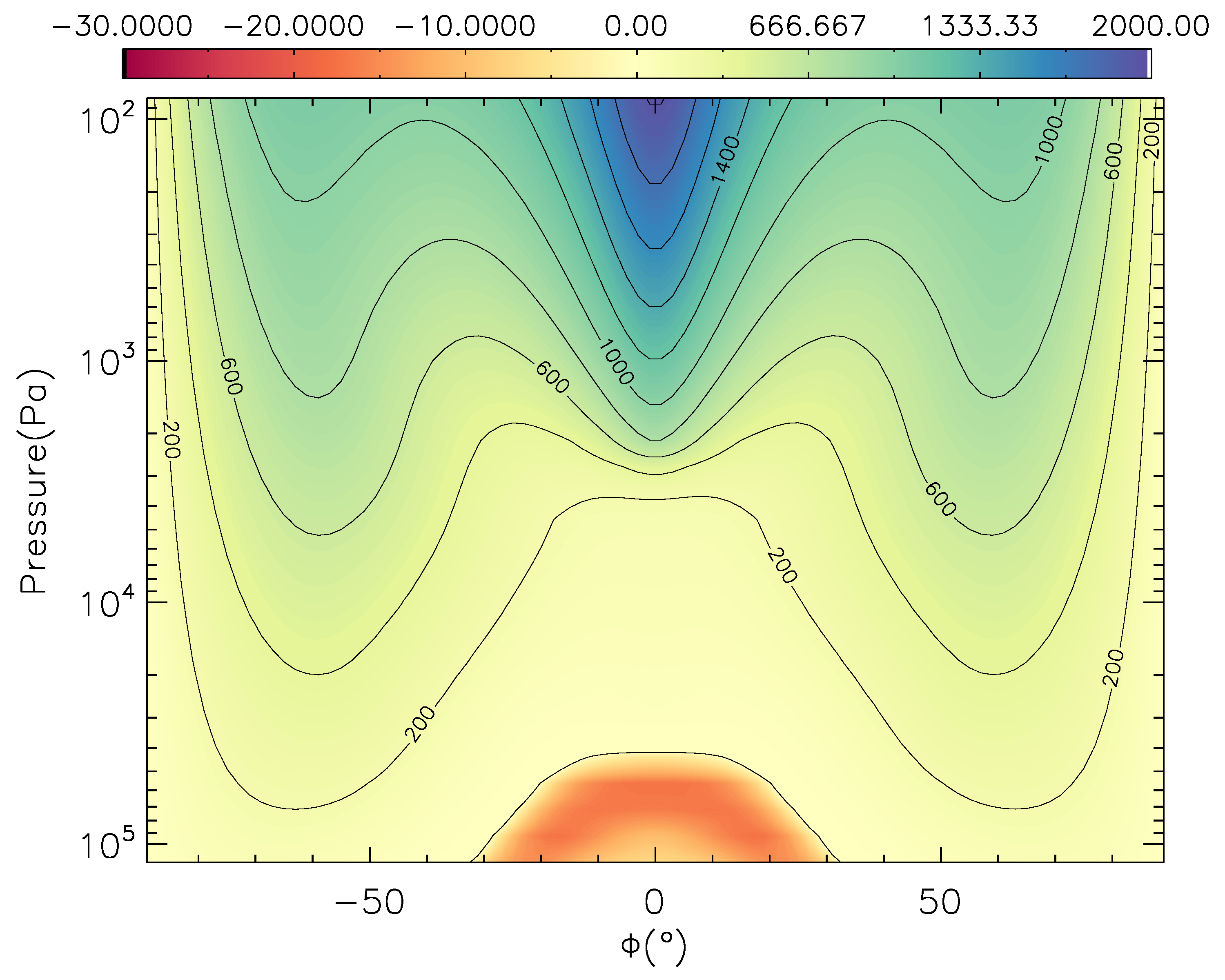}\label{dt_full_800_1000_uvel_bar}}
\end{center}
\caption{Figure similar to Figure \ref{std_uvel_bar}
  but for the case with an increased day--night temperature contrast, dT$+$ simulation but including the version solving the ``deep'' equations (i.e. ``full''
  but with constant gravity, see Table \ref{model_names} for
  explanation of simulation names). \label{dt_uvel_bar}}
\end{figure}

\begin{figure*}
\begin{center}
  \subfigure[dT$+$ Full: 100\,Pa]{\includegraphics[width=8.5cm,angle=0.0,origin=c]{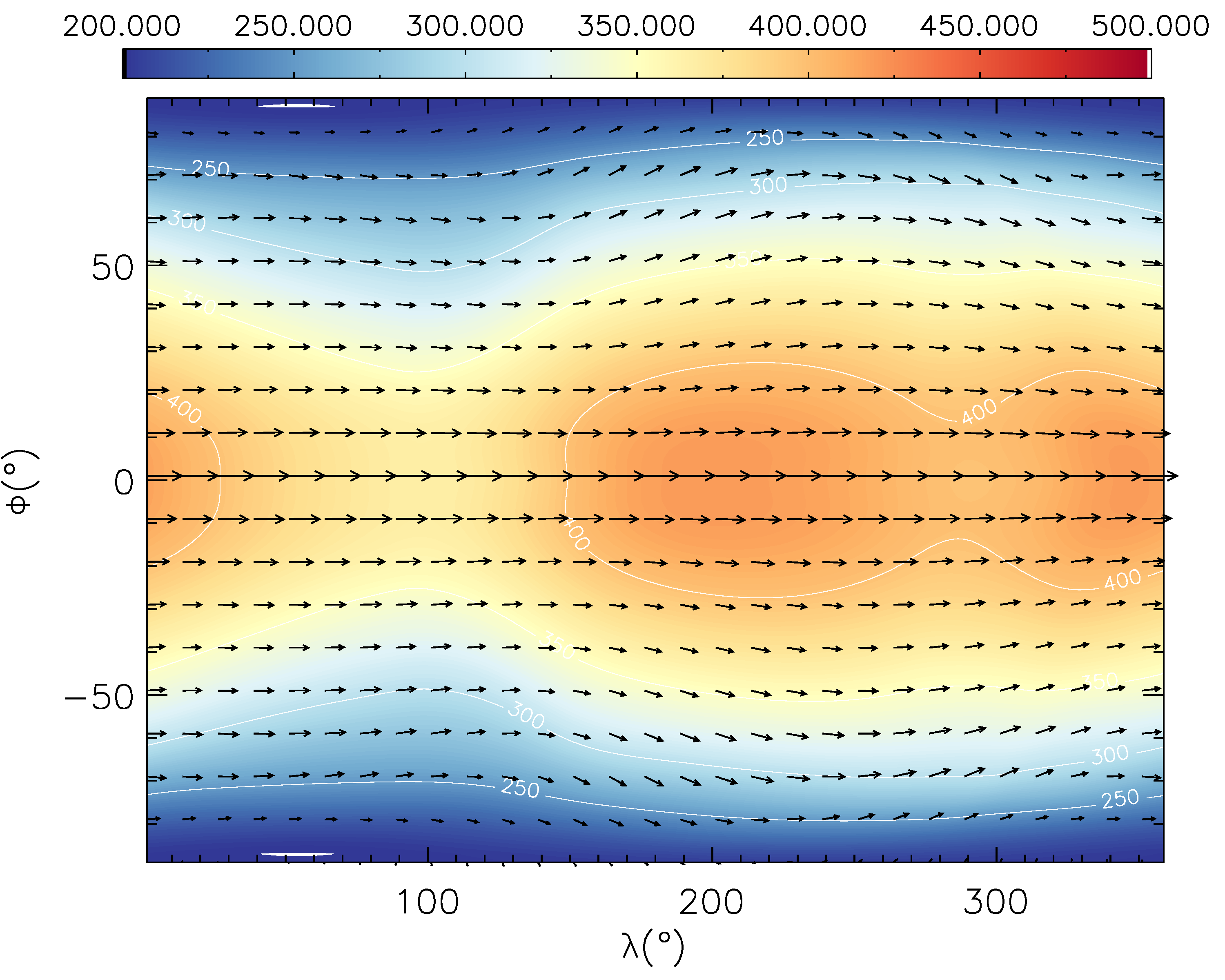}\label{dt_full_1000_100_slice}}
  \subfigure[dT$+$ Full: 3\,000\,Pa]{\includegraphics[width=8.5cm,angle=0.0,origin=c]{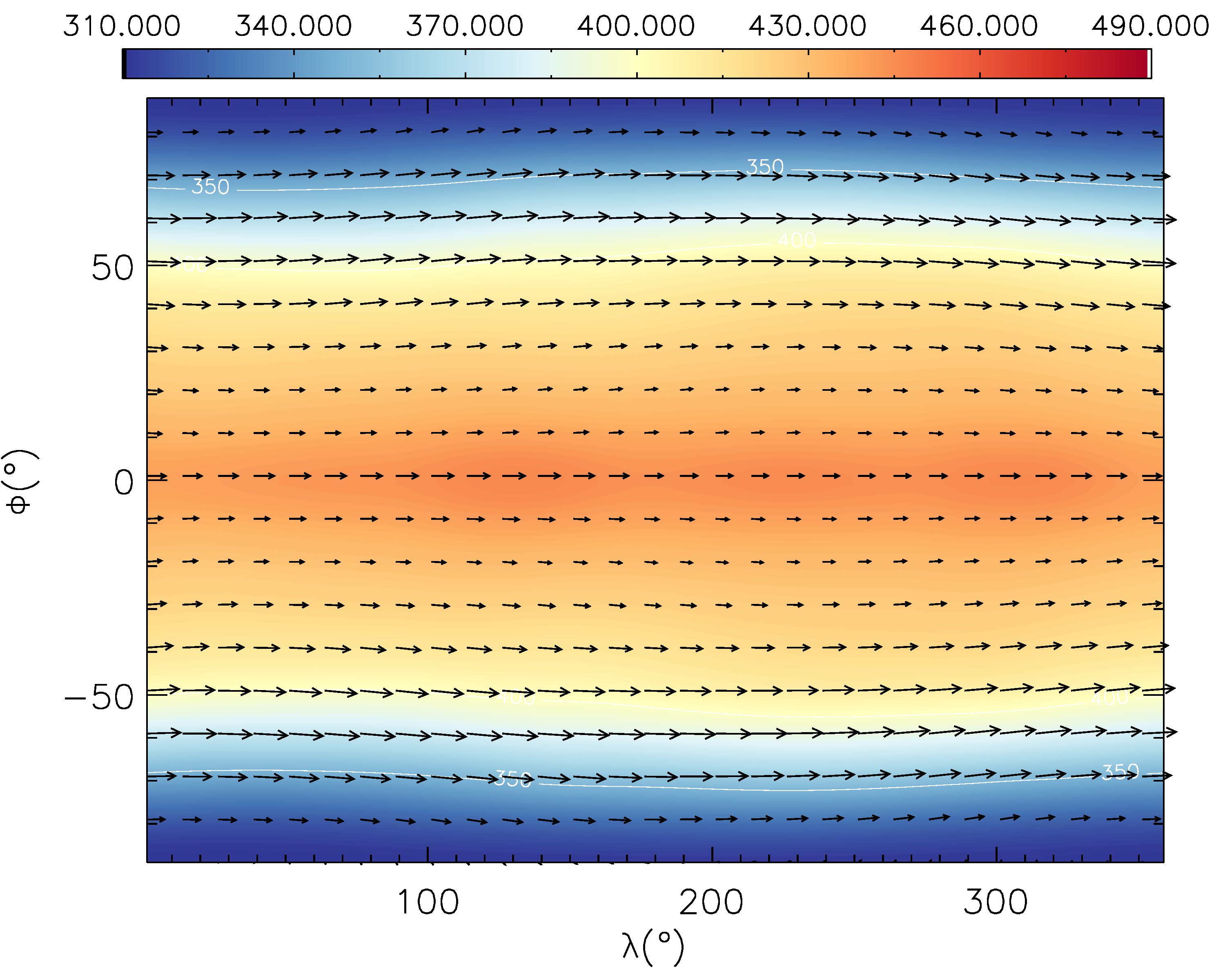}\label{dt_full_1000_3000_slice}}
  \subfigure[dT$+$ Full-Prim: 100\,Pa]{\includegraphics[width=8.5cm,angle=0.0,origin=c]{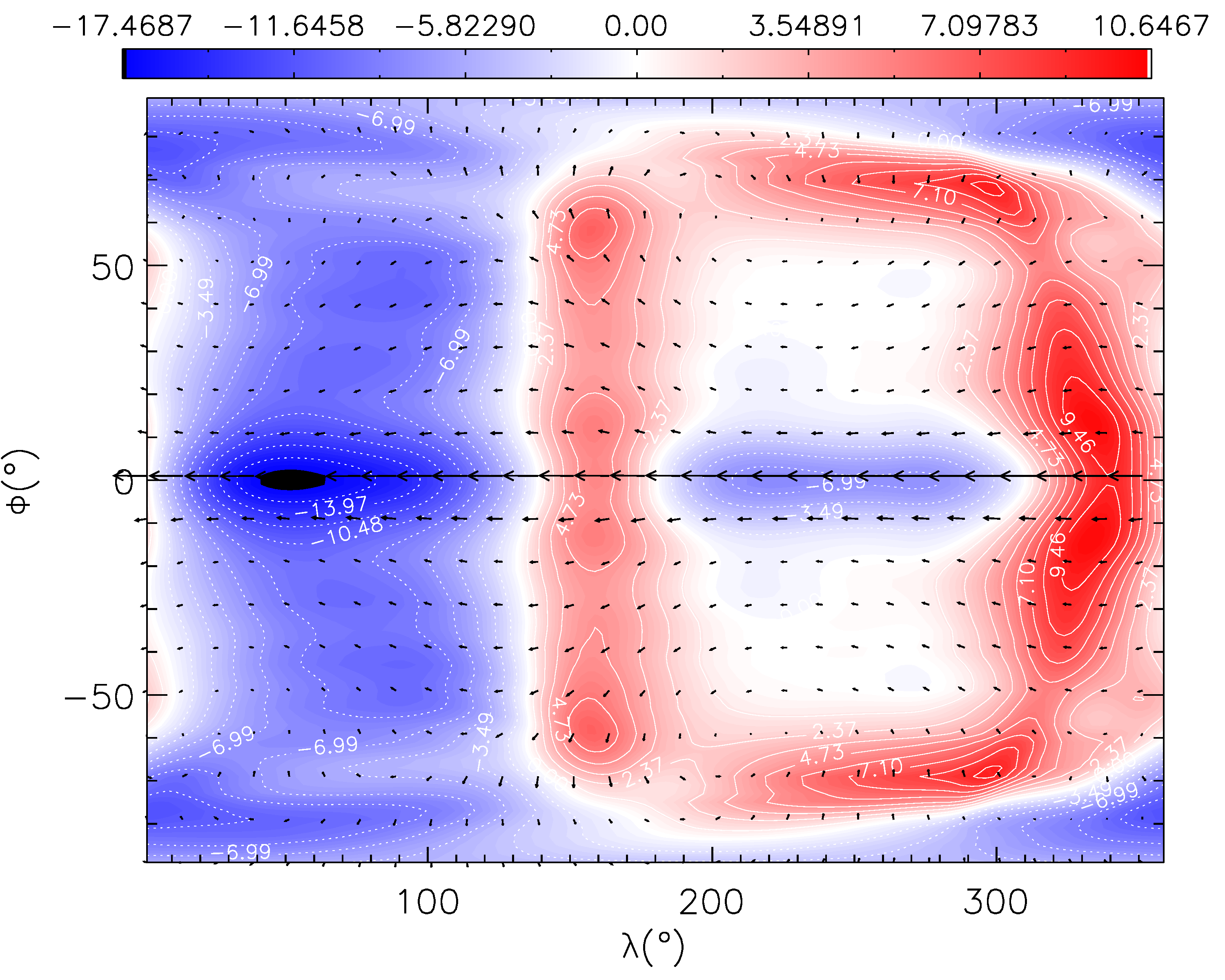}\label{dt_diff_1000_100_slice}}
  \subfigure[dT$+$ Full-Prim: 3\,000\,Pa]{\includegraphics[width=8.5cm,angle=0.0,origin=c]{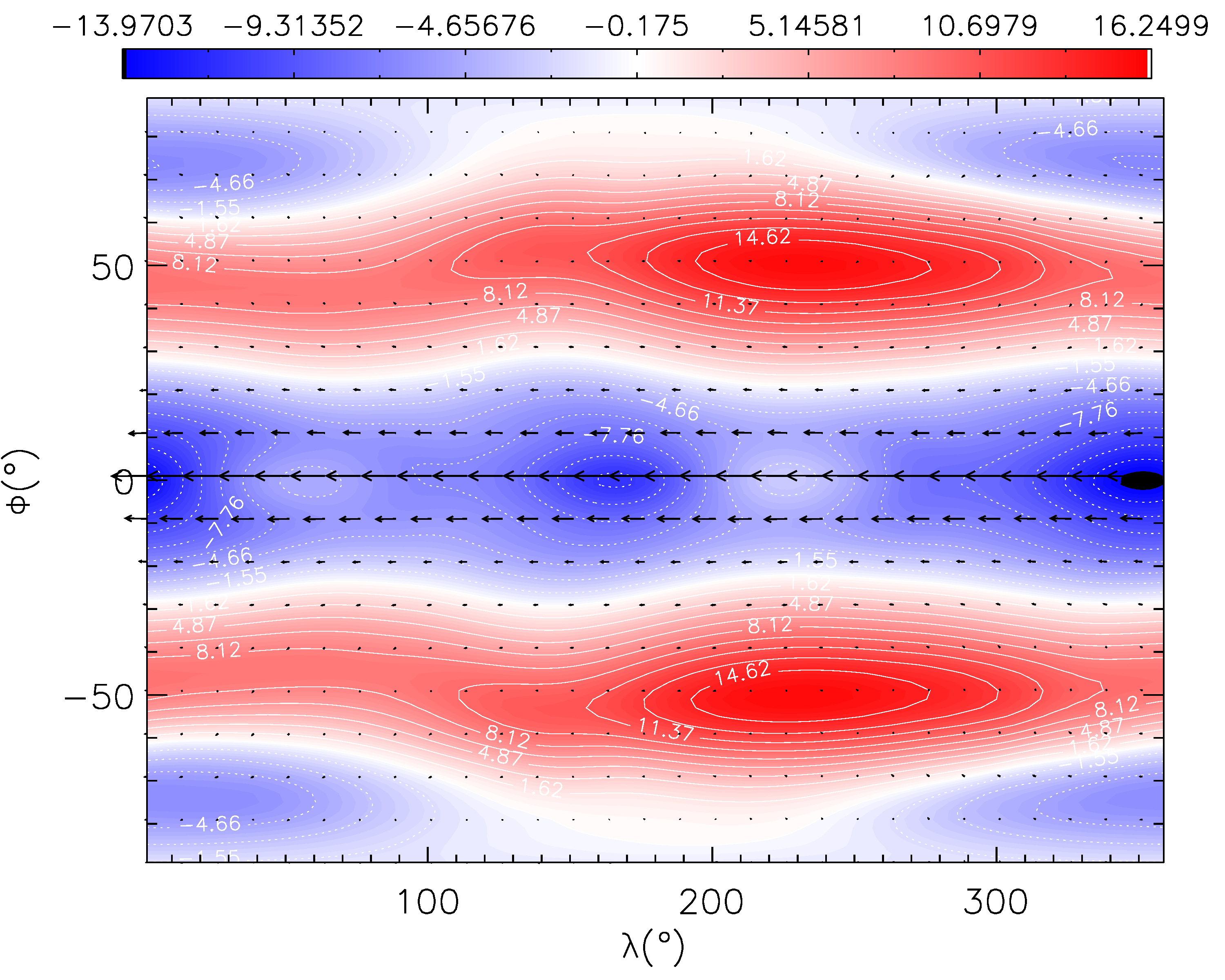}\label{dt_diff_1000_3000_slice}}
  \subfigure[dT$+$ Prim/Full: 100\,Pa]{\includegraphics[width=8.5cm,angle=0.0,origin=c]{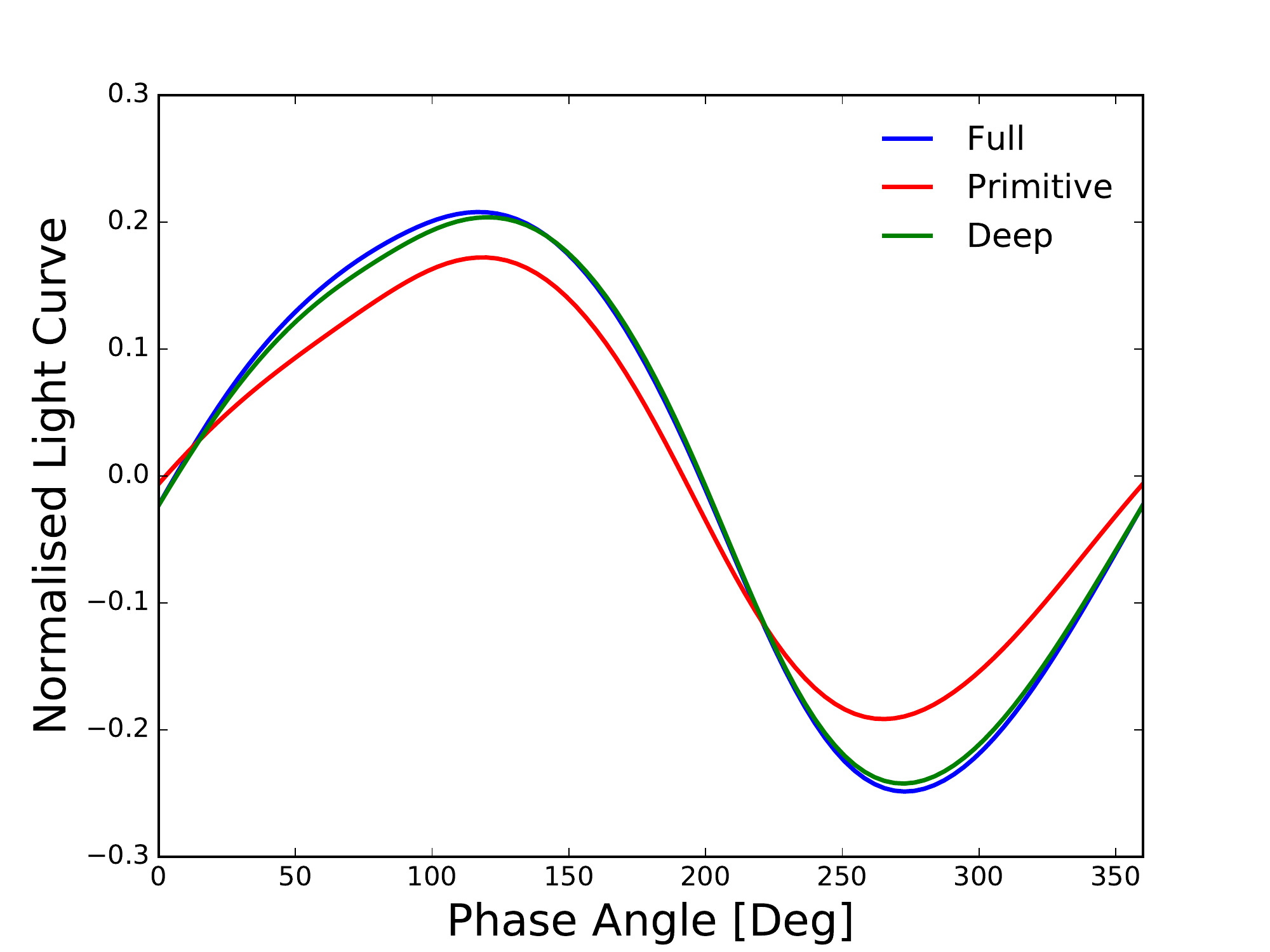}\label{phase_dt_100}}
  \subfigure[dT$+$ Prim/Full: 3\,000\,Pa]{\includegraphics[width=8.5cm,angle=0.0,origin=c]{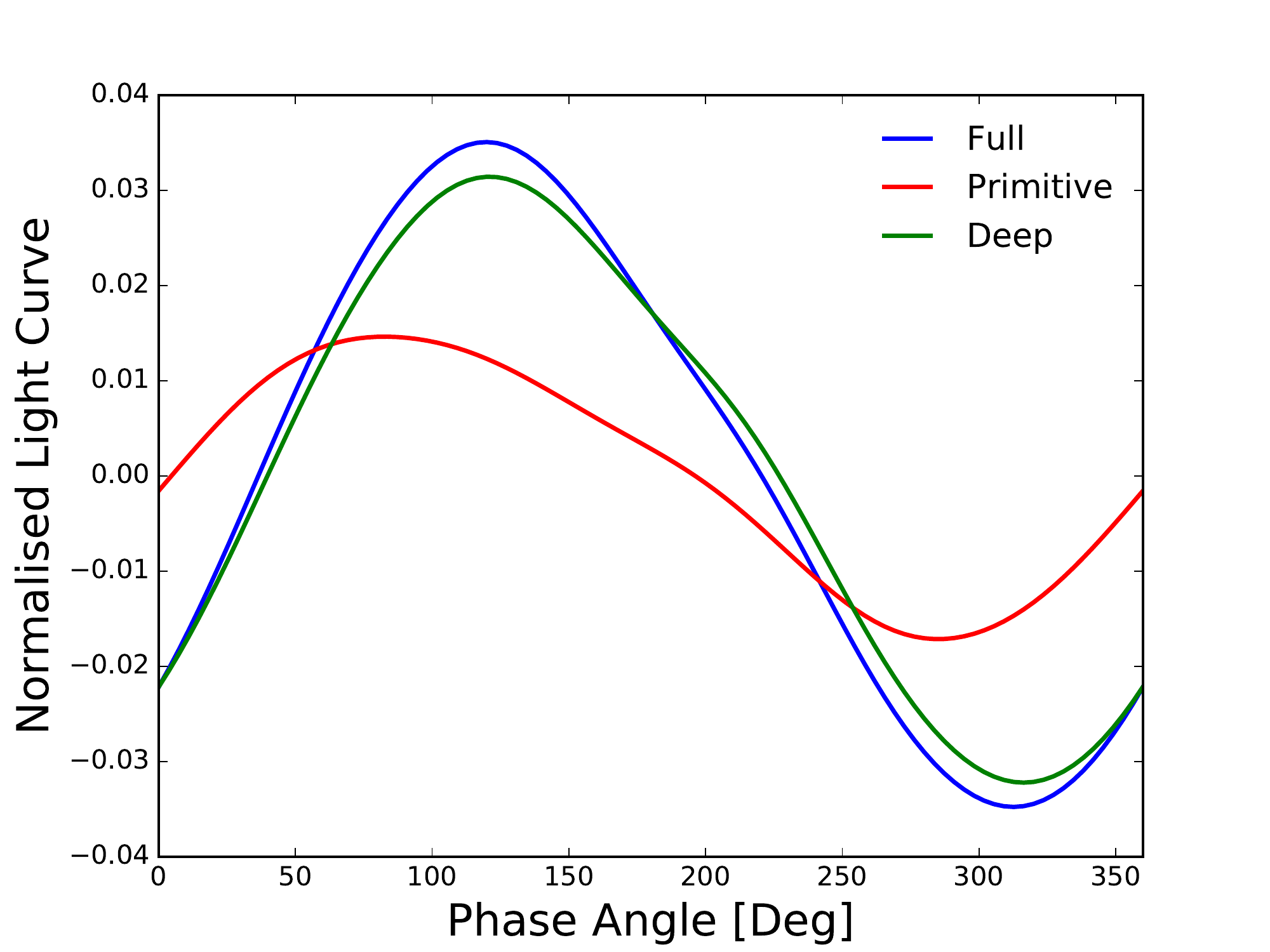}\label{phase_dt_3000}}
\end{center}
\caption{As Figure \ref{thermal_std} but for the dT$+$
  simulations (see Table \ref{model_names} for explanation of
  simulation names). The phase curves include the additional deep
  simulation. Note the change in the vertical
  axes for the \textit{bottom panels}. \label{thermal_dt}}
\end{figure*}

Figure \ref{dt_uvel_bar} shows the wind structure with an increased
planetary temperature contrast (thereby increasing the zonal wind
speed).  The difference between the primitive and full simulations is
enhanced when the temperature contrast is increased and the resulting
zonal flow velocity increased, thereby moving the flow to a more
advectively dominated regime (\textit{bottom panel}). Figure
\ref{dt_deep_800_1000_uvel_bar} then shows the same setup as the dT$+$
Full and dT$+$ Prim cases but enforcing a constant gravity with height
\citep[termed the `deep' equations, see][]{mayne_2014b}. Importantly,
the resulting flow matches, more closely, the full equation case
indicating that the changes are independent of the treatment of
gravity. This is perhaps unsurprising given that the maximum variation
in $g$ between the constant and varying case, found at the top of the
atmosphere, is only $\sim30$\%, with the difference rapidly reducing
with altitude, $\propto 1/r^2$. Additionally, the $z_{\rm top}$, and
therefore vertical resolution in \emph{height} is identical between
the dT$+$ Prim and dT$+$ Deep, meaning that as the flow from the dT$+$
Deep simulation matches that of the dT$+$ Full simulation the vertical
\emph{height} resolution is not the cause of the differences between
the primitive and full setups. Figure \ref{thermal_dt} then shows the
temperature structure and phase curves, as in Figure \ref{thermal_std}
but for the dT$+$ case. A similar picture emerges, and it is important
to note that the dT$+$ Deep simulation matches the dT$+$ Full
simulation versions very closely as one would expect from Figures
\ref{std_uvel_bar} and \ref{dt_uvel_bar}.

Figures \ref{thermal_dt} shows the resulting thermal structure, and simplified thermal phase curve for the dT$+$ Prim and Full simulations (the simulation solving the deep equations are omitted due to its similarity with the full case) in the same format as Figure \ref{thermal_std}. As in the standard case Figure \ref{thermal_dt} show significant
changes in the thermal structure, and subsequent phase curve between
the primitive and full equations for our setups with
enhanced temperature contrast. The changes in the zonal flow (see
Figure \ref{dt_uvel_bar}) clearly translate to alterations in the
temperature structure, and the changes shown in Figure \ref{thermal_dt} are similar to those found in the standard setup (Figure \ref{thermal_std}).

In summary, the changes in the atmospheric dynamics are found when moving from solving the
primitive to full dynamical equations for cases where the flow is
strong (and rotation weak), and the atmosphere is thicker in relation
to the planetary radius. These changes in the advection, result in changes to the temperature
structure. Specifically, we have shown, using simplified phase curves,
that moving from the primitive to full equations can affect the
day-night temperature contrast and the location of the hot spot. This
indicates a net change in the overall heat redistribution efficiency
of these atmospheres when modeled using dynamical equations of
differing simplicity. Essentially, the validity and applicability of the primitive equations of dynamics becomes questionable as the atmosphere becomes thicker (i.e. $z_{U}/R_{\rm p}$ becomes larger, shallow--fluid approximation) and $w \ll v\tan\phi$ is violated (the traditional approximation). In Appendix \ref{app:equations}, Figure \ref{fig:break} we show the direct violation of this limit from our simulations (namely the Std, $R_{\rm p}+$, dT$+$ and $\Omega+$ simulations). These conditions are well known, but we have demonstrated that an important class of exoplanet could well exist within the regime where the primitive equations are no longer valid namely warm/hot small Neptunes or Super Earths.

\section{Assumptions and Limitations} \label{sec:assumptions}


The inner boundary is modelled as a friction free impermeable boundary
with no heat or mass exchange \citep[aside from the prescribed
internal heat flux in the radiative transfer simulations presented in
Appendix \ref{app_sec:radiative}, see][]{amundsen_2016}. Therefore, we
have effectively modeled GJ~1214b as a gas giant planet, without a
treatment of a potential rocky (or ocean) surface nor inclusion of a
`drag' \citep[see discussion in][]{mayne_2014b}. However, as our inner
boundary pressure is $200\times 10^5$\,Pa, and the atmosphere
vertically extended, the inclusion of a surface treatment is unlikely
to alter our results as radiation does not significantly penetrate to
the surface itself. Additionally, the near surface layers of the
atmosphere are not flowing rapidly, and will therefore not be strongly
affected by frictional effects. Furthermore,
\citet{mayne_2014b,mayne_2017} show that simulations without a bottom
`drag' return the same qualitative flow as those
without. Additionally, \citet{tort_2015}, have already shown that the
traditional approximation weakens, for terrestrial planets, as
$\Omega$ is increased but $\Omega R_{\rm p}$ is held constant, (see
discussion in Section \ref{subsubsec:trad_approx}), where we have
focused on varying the atmospheric extent.

Our model assumes that the self--gravity of the atmosphere is
negligible, i.e. the mass of the modeled atmosphere, $M_{\rm atm}$ is
negligible compared to the mass of the bulk planet comprising the
unmodeled material within the inner boundary, $M_{\rm p}$. For our
basic setup the adopted surface gravity, $g_{\rm p}=$12.20\,ms$^{-2}$
and planetary radius, $R_{\rm p}=1.45\times10^7$\,m imply
$M_{\rm p}\sim 3.8\times 10^{25}$\,kg and the standard simulations
indicate $M_{\rm atm}\sim 5.1\times 10^{21}$\,kg, suggesting that
this is a reasonable assumption. For the purposes of this study we
have simply altered the planetary radius without a commensurate change
in the gravity, while the atmospheric extent has also been altered to
preserve a similar pressure range between simulations. For the reduced
and increased planetary radii simulations
$M_{\rm atm}\sim 9.7\times 10^{20}$ and $2.3\times 10^{23}$\,kg,
respectively, and the assumed planetary masses (as the surface gravity
is not altered) are, $M_{\rm p}\sim 6.58\times 10^{24}$ and
$1.83\times 10^{27}$\,kg, respectively, indicating that neglecting
self--gravity remains a reasonable approximation for all
simulations. Additionally, although vertically extended the total
atmospheric mass, in our modelled domain
($\sim 200\times 10^5$--$10$\,Pa), remains a plausible fraction of the
total planetary mass, i.e. less than 1\% \citep{elkins_2008,
  lopez_2014}.

As discussed the vertical extent of our atmosphere is set to retain a
similar pressure range across all simulations, corresponding to
$\sim 200\times 10^5$\,Pa to $\sim 10$\,Pa, allowing both
inter--comparison of our simulations, and comparison with the results
of \citet{zhang_2017}. The model calculates hydrostatic balance for an
input temperature--pressure profile, which is derived using a 1D
radiative transfer code \citep[selected as on the mean $T_0$ profile
from][]{zhang_2017}, an initial inner boundary pressure (see Table
\ref{basic_par}) and the selected height. The selected height is set
so that a similar pressure range for all simulations is achieved. The
pressure structure will then evolve, including that at the inner and
outer boundary, however the evolution does not result in a
significantly different maximum or minimum pressures. This means that the height
of our atmosphere is effectively set by the gravity and inner boundary
pressure we select, where the gravity is approximately consistent with
the measured planetary mass.

One of our main conclusions rests on the modeled atmospheric height becoming
comparable to the defined planetary radius. Therefore, could the
differences be erased by selecting a different planetary radius and
atmospheric height? For our simulations the inner boundary is placed at pressures high enough so as to capture the entire `dynamically active'
region of the atmosphere. The atmospheric flow we are simulating often extends across the entire pressure or
height range we simulated, with only small quiescent regions close to the inner
boundary, most importantly for the Std Full and Std Prim cases. This means that
restricting our height range will likely alter the dynamics, and not
capture the atmospheric flow correctly. Our inner boundary pressure,
chosen to be consistent with \citet{amundsen_2016} is also consistent
with the typical radiative--convective boundary for small Neptune
sized planets at a few Gyrs which is $10^7-10^8$\,Pa
\citep{lopez_2014}. The planetary radius is usually measured using the
optical transit, and can be estimated to be the radius at 20\,mbar or
$2\times 10^3$\,Pa \citep{hubbard_2001}. Here we have set this radius
as our inner boundary at a pressure of $\sim 200\times 10^5$\,Pa, with
the standard simulation reaching $2\times 10^3$\,Pa at
$\sim 2\times 10^6$\,m. Therefore, our actual $R_{\rm p}$ could
plausibly be reduced from $1.4\times 10^7$ to $\sim 1.2\times 10^7$\,m for the standard setup i.e. fixing the radial distance out to a pressure of $2\times 10^3$\,Pa as the measured value of $1.4\times 10^7$\,m, and therefore, deriving a new value for our inner boundary location such that the height above the inner boundary of $2\times 10^3$\,Pa matches our simulation value ($\sim 2\times 10^6$\,m.). This reduction in assumed planetary radius, or radial distance to the inner boundary would act to
enhance the differences we have found between the primitive and full
equations.

An estimate of the sound speed is $c_{\rm s}\sim\sqrt{\gamma RT}$, with $\gamma=1.5$ and temperatures in the jet region of $\sim 500$\,K this yields $c_{\rm s}\sim 1600$\,ms$^{-1}$ for our simulations ($\sim 400$\,ms$^{-1}$ for the CO$_2$ simulations). Therefore the approximate Mach number in our simulations is less than unity throughout the atmosphere except, in some cases, the jet core, where it can reach $\sim 1.5$ in isolated regions. It is possible shocks might occur in these regions, dissipating energy from the flow, a mechanism which is not captured in our model. However, this study is concerned with the \textit{relative} changes in flow between the primitive and full simulations, and the maximum wind speed for our simulations can effectively be set by the level of dissipation, which is a free parameter. Additionally, studies of the effect of shocks in the atmospheres of hot Jupiters, sharing several characteristics with the targets of this paper, have shown their effects to be negligible on the large--scale dynamics of the atmospheres \citep{fromang_2016}.

\section{Conclusions} \label{sec:conclusions}

Modeling the 3D dynamics of planetary atmospheres can be performed at
varying levels of simplification. Much progress has been made using
models adopting the primitive equations, but more recent results have
been derived using models adopting the more complete ``full''
equations \citep[following the nomenclature of][]{mayne_2014b}.

In this work we have demonstrated with a set of self--consistent
simulations that the dominant zonal flow recovered adopting the
primitive or full dynamical equations differs markedly for
GJ~1214b. Additionally, we have demonstrated that the change in the
zonal--mean, zonal flow results in changes to the thermal structure,
via a change in the efficiency of the heat redistribution. The
day-night temperature contrast and location of the hot spot differ
markedly between pairs of simulations, using the simplified and more
complete dynamical equations for several cases. In other words the
simplifications involved in the derivation of the primitive equations
begin to break down, specifically the so--called `shallow--fluid' approximation
where the atmosphere is assumed to be small in vertical
extent compared to the radius of the bulk planet, and the traditional approximation. We have shown that
these differences are most apparent for slowly rotating planets, where
advective terms dominate over the Coriolis terms, i.e. planets in
short--period orbits which are strongly heated driving fast
winds. Here we stress that our standard simulations (i.e. Std Full,
Std Prim, Std Hires Full \& Std Hires Prim) which are setup to be
physically based on GJ~1214b, clearly exhibit this effect. Our
remaining simulations which are less physically--based are then used
to explore the parameter space. 

Although the validity of the `shallow--fluid' and traditional
approximations is not a new consideration \citep[see][and references
therein in the context of Earth]{gerkema_2008,tort_2015}, we have
shown that for an exceedingly important class of exoplanet i.e., Super
Earths or small Neptunes, with potentially large, vertically extended
atmospheres in tidally--locked orbital configurations, the dynamical
solutions obtained from treatments adopting the primitive equations
are likely to become inaccurate. Indeed our conclusions complement
those of \citet{tort_2015} who find the traditional approximation
weakens for faster rotating terrestrial planets on the condition that
$\Omega R_{\rm p}$ is held constant (see Section
\ref{subsubsec:trad_approx}). Importantly, our numerical setups are
practically identical between pairs of simulations solving the
primitive and full equations. The fact that the hydrostatic
approximation appears to hold for the atmospheres we have simulated
also suggests that the quasi--hydrostatic deep atmosphere--equations
\citep[e.g.,][]{white_1995,tort_2015} should also yield comparable
results to the `full' equations.

Given the potential for these planets to show departures from chemical
equilibrium driven by the advection
\citep{madhusudhan_2016,venot_2017}, inaccuracies in the derived
dynamical structure of the atmosphere could have significant impacts
on the inferred underlying chemical and thermodynamic structures. This
class of objects is also important as it represents the regime
bridging giant planets with terrestrial planets \citep{lopez_2014}. 

The next steps for this study would be to move to a full radiative
transfer solution \citep[as used in][]{amundsen_2016}, relax the
assumption of chemical equilibrium \citep{drummond_2018b,drummond_2018c} and include
the cloud scheme adopted in \citet{lines_2018}. These next steps would
allow us to investigate the effect of the differences in the dynamics
on both the chemistry and cloud structures, but also to derive more
accurate synthetic observations for comparison with real data.

\acknowledgments

NJM is part funded by a Leverhulme Trust Research Project Grant and
gratefully acknowledges their support. BD acknowledges funding from
the European Research Council (ERC) under the European Unions Seventh
Framework Programme (FP7/2007-2013) / ERC grant agreement no.  336792.
JM and IAB acknowledge the support of a Met Office Academic
Partnership secondment. Material produced using Met Office
Software. This work used the DiRAC Complexity system, operated by the
University of Leicester IT Services, which forms part of the STFC
DiRAC HPC Facility (www.dirac.ac.uk ). This equipment is funded by BIS
National E-Infrastructure capital grant ST/K000373/1 and STFC DiRAC
Operations grant ST/K0003259/1. DiRAC is part of the National
E-Infrastructure. This research made use of the ISCA High Performance
Computing Service at the University of Exeter. We acknowledge the
comments from the anonymous reviewer which improved this manuscript.



\appendix

\section{Radiative Transfer}\label{app_sec:radiative}

\begin{figure*}
\begin{center}
  \subfigure[Radiative Transfer Prim: 400-500\,days]{\includegraphics[width=8.5cm,angle=0.0,origin=c]{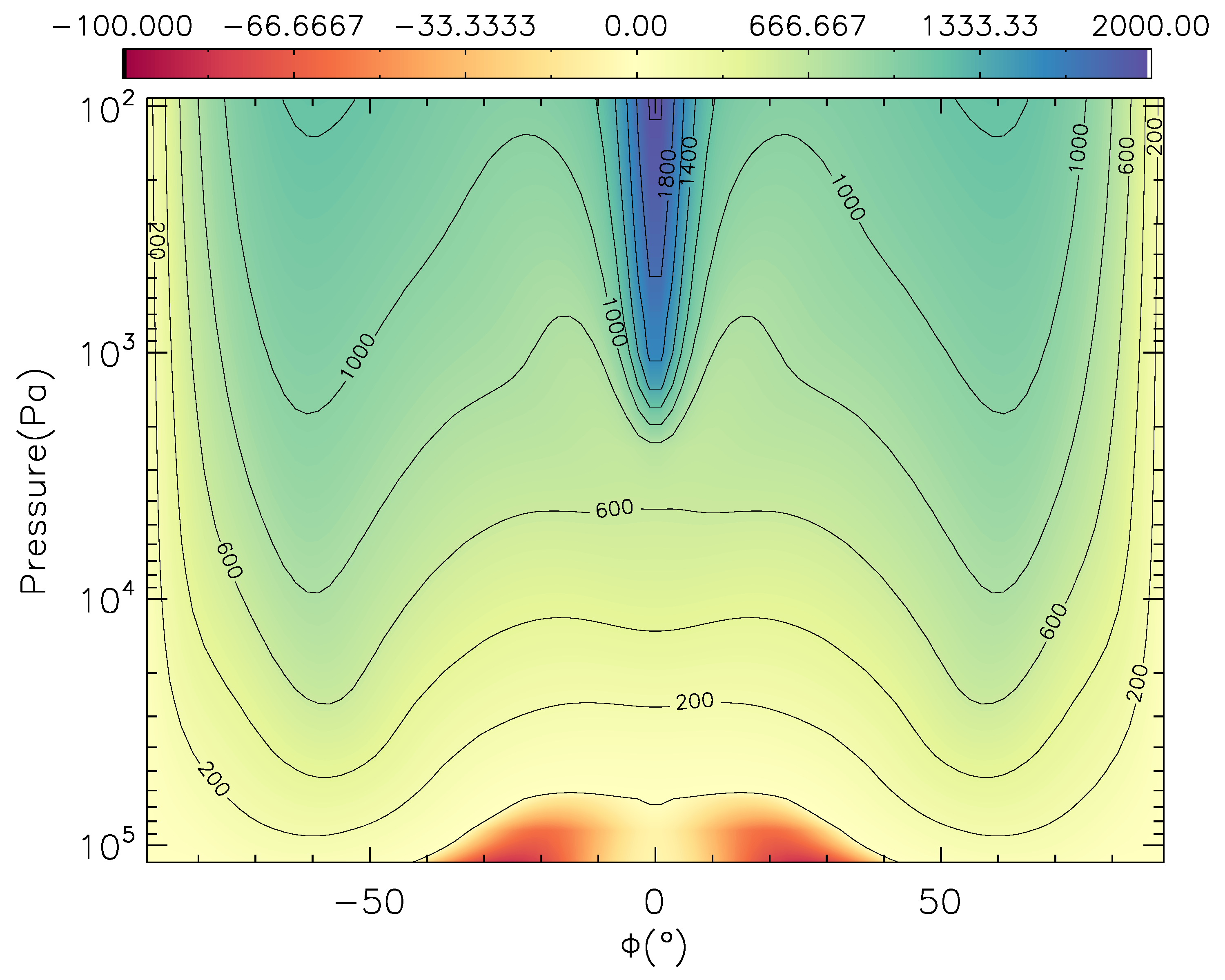}\label{rt_prim_400_500_uvel_bar}}
  \subfigure[Radiative Transfer Full: 400-500\,days]{\includegraphics[width=8.5cm,angle=0.0,origin=c]{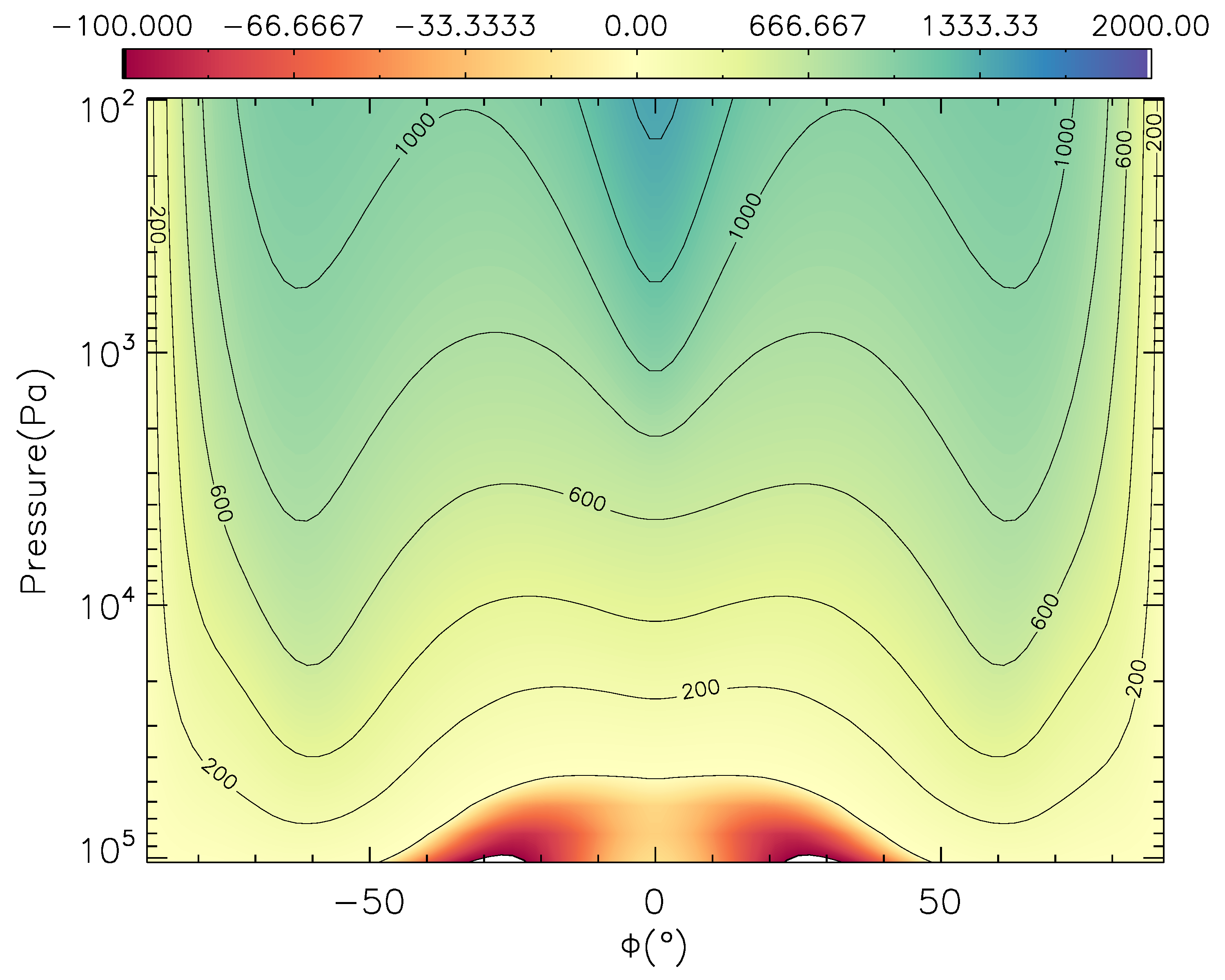}\label{rt_full_400_500_uvel_bar}}
\end{center}
\caption{Figure similar to Figure \ref{std_uvel_bar} for simulations
  matching the Std Prim and Std Full but including full radiative
  transfer \citep[as described in][]{drummond_2018} as opposed to
  temperature forcing (see Table \ref{model_names} for explanation of
  simulation names). Note the shorter simulation time and averaging
  period. \label{rt_uvel_bar}}
\end{figure*}

All of our simulations analysed in this work, and listed in Table
\ref{model_names} are performed using temperature forcing
\citep[see][for implementation]{mayne_2014}. Such temperature forcing
schemes have proved exceedingly useful in analysing the flow regimes
of planetary atmospheres, and are extremely computationally
efficient. However, this approach is certainly less physically
accurate than the use of more complete radiative transfer scheme
\citep[see discussion
in][]{showman_2009,amundsen_2014,amundsen_2016}. To ensure that our
conclusions are not affected by the use of a simplified treatment of
the heating, we have run two additional simulations for a shorter
period of 500\,days employing the full radiative transfer scheme
detailed in \citet{amundsen_2014,amundsen_2016}. These simulations are
setup to match the Std Prim and Std Full simulations, with the
complete setup (including radiative transfer elements) detailed in
\citet{drummond_2018}. The resulting temporal and zonal mean of the
zonal wind for both these simulations are shown in Figure
\ref{rt_uvel_bar}. 

As shown in Figure \ref{rt_uvel_bar}, the difference between the
primitive and full equations is also apparent when using a radiative
transfer scheme, and qualitatively similar to that found in our
temperature forced setups (see Figure \ref{std_uvel_bar}). A full
analysis of the radiative transfer simulations is beyond the scope of
this paper, and we reserve this for an upcoming work which will focus
on the implications for the atmospheric chemistry and observations
(requiring radiative transfer) of our findings.




\section{Verification of Approximations}
\label{app:equations}

\begin{figure*}[ht!]
\subfigure[Std Full]{
  \includegraphics[width=8.5cm]{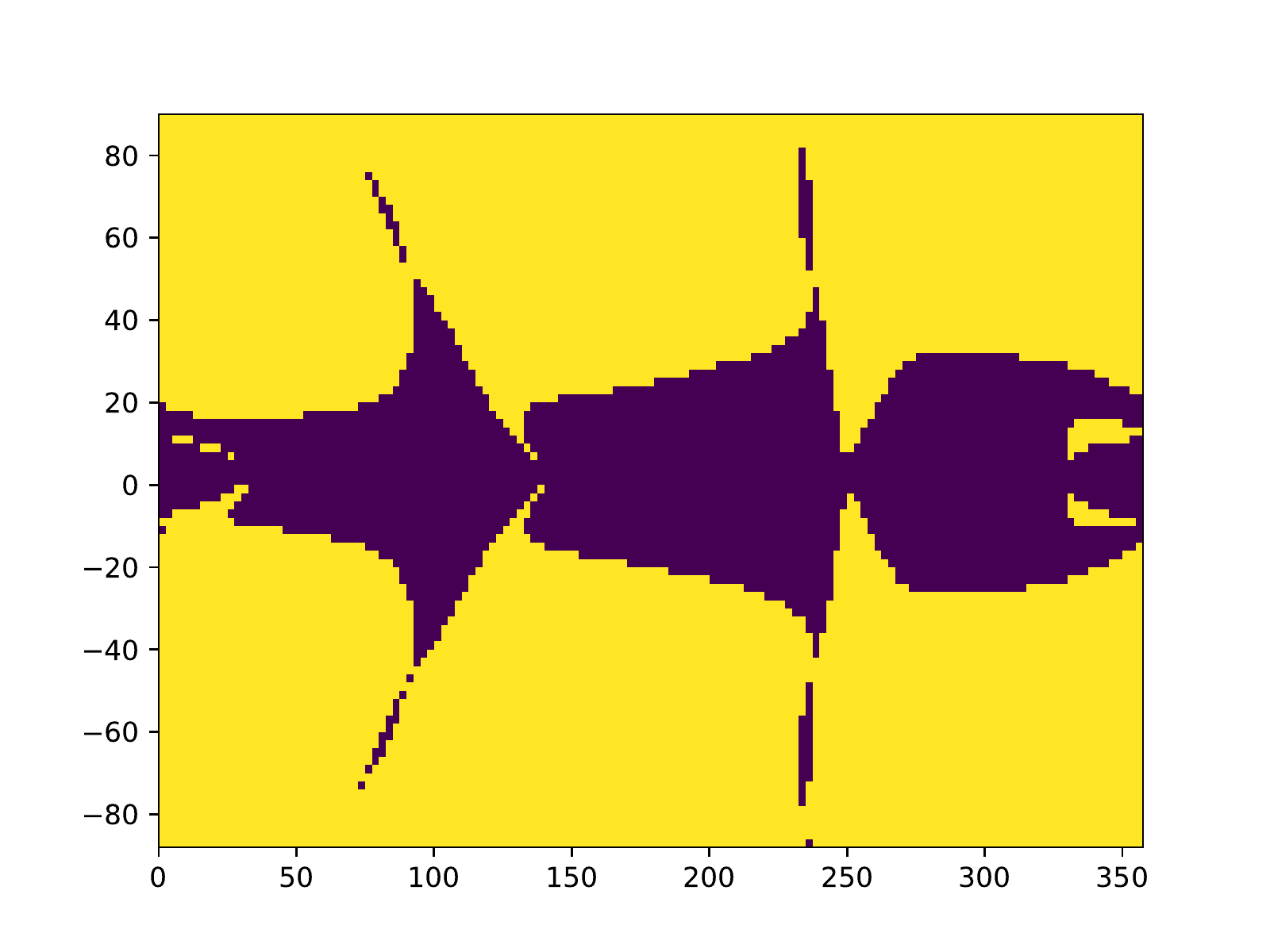}\label{fig:break_std}}
  \subfigure[$R_{\rm p}+$ Full]{
  \includegraphics[width=8.5cm]{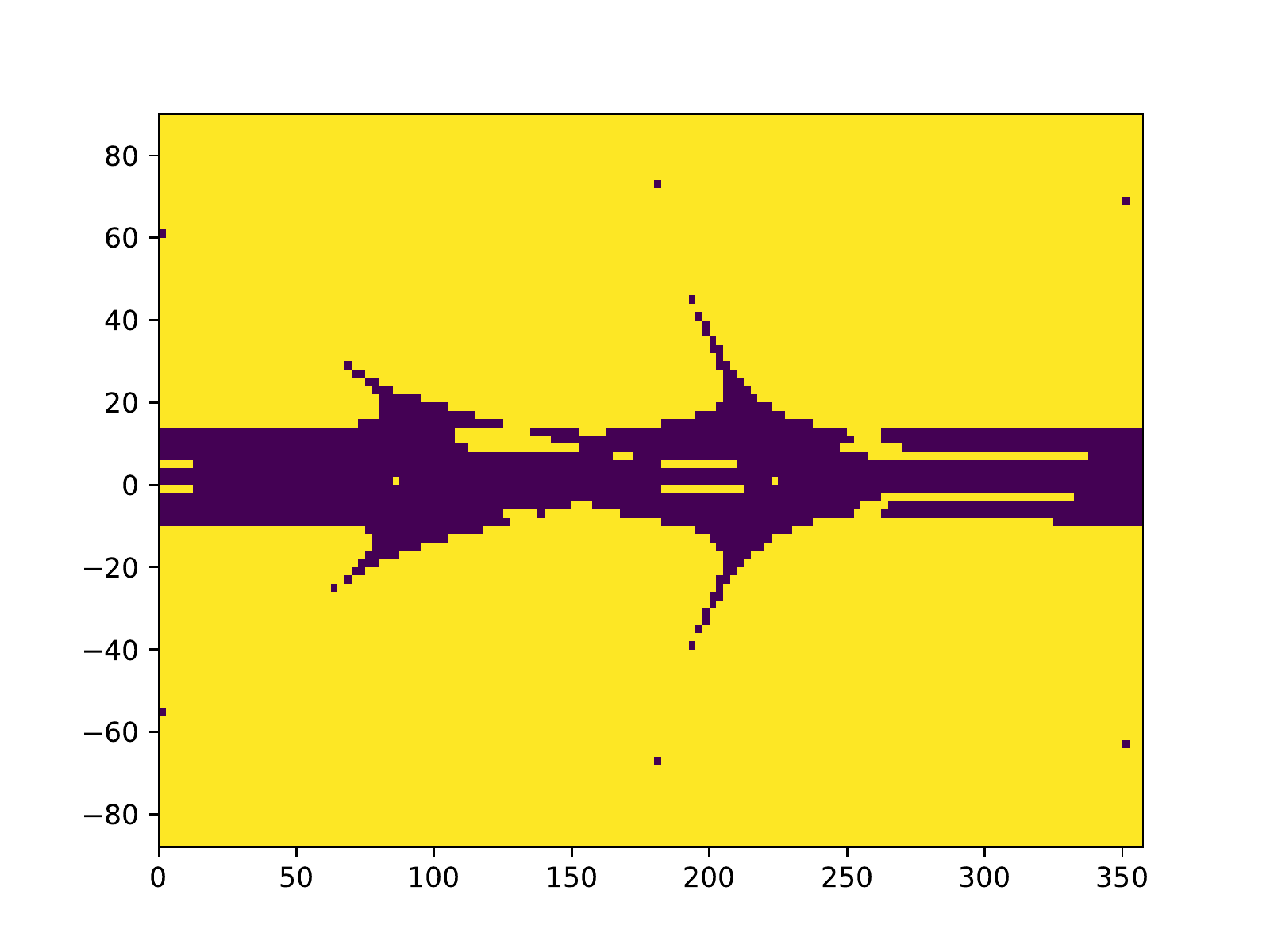}
  \label{fig:break_Rp}}
  \subfigure[dT$+$ Full]{
  \includegraphics[width=8.5cm]{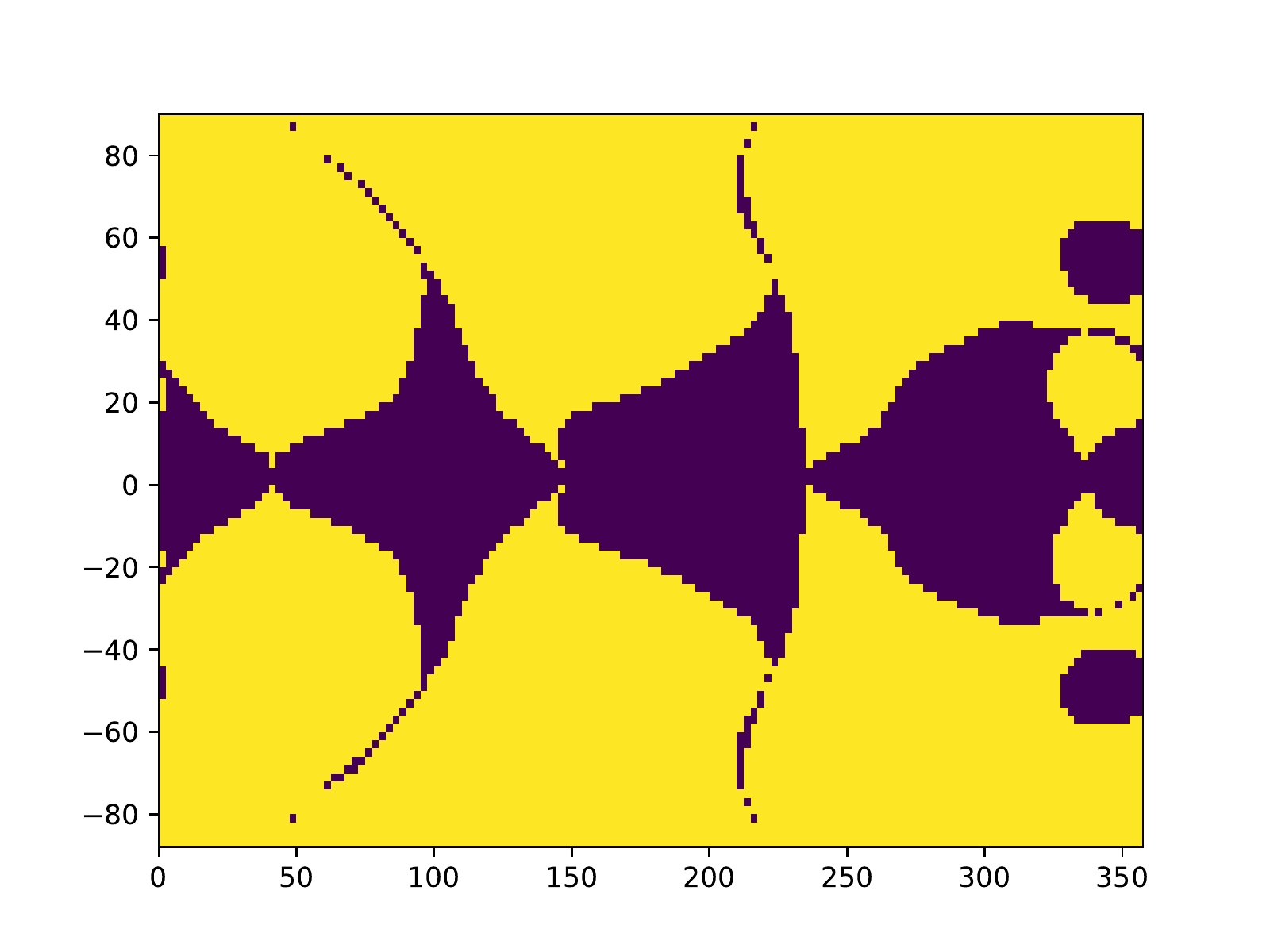}
  \label{fig:break_dt}}
 \subfigure[$\Omega+$ Full]{
  \includegraphics[width=8.5cm]{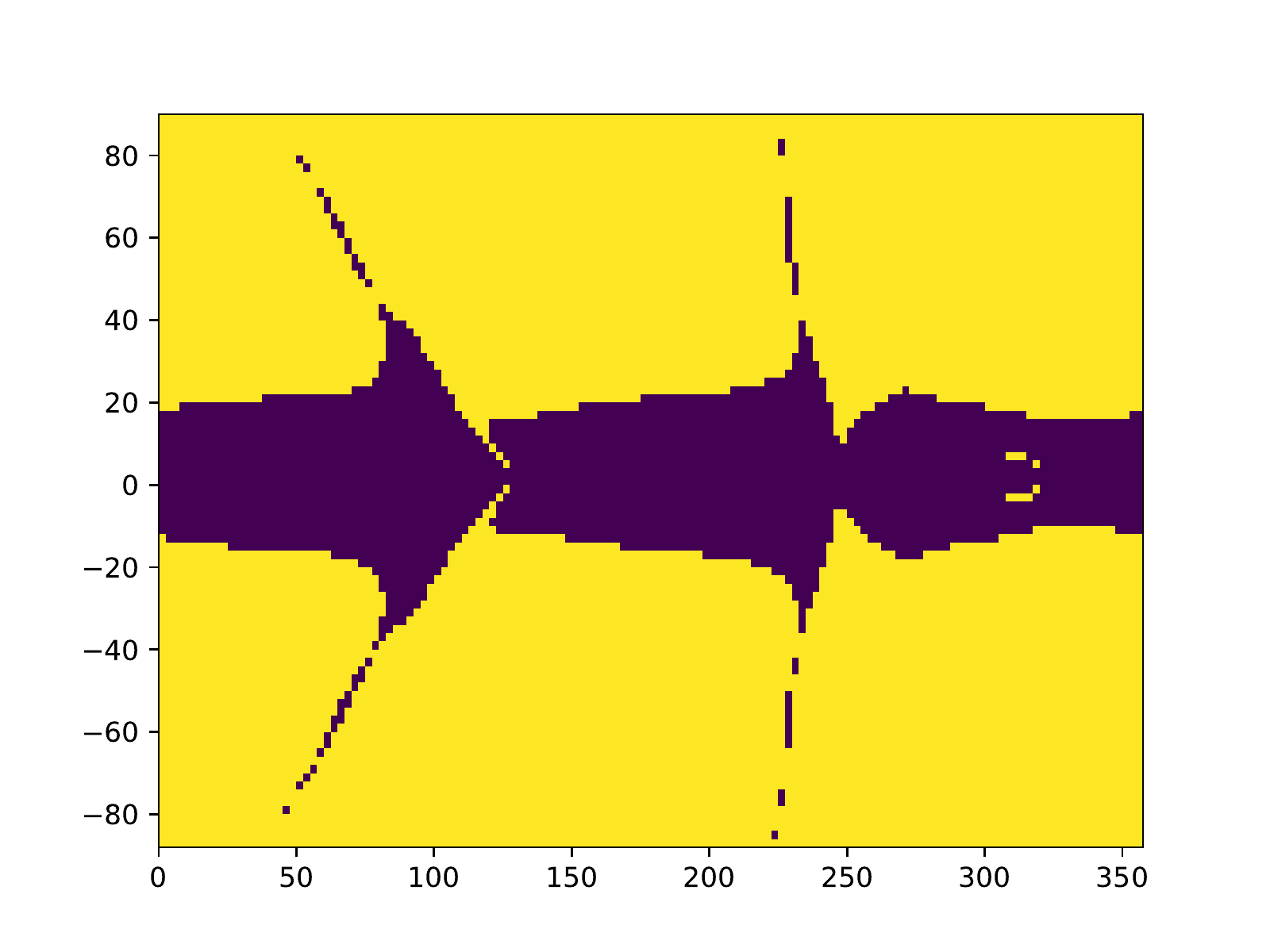}
  \label{fig:break_omega}}
\caption{Figure showing the sign of $\frac{v \tan \phi}{10 w} -1$ as a function
of longitude and latitude at a height corresponding to an equatorial
pressure of $\sim$100 Pa for four simulations, after 1\,000\,days. Yellow and purple regions show positive ($w \ll v \tan \phi$) and negative ($w \gtrsim v \tan \phi$) values.}
\label{fig:break}
\end{figure*}

As discussed in Section \ref{subsubsec:trad_approx} the traditional
approximation, invoked to derive the primitive equations, requires
$w \ll v \tan \phi$ to be valid. Also as discussed in Section
\ref{subsubsec:trad_approx} this condition can not be met at the
equator, and a critical or limiting latitude for the condition can be
defined $\phi_c$. Where $|\phi| < \phi_c$, $w \gtrsim v \tan \phi$,
and $w \ll v \tan \phi$ elsewhere. Clearly, the smaller $\phi_c$ the
more reliable the flow resolved using the primitive equations
becomes. Our estimates from Eq.\eqref{eq:estimate_v} and
\eqref{eq:estimate_w} allow us to understand the parameters
controlling the value of $\phi_c$: the larger $W$ compared to
$V \tan \phi$, the larger $\phi_c$ is expected to be. In Figure
\ref{fig:break} we verify our estimations by showing the sign of
$\frac{v \tan \phi}{10 w} -1$ at $p \approx 100 Pa$, where yellow and
purple denote regions of $v \tan \phi \gg w$ and
$v \tan \phi \lesssim w$, respectively. Figure \ref{fig:break}
presents values for four of our simulations solving the \emph{full}
equations of motion, namely the Std Full, $R_{\rm p}+$ Full, dT$+$
Full and $\Omega+$ Full simulations (see Table \ref{model_names} for
explanation) after 1\,000\,days.

Figure \ref{fig:break} shows a broad region extending to $40$ degrees in latitude where $v \tan \phi \gg w$ for the Std Full and dT$+$ Full simulations, where the flow will be resolved differently in the full and primitive cases. However, as expected, when the rotation rate is increased ($\Omega+$ Full) this region is reduced somewhat as the rotational terms become stronger. Finally, the $R_{\rm p}$ simulation results in a significantly reduced region violating the $w \ll v \tan \phi$ consistent with our results, and interpretation. 

\bibliography{references}



\end{document}